# Interplay of polar order and positional order in liquid crystals – observation of re-entrant ferroelectric nematic phase

Grant Strachan[a], Shona Ramsay[b], Marijus Juodka[b], Damian Pociecha[a], Jadwiga Szydlowska[a], John M.D. Storey[b], Natasa Vaupotic[c], Rebecca Walker[b], Ewa Gorecka[a]

[a] Grant Strachan, Jadwiga Szydlowska, Damian Pociecha*, Ewa Gorecka
Faculty of Chemistry
University of Warsaw
Zwirki i Wigury 101, Warsaw 02–089, Poland

E-mail: pociu@chem.uw.edu.pl

[b] Shona Ramsay, Marijus Juodka, John M.D. Storey, Rebecca Walker
School of Natural and Computing Sciences, University of Aberdeen, Aberdeen, Great Britain

[c] Natasa Vaupotic
Faculty of Natural Sciences and Mathematics, University of Maribor, Maribor, Slovenia; Jozef Stefan Institute, Ljubljana, Slovenia

Supporting information for this article is given at the end of the document.

**Abstract:** We show that development of polar order may spontaneously destroy the lamellar structure of a liquid crystal. This results in an unusual sequence of phases with the ferroelectric nematic phase appearing below a non-polar smectic phase. The effect is related to unfavourable dipole interactions within the smectic layers and can be explained by Landau theory in which the temperature dependent term is non-monotonic as it is renormalized by spontaneous electric polarization.

Re-entrant behaviour refers to a situation where a system transitions from one phase to another, and upon further variation of an external parameter - typically temperature - it reverts back to the original phase. The re-entrant phase phenomenon is unusual behaviour, as in most systems the degree of order increases monotonically with decreasing temperature. It is driven by a competition between ordering forces (enthalpic contributions) and disordering tendencies (entropic effects). In liquid crystals (LCs), such competing effects may arise from dipolar interactions, steric hindrance, or molecular geometry. Re-entrancy has been often reported in systems composed of rod-like molecules with strongly interacting terminal groups and attributed to competing forces that favour bilayer packing in the higher temperature smectic phase and monolayer packing at lower temperatures, with a nematic phase obtained in a temperature range between these smectics. [1-3] The phenomenon has also been observed in materials with mesogenic cores that deviate from typical rod- or disc-like shapes. [4,5] In these cases, packing frustration may destabilize ordered phases, leading the system to revert to a less ordered (e.g., nematic or even isotropic) phase upon cooling, before going again to a more ordered structure. Other re-entrant examples are related to the reappearance of a non-tilted smectic phase below the tilted one, [6] or ferroelectric order below the phase with antiferroelectric order.[7] Here we studied re-entrant phenomena for new class of mesogens able to form proper ferroelectric phases. In contrast to traditional ferroelectric LCs—where chirality or bent molecular shape is essential—the proper ferroelectric systems derive their polar properties directly from the dipole-dipole interactions. Liquid crystals with proper ferroelectric order display spontaneous polarisation equalling that of crystalline materials, in combination with fluidity, and varying degrees of symmetry depending on the type of LC phase. The first example of these, the ferroelectric nematic ($N_F$) phase, was discovered in 2017,[8-10] and since then the realm of proper polar LC phases has expanded to include orthogonal and tilted layered phases ($SmA_F$ [11-13] $SmA_{AF}$ [14] and $SmC_F$ [15,16]), heliconical phases ($N_{TBF}$ [17,18], $SmC^P_H$ [19,20]) and the antiferroelectric nematic phase ($N_X$/$M_{AF}$/$SmZ_A$ [21,22]).

We have studied two homologous series of compounds featuring a highly fluorinated mesogenic core and thus having strong dipole moment along the molecular long axis (~11D): the *SR-n-Re series*, terminated with a dioxane unit, and the *GS-n-Re* series, terminated with a phenyl group. In both series *n* stands for the number of carbon atoms in the terminal chain. Both series show similar trends — the behaviour of short homologues is dominated by broad temperature range of nematic phases and the longest homologues form smectic phases (Figure 1). As the terminal chain length increases, the temperature at which lamellar order appears rises. This behaviour is typical: elongation of the terminal chain increases the tendency for self-segregation between alkyl chains and mesogenic cores, promoting the formation of a lamellar structure. However, for the studied materials the lamellar structure appears for longer homologues than in other series of similar mesogens, suggesting that propensity for lamellar structure is rather weak.[18] Simultaneously, elongation of the terminal chain decreases the onset temperature of the polar order, that is attributed to the increasing longitudinal distance between dipoles, which are mainly located in the mesogenic cores. What is exceptional for both series is that the appearance of polar order seems to weaken the lamellar structure, as a result for intermediate homologues—specifically *SR-6-Re, SR-7-Re* and *GS-5-Re* - the unusual phase sequence is found with a re-appearance of a nematic phase below a smectic phase. On cooling, these homologues undergo a transition from the isotropic liquid to a non-polar nematic phase, followed by the formation of an orthogonal non-polar smectic A (SmA) phase; a nematic phase reappears at lower temperatures, but now with ferroelectric order (reN$_F$), which in turn upon further cooling is followed by transition to a tilted ferroelectric smectic phase (SmC$_F$). Apparently in these systems, building up of polar order can destabilize the lamellar structure, giving rise to a polar nematic phase below an apolar smectic phase.

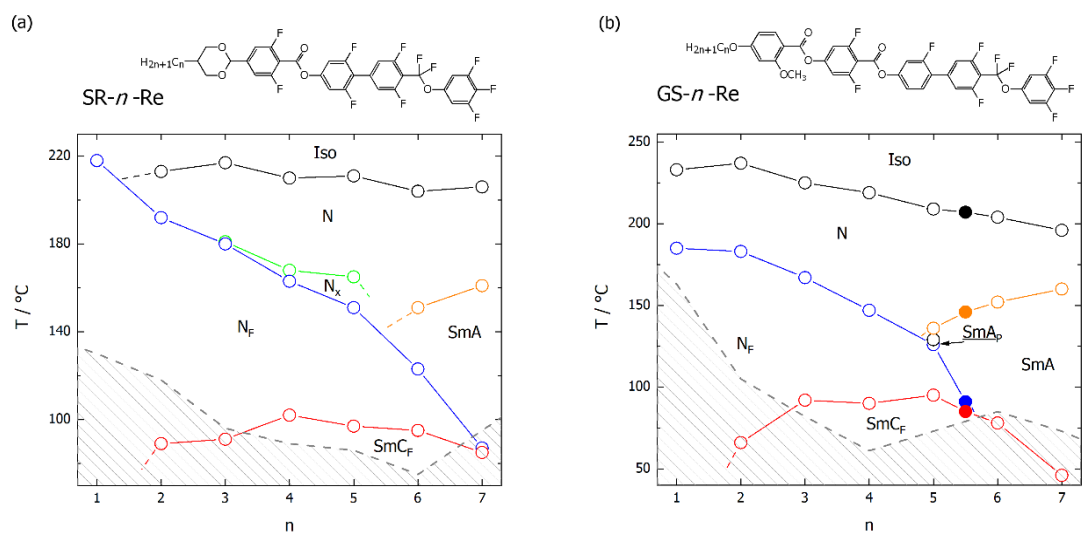

**Figure 1.** Phase diagrams for *SR-n-Re* and *GS-n-Re* homologue series, with their general molecular structure. The transition temperatures for the mixture of homologues *GS-5-Re* and *GS-6-Re* were marked by the filled points.

The X-ray scattering and optical studies support the assigned phase sequences (Figure 2). For *SR-6-Re*, the re-entrant N$_F$ phase appears between two smectic phases; in the smectic phases the diffraction signal related to the layer thickness is of instrumental resolution. In the reN$_F$ phase as well as in N phase it broadens and the corresponding longitudinal correlation length deduced from the signal width is of order of molecular length, 2-3 nm. In the SmA phase the position of the signal is constant and corresponds to the molecular length, at the transition to the re-entrant nematic phase the position of this signal, that reflects the longitudinal distance between molecules, moves to higher angles. In the SmC$_F$ phase the layer spacing is smaller than in the SmA phase due to the molecular tilt. In *GS-5-Re* the temperature range of the smectic phase separating N and reN$_F$ phases is considerably narrower, and the changes of X-ray diffraction signal width are more gradual on approaching the smectic phase from the nematic phases, above and below. Regarding optical studies, in a cell (parallel rubbing of polymer at both surfaces), the director is perfectly oriented along the rubbing direction in nematic phase and the transition to the smectic phase is observed only as suppression of the Brownian motions of the director (Figure S1 and Figure S2). At the transition to the SmC$_F$ phase, the sample loses alignment, and a non-characteristic stripy texture is formed. Optical birefringence measurements show small steps at both N-SmA and at SmA-reN$_F$ phase transitions, reflecting the small increase of orientation order associated with both lamellar and polar alignment of molecules (Figure S3). In *GS-5-Re*, two distinct thermal events in Δn vs. T were detected (Figure S3b), preceding the reN$_F$ phase, suggesting the existence of two smectic phases between reN$_F$ and N phases, the upper temperature one being apolar SmA phase, while the exact nature of lower temperature SmA phase (ferroelectric or possibly antiferroelectric) is difficult to resolve due to the narrow temperature window of the phase.

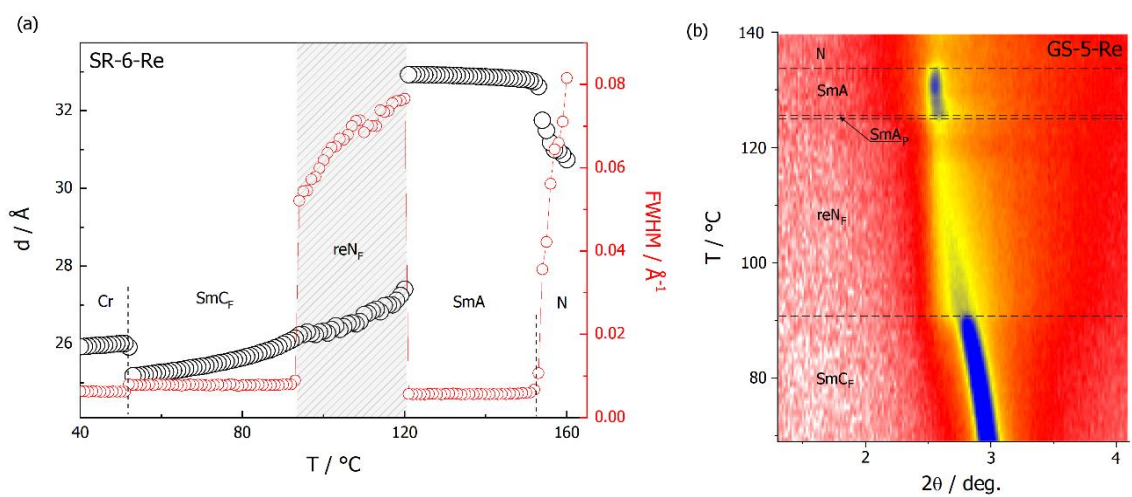

**Figure 2.** (a) The layer spacing (in smectic) or longitudinal distance between molecules (in nematic) deduced from the position of low angle X-ray diffraction signal (black circles) and the full width at half maximum of diffraction signal, reflecting the positional correlation length (red squares) for *SR-6-Re* and (b) the 2D plot of low angle diffraction signal intensity vs. scattering angle and temperature for *GS-5-Re*.

The polar character of the re-entrant N$_F$ phase is confirmed by its response to an electric field, a clear polarization switching current peak is observed under application of a triangular voltage. The determined value of spontaneous polarization (Figure 3) indicates nearly complete alignment of molecular dipoles in the reN$_F$ phase. In *SR-6-Re* the evolution of polarization in the reN$_F$ phase on approaching the SmA phase suggests discontinuous changes characteristic for first order phase transition, while for its shorter homologue *SR-4-Re* as well as for *GS-5-Re* the polarization in the N$_F$ phase gradually decreases on approaching the non-polar phase, suggesting nearly continuous phase transition. The reN$_F$ phase shows a strong second harmonic generation (SHG) activity, which confirms the non-centrosymmetric, polar nature of the phase (Figure S4). Also, a dramatic increase in the dielectric response upon entering reN$_F$ from SmA phase is consistent with ferroelectric character of reN$_F$ phase (Figure S5).

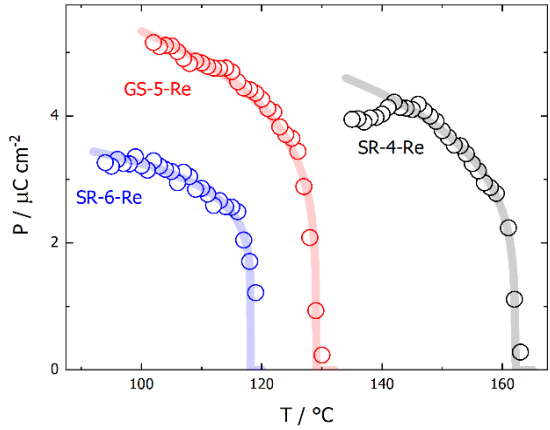

**Figure 3.** Spontaneous polarization vs. temperature for *SR-4-Re*, *SR-6-Re* and *GS-5-Re*. In the inset the SHG image in reN$_F$ and SmC$_F$ phases.

Further evidence of polar order in reN$_F$ phase comes from the observation of optical textures (Figure 4). Although cells with planar anchoring show nearly ideal alignment of the director a few air bubbles can be produced as the cell is filled with LC material. The observation of director field around such bubbles allows us to follow the changes in the polar character of the phases. In the upper temperature, non-polar nematic phase the director is radially aligned around air bubbles, whereas in the ferroelectric reN$_F$ phase, it adopts a tangential configuration. The latter is a characteristic behaviour of polar fluids that minimizes surface-bound charges. Interestingly, in *GS-5-Re*, an asymmetric defect pattern, tangential on one side and radial on the other, appears in reN$_F$ near the transition to the smectic phase, suggesting that the smectic phase directly above the reN$_F$ phase might be antiferroelectric.

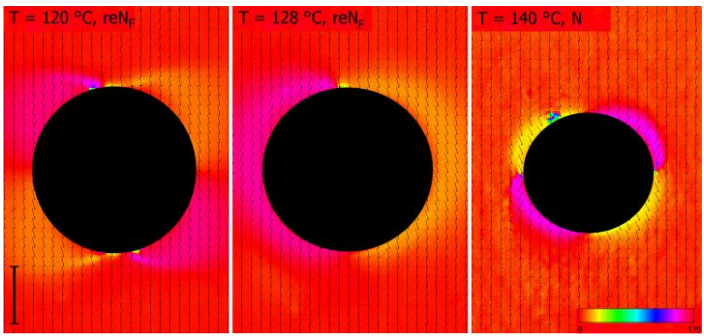

**Figure 4.** PolScope textures showing director field in N and reN$_F$ phases around air bubble for compound *GS-5-Re* observed in the cell with planar anchoring and rubbing parallel on both surfaces. Colour scale codes azimuthal orientation of the director and a scale bar, 20 μm, is placed along the rubbing direction.

These experimental findings highlight a unique interplay between polar order and positional order in liquid crystalline phases exhibiting proper ferroelectric behaviour. We suggest that in such systems there are competing tendencies for molecular arrangement: dipole-dipole interactions favour ferroelectric order along director, while within well-defined smectic layers the antiparallel packing of dipoles is preferred. Thus, upon the onset of the polar phase the system may prefer to avoid a layer structure, favouring instead a nematic phase, in which 'transverse interactions' between molecular dipoles are less energetically unfavourable (Figure 5a).

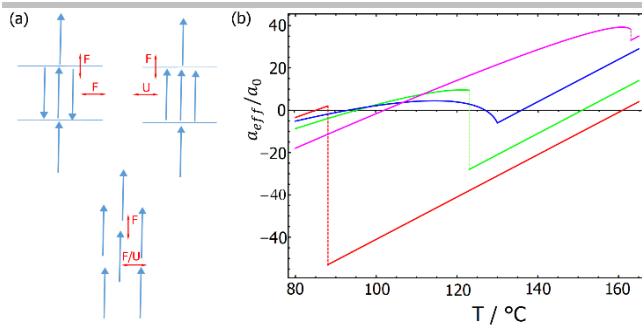

**Figure 5.** (a) Considering nearest neighbour interactions of longitudinal molecular dipole moments, an antiparallel orientation of the neighbouring molecules within the smectic layer results in their favourable interactions (F), while the parallel alignment leads to unfavourable (U) interactions. Parallel arrangement of molecular dipoles in successive layers is always preferred. A displacement of the neighbouring dipoles in a nematic phase makes the transverse interactions between parallel dipoles less unfavourable. (b) The temperature dependent Landau parameter a$_{eff}$ for material *SR-4-Re* (magenta), *SR-6-Re* (green), *SR-7-Re* (red), and *GS-5-Re* (blue), materials. In the range in which a$_{eff}$ parameter is positive the nematic phase becomes more stable than the smectic one.

To describe the observed phase transitions, we apply a model analogous to the Landau theory of the re-entrant nematic – smectic A phase transition, developed by Pershan and Prost. [23] Assuming a second order phase transition from the nematic to smectic phase, the energy density difference between the smectic and nematic phase, $f_S$, is expressed as a power series in the smectic order parameter, $\psi$, as

$$f_S = \frac{1}{2} a |\psi|^2 + \frac{1}{4} b |\psi|^4 , \qquad (1)$$

where $a$ and $b$ are Landau parameters. Parameter $a = a_0(T - T_{NS})$ is temperature dependent and changes sign at $T_{NS}$, the transition temperature to the smectic phase. We assume that the smectic order parameter couples to polarisation ($P$) at temperatures $T < T_P$, where $T_P$ is the temperature at which polar order becomes favourable. Following Pershan and Prost [23], the simplest approach is to add an additional term of the form $g(P)|\psi|^2$ to the free energy density. If a term $\frac{1}{2} g_2 P^2 |\psi|^2$ is added to eq. (1), the parameter $a$ changes to $a_{eff}$

$$a_{eff} = a_0 \left( T - T_{NS} + \frac{g_2}{a_0} P(T)^2 \right) . \qquad (2)$$

With $g_2 > 0$, the transition temperature to the smectic phase will be shifted to some lower value $T_{PS}$ given by

$$T_{PS} = T_{NS} - \frac{g_2 P(T_{PS})^2}{a_0}. \qquad (3)$$

For the second order phase transition from the nematic to the smectic phase, the sign of $a_{eff}$ defines whether the system is in the nematic ($a_{eff} > 0$) or smectic phase ($a_{eff} < 0$). At the onset of the polar order, parameter $a_{eff}$ can jump from a negative to a positive value if polarization is high enough. To have a discontinuous change in $a_{eff}$ the phase transition to the polar phase needs to be first order. We can express the free energy density $f_P$ related to the polar order as

$$f_P = \frac{1}{2} a_P P^2 + \frac{1}{4} b_P P^4 + \frac{1}{6} c_P P^6 , \qquad (4)$$

where $P$ is polarization and $a_P$, $b_P$ and $c_P$ are Landau parameters. Parameter $a_P$ changes sign at temperature $T^*$, so we express it as $a_P = a_{P0}(T - T^*)$. For the first order phase transition $b_P < 0$ and $c_P$ is positive. Neglecting the term with polarization in $a_{eff}$ in the smectic free energy density, the phase transition to the polar phase is found by setting $f_P = 0$ at $P \neq 0$ and one finds the transition temperature $T_P$:

$$T_P = T^* + \frac{3 b_P^2}{16 a_{P0} c_P} . \qquad (5)$$

At this temperature, polarization, which is obtained by satisfying the condition $\partial f_P / \partial P = 0$, is

$$P^2(T_P) = \frac{3|b_P|}{4c_P} \quad . \tag{6}$$

With further lowering of the temperature, polarization increases as

$$P^2 = \frac{1}{2c_P}\left(|b_P| + \sqrt{\frac{b_P^2}{4} + 4c_P a_{P0}(T_P - T)}\right) \quad . \tag{7}$$

At temperature $T_P$, the parameter $a = a_0(T_P - T_{NS})$ is negative. At the onset of polar order it jumps to the value $a_{eff}(T_P) = a_0\left(T_P - T_{NS} + \frac{g_2}{a_0}P(T_P)^2\right)$, which can be either negative or positive, depending on the magnitude of polarization (given by eq. (6)). If $a_{eff}(T_P) > 0$, then the smectic phase is not stable anymore and one observes a phase transition to the polar nematic phase, N$_F$. With further lowering of the temperature polarization increases as given by eq.(7), however the increase is slower than the decrease of the term $(T - T_{NS})$ in the expression of $a_{eff}$, so eventually $a_{eff}$ becomes negative, allowing transition back to the smectic phase.

If $a_{eff}(T_P) < 0$, then a transition to a polar smectic phase is observed. With further reduction of temperature, there are two possibilities. Even though $a_{eff}$ approaches 0 with decreasing temperature, it might not reach it, so there is no formation of a nematic phase. On the other hand, if it becomes positive, one obtains a reentrant (polar) nematic phase below the polar smectic phase.

If $b_P$ in eq. (4) is positive, the phase transition to the polar phase is of the second order. In this case transition to the polar phase occurs at temperature $T^*$ and with reduction of temperature polarization increases as

$$P^2 = \frac{1}{2c_P}\left(-b_P + \sqrt{b_P^2 + 4c_P a_{P0}(T^* - T)}\right) \quad . \tag{8}$$

At $T = T^*$, $a_{eff}$ is still negative, but it increases towards zero with decreasing temperature. At temperature $T_{rN}$, where the condition

$$T_{rN} - T_{NS} + \frac{g_2}{a_0}P(T_{rN})^2 = 0 \tag{9}$$

is fulfilled, $a_{eff}$ becomes 0 and then positive with further lowering of temperature. This means that at temperatures $T < T_{rN}$ the smectic phase is not stable anymore and we have a transition to the (reentrant) nematic phase, which is polar. With further lowering of the temperature, the value $a_{eff}$ will eventually start to decrease and become 0 at temperature $T_{rS}$ given by

$$T_{rS} - T_{NS} + \frac{g_2}{a_0}P^2(T_{rS}) = 0 \tag{10}$$

So, at $T < T_{rS}$ one observes a restoration of polar smectic phase.

We estimated the magnitudes of Landau parameters for the studied materials. For this purpose, we take $P$ as a polar order parameter, with values between 0 and 1, defined as the ratio between the polarization at some temperature and maximum polarization. The phase behaviour of the *SR-n-Re* compounds can be explained by a weak first order phase transition to the polar phase and behaviour of material *GS-5-Re* by a second order phase transition. The temperature dependence of $a_{eff}/a_0$ for these materials is shown in Figure 5b. Not all the parameters can be calculated from the measured transition temperatures, so some values had to be estimated. For the *SR-n-Re* series, we assume that $P^2(T_P) = 0.075$, which means $c_P = 10b_P$. Because for homologues $n = 6$ and 7 a N → SmA → N$_F$ sequence is observed, $a_{eff}(T_P) > 0$ and $T_{NS} > T_P$. From this condition we find that $g_2 > 374a_0$ for $n = 6$ and $g_2 > 974a_0$ for $n = 7$. We chose $g_2 = 500a_0$ and $g_2 = 1000a_0$ for $n = 6$ and $n = 7$, respectively. At temperature of the re-entrant smectic phase $a_{eff}(T_{rS}) = 0$, which gives $a_{P0} = 1.15 \cdot 10^{-3}b_P$ for $n = 6$ and $a_{P0} = 1.7 \cdot 10^{-4}b_P$ for $n = 7$. To model the phase sequence of $n = 4$, we estimate $T_{NS}$ by a linear extrapolation of $T_{NS}$ for $n = 6$ and 7. For this material $T_P$ is the temperature at which a transition from N to N$_F$ is observed. At $T = T_P$, $a = a_0(T - T_{NS})$ is still positive, so there is a direct transition from the nematic to the ferroelectric nematic phase. We assume that $g_2/a_0$ also reduces approximately linearly with reducing $n$, and choose $g_2 = 70a_0$. By requiring $a_{eff}(T_{rS}) = 0$, we obtain $a_{P0} = 0.033b_P$. The transition temperatures taken to estimate the Landau parameters are given in Table S1.

For material *GS-5-Re*, we require $a_{eff}(T_{rN}) = 0$ and $a_{eff}(T_{rS}) = 0$. From these two conditions we find $a_{P0}c_P = 0.413b_P^2$. If we assume that $P^2(T_{rN}) = 0.1$, we find also $c_P = 0.014b_P$ and $a_{P0} = 30b_P$ and $g_2 = 0.17a_0$.

In summary, in the field of liquid crystals, re-entrant phenomena—where the nematic phase reappears below the smectic phase as the temperature is lowered—are not unique and have been observed in many systems with competing interactions. However, in the materials studied here, this behaviour is rather unusual, as the melting of smectic layers is induced by polar order. The re-entrant phenomenon can be explained within the framework of Landau theory, where the typical linear temperature dependence of the coefficient *a* is violated and becomes non-monotonic near the transition to the polar phase. This non-monotonicity is the reason for the re-emergence of nematic order.

**Acknowledgements**

The research was supported by the National Science Centre (Poland) under the grant no. 2021/43/B/ST5/00240

**Keywords:** proper ferroelectricity, re-entrant phenomena, nematic

# Supporting Information

# 1. Experimental methods

***Calorimetric studies*:** transition temperatures and the associated enthalpy changes were measured by differential scanning calorimetry using either TA DSC Q200 or Mettler Toledo DSC3 instrument. Measurements were performed under a nitrogen atmosphere with a heating/cooling rate of 5 or 10 K min$^{-1}$.

***Optical Studies*:** Observations of optical textures of liquid crystalline phases was carried out by polarised-light optical microscopy using a Zeiss AxioImager.A2m microscope equipped with a Linkam heating stage. For the determination of optical axis direction and birefringence variation in the sample PolScope Abrio system mounted on the Zeiss microscope was used. Optical birefringence as a function of temperature was measured with a setup based on a photoelastic modulator (PEM-90, Hinds) working at a modulation frequency f = 50 kHz; as a light source a halogen lamp (Hamamatsu LC8) equipped with narrow bandpass filters was used. The transmitted light intensity was monitored with a photodiode (FLC Electronics PIN-20) and the signal was deconvoluted with a lock-in amplifier (EG&G 7265) into 1f and 2f components to yield a retardation induced by the sample. Knowing the sample thickness, the retardation was recalculated into optical birefringence. Samples were prepared in 1.6-μm-thick cells with planar anchoring. The alignment quality was checked prior to measurement by inspection under the polarised-light optical microscope.

***X-ray diffraction studies:*** 2D-XRD patterns were registered using a Bruker D8 GADDS system, equipped with micro-focus-type X-ray source with Cu anode and dedicated optics and VANTEC2000 area detector. For precise measurements of layer structure small angle diffraction experiments were performed on a Bruker Nanostar system (IμS microfocus source with copper target, MRI heating stage, Vantec 2000 area detector).

***Second Harmonic Generation:*** The SHG response was investigated using a home-made microscopic setup based on a solid-state laser EKSPLA NL202. Laser pulses (9 ns) at a 10 Hz repetition rate and max. 2 mJ pulse energy at λ=1064 nm were applied. The pulse energy was adjusted for each sample to avoid its decomposition. The infra-red beam was incident onto cells with planar anchoring at both surfaces, the cell thickness was ~5 μm. An IR pass filter was placed at the entrance to the sample and a green pass filter at the exit. For controlling temperature Linkam stage was used. Optical SHG images were recorded with homemade microscopic setup.

***Spontaneous Polarization Measurements:*** spontaneous electric polarisation was determined by integration of the current peaks recorded during polarization switching upon applying a triangular-wave voltage (60Hz, 200V$_{pp}$). The repolarization current was measured by monitoring the voltage drop on the 100 kOhm resistor in parallel connection with the LC cell. The cell with in-plane electric field was used, the ITO electrodes were deposited on one surface, the distance between electrodes was 2 millimetres. The thin polymer layers deposited on both surfaces ensured planar alignments of molecules. The cell thickness was 4 µm. The obtained value of the polarization should be treated with some caution as the electric field is not ideally uniform between electrodes, and the switching occurs also on the electrodes.

***Dielectric spectroscopy***: The complex dielectric permittivity was measured in the 1 Hz–10 MHz frequency range using a Solartron 1260 impedance analyzer. The material was placed in 10-µm-thick glass cell with gold electrodes (without surfactant to avoid the influence of the high capacitance of a thin polymer layer). The amplitude of the measuring ac voltage, 20 mV, was low enough to avoid Fréedericksz transition in ferroelectric nematic phases.

## 2. Additional results

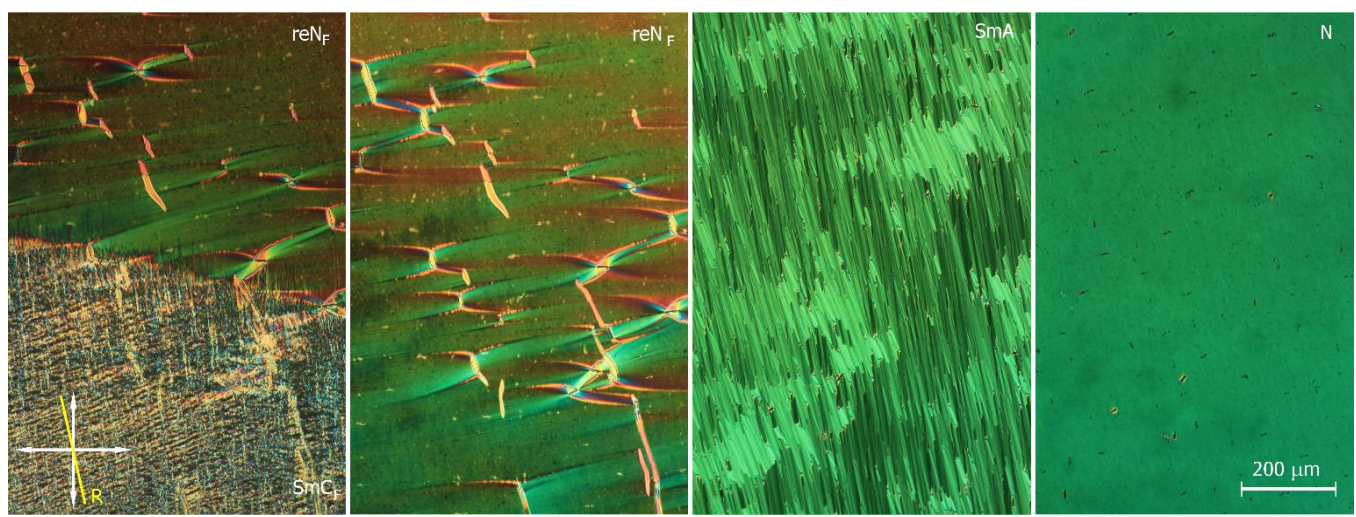

*Figure S1: Textures of mesophases for SR-6-Re observed in 5-µm-thick cell with planar anchoring (rubbed at both surfaces unidirectionally). Polarizer and analyser were crossed (white arrows) while the rubbing direction (yellow line) was slightly inclined to polarizers. Note, that in left panel the transition from $SmC_F$ to $reN_F$ phase is observed.*

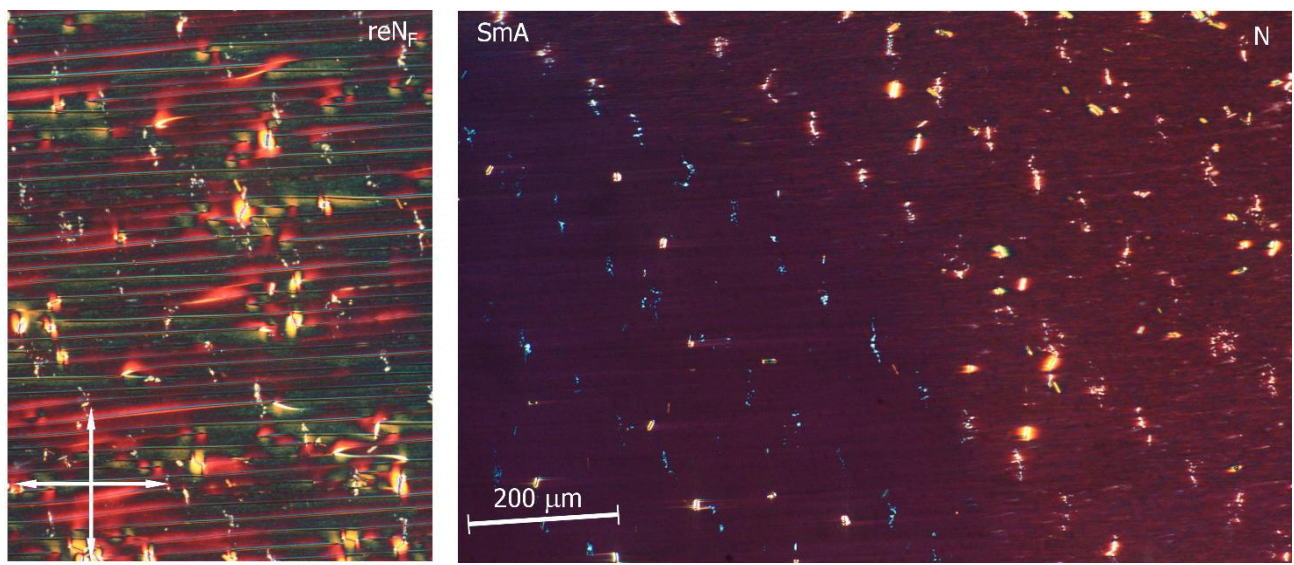

*Figure S2: Textures of mesophases for GS-5-Re observed in 3-µm-thick cell with planar anchoring (rubbed at both surfaces unidirectionally). Right panel shows transition from N to SmA phase. Polarizer and analyser were crossed (white arrows), scale bar shows rubbing direction while the rubbing direction that was slightly inclined to polarizers.*

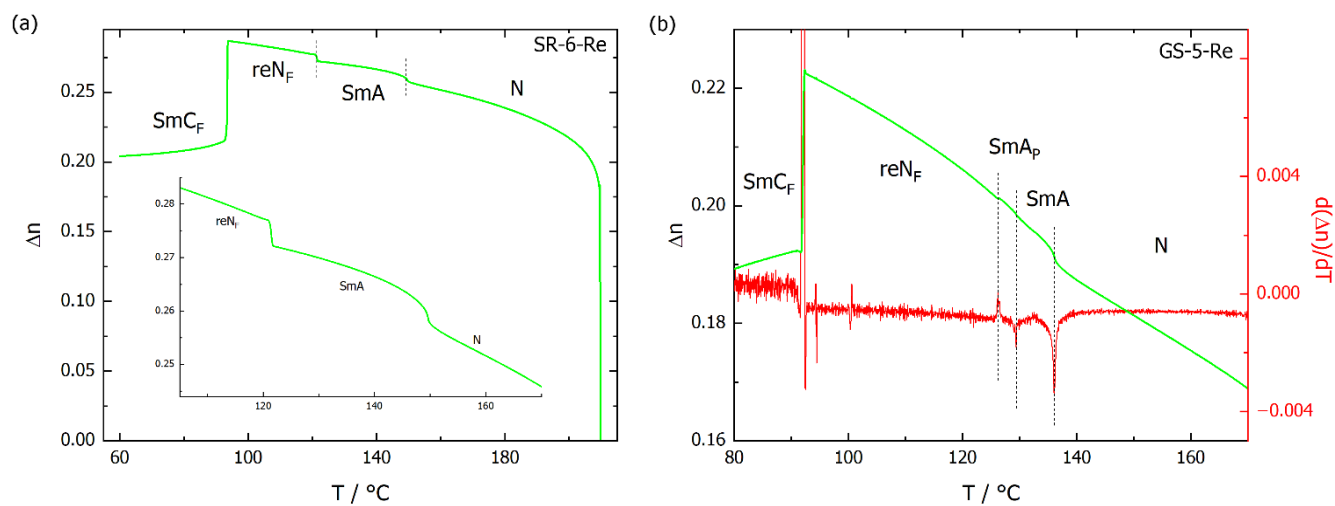

*Figure S3: Temperature dependence of optical birefringence (green line) measured for green light (λ=532 nm) for compounds (a) SR-6-Re and (b) GS-5-Re. For GS-5-Re the derivative of Δn(T) is also shown (red line) to indicate the phase transition temperatures more clearly.*

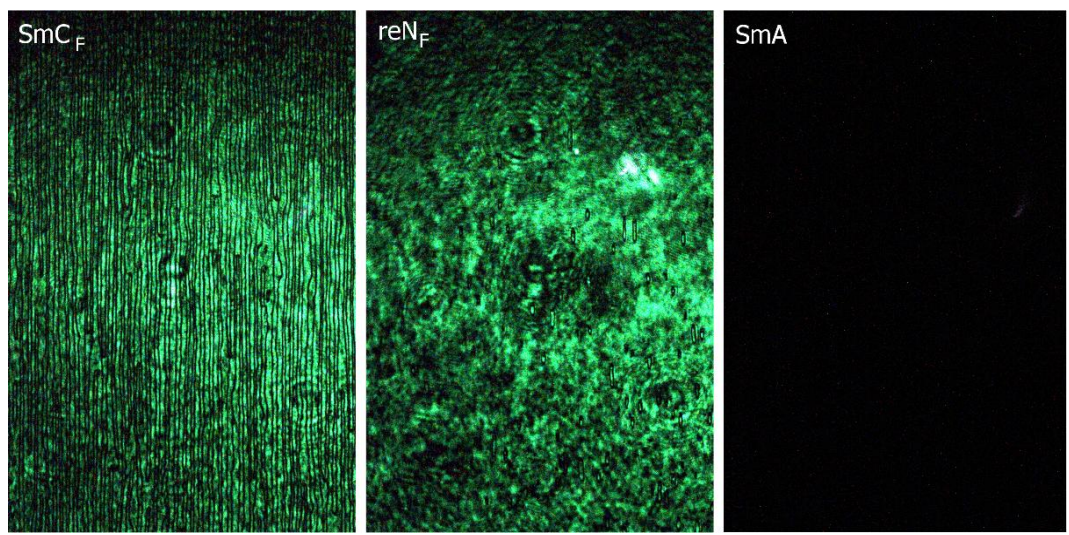

*Figure S4: SHG microscopic images taken for GS-5-Re, SmA is SHG silent showing that the phase is non-polar.*

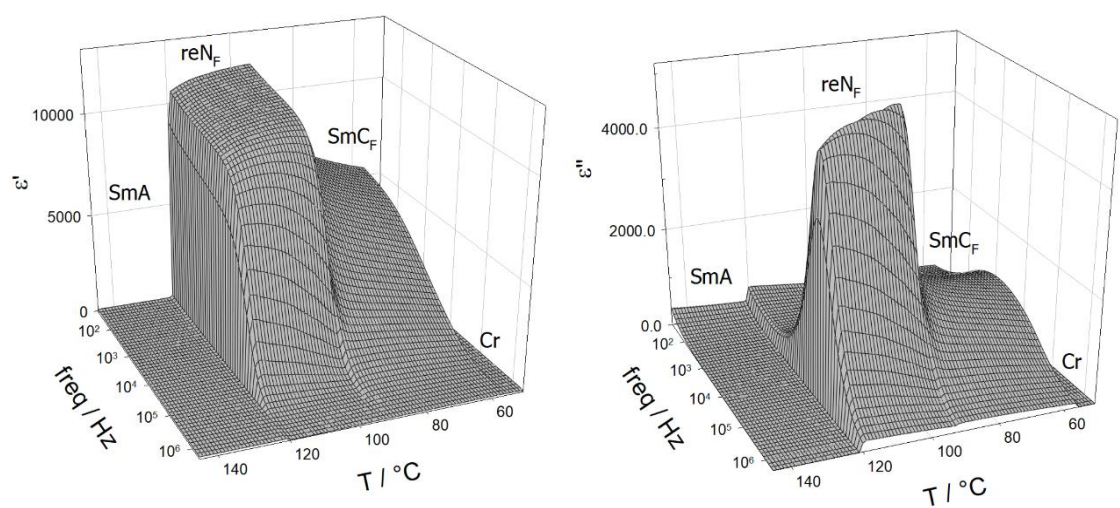

*Figure S5: Apparent dielectric permittivity (real and imaginary part of Δε) vs. frequency and temperature for compound SR-6-Re. The strong dielectric response in reN$_F$ that appears below SmA shows the polar nature of the phase.*

Table S1: Transition temperatures taken to model $a_{eff}$ for materials shown in Figure 5; $T_{NS}$ transition from the apolar nematic to the smectic phase, $T_P$ to the polar phase, $T_{rN}$ to the re-entrant nematic phase and $T_{rS}$ to the re-entrant smectic phase

| material | $T_{NS}$ [°C] | $T_P$ [°C] | $T_{rN}$ [°C] | $T_{rS}$ [°C] |
|---|---|---|---|---|
| SR-4-Re | 130 | 163 | / | 102 |
| SR-6-Re | 151 | 123 | 123 | 95 |
| SR-7-Re | 161 | 88 | 88 | 85 |
| GS-5-Re | 136 | 130 | 127 | 93 |

## 3. Synthetic Procedures and Structural Characterisation

Unless otherwise stated, all materials were obtained from commercial sources and used without further purification. Reactions were monitored using thin layer chromatography (TLC) using aluminium-backed plates with a coating of Merck Kieselgel 60 F254 silica and an appropriate solvent system. Spots were visualised using UV light (254 nm). Flash column chromatography was carried out using silica grade 60 Å 40-63 micron. $^1$H, $^{19}$F, and $^{13}$C NMR spectra were recorded on a 400 MHz Agilent NMR spectrometer using either CDCl$_3$ or DMSO-$d_6$ as solvent and using residual non-deuterated trace solvents as reference. Chemical shifts (δ) are given in ppm relative to TMS (δ = 0.00 ppm). Coupling constants (J) are given in Hz and are $^3J_{HH}$ unless otherwise stated. Mass spectroscopy was conducted on a Micromass LCT instrument.

### 3.1 Synthesis of the SR-*n*-Re series

The synthetic route followed for series SR-*n*-Re is summarised in Scheme 1. Intermediates 1.1-4 were synthesised according to procedures detailed in Juodka *et al.*[1] ; intermediates 1.5 and 1.6 and final products followed procedures as in reference [2].

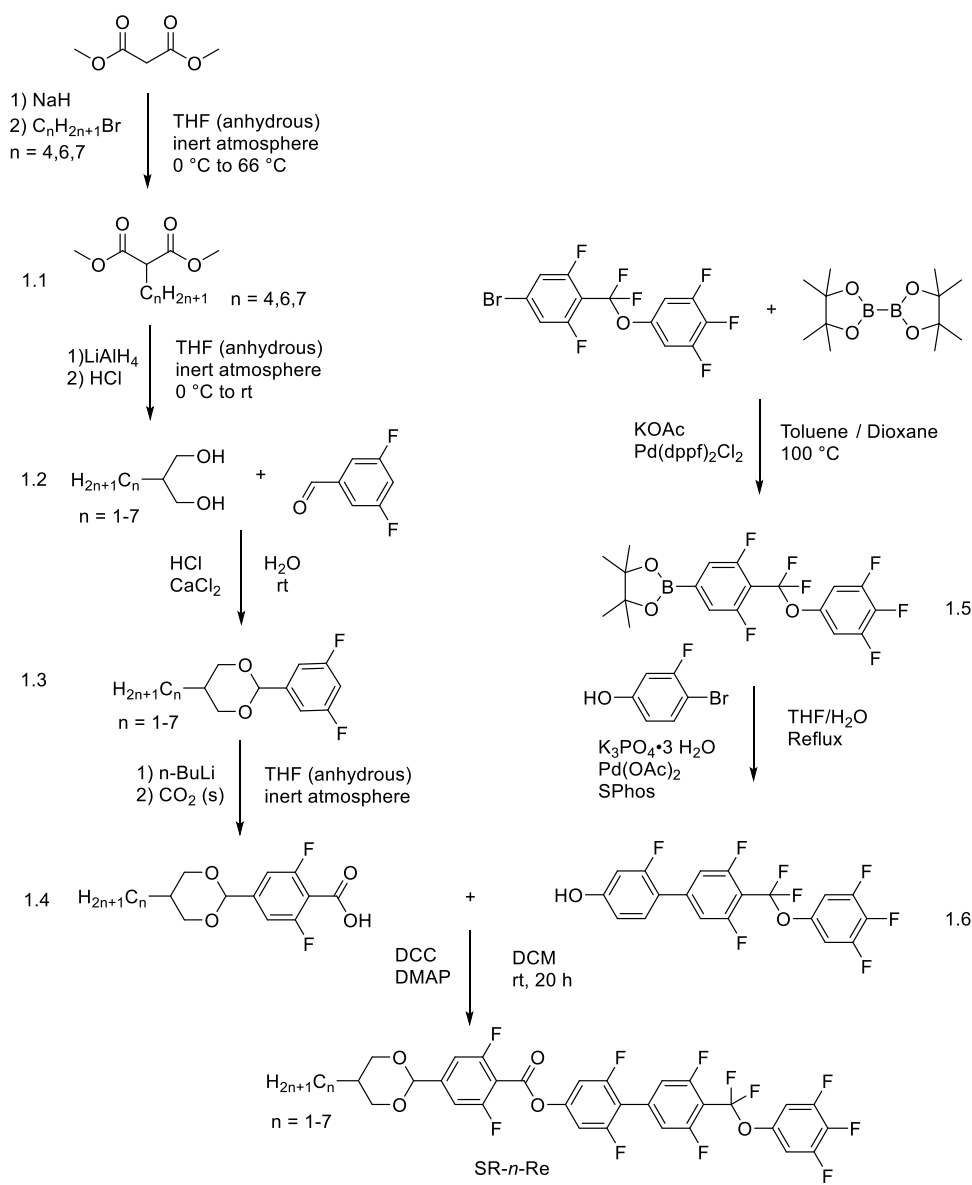

Scheme 1. Synthetic route to SR-n-Re, where n = 1-7 carbons.

**General procedure for esterification SR-*n*-Re**

The required 4-(5-*alkyl*-1,3-dioxan-2-yl)-2,6-difluorobenzoic acid (1.4) (1 eq) and DCC (1.3 eq) were dissolved in DCM (10 mL) and stirred until the mixture turned opaque. 1.6 (1.1 eq) and DMAP (0.13 eq) were added and the reaction mixture stirred for 18 hours. The solvent was removed *in vacuo* and the crude product purified by flash column chromatography (DCM, followed by 1:9 ethyl acetate:petroleum ether (40-60)) and subsequent recrystallisation in ethanol to yield the title compound as a white solid.

*Table S2. Quantities of reagents used in the synthesis of SR-n-Re where n = 1-7.*

|   | 4-(5-*alkyl*-1,3-dioxan-2-yl)-2,6-difluorobenzoic acid 1.4 | | 4'-(difluoro(3,4,5-trifluorophenoxy)methyl)-2,3',5',6-tetrafluoro-[1,1'-biphenyl]-4-ol 1.6 | |
|---|---|---|---|---|
| *n* | Mass (g) | Moles (mmol) | Mass (g) | Moles (mmol) |
| 1 | 0.059 | 0.228 | 0.101 | 0.230 |
| 2 | 0.056 | 0.206 | 0.100 | 0.228 |
| 3 | 0.059 | 0.206 | 0.100 | 0.228 |
| 4 | 0.065 | 0.216 | 0.100 | 0.228 |
| 5 | 0.069 | 0.220 | 0.101 | 0.230 |
| 6 | 0.068 | 0.207 | 0.102 | 0.233 |
| 7 | 0.075 | 0.219 | 0.099 | 0.226 |

**4'-(difluoro(3,4,5-trifluorophenoxy)methyl)-2,3',5',6-tetrafluoro-[1,1'-biphenyl]-4-yl 2,6-difluoro-4-(5-methyl-1,3-dioxan-2-yl)benzoate (SR-1-Re)**

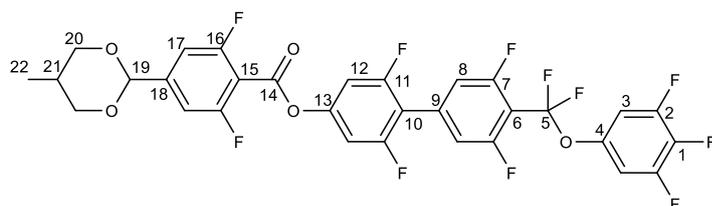

Yield: 0.045g (29%) R$_F$: 0.28 (1:9 ethyl acetate: petroleum ether (40-60))

m.p. 130 °C, T$_{NF-I}$ 208 °C

$^1$H NMR (400 MHz, CDCl$_3$) δ ppm: 7.21 (d, $^3J_{HF}$ = 9.4 Hz, 2H, Ar-**H**, H$_{17}$), 7.16 (d, $^3J_{HF}$ = 10.5 Hz, 2H, Ar-**H**, H$_{12}$), 7.08 – 7.03 (m$_{overlapping\,d}$, 2H, Ar-**H**, H$_8$), 7.02 – 6.97 (m$_{overlapping\,d}$, 2H, H$_3$), 5.41 (s, 1H, R-CH-(CH$_2$)$_2$-(O)$_2$-**CH**-Ar, H$_{19}$), 4.28 – 4.17 (m, 2H, R-CH-(**CH$_2$**)$_2$-(O)$_2$-CH-Ar, H$_{20}$), 3.59 – 3.45 (m, 2H, R-CH-(**CH$_2$**)$_2$-(O)$_2$-CH-Ar, H$_{20}$), 2.30 – 2.14 (m, 1H, R-**CH**-(CH$_2$)$_2$-(O)$_2$-CH-Ar, H$_{21}$), 0.80 (d, $^3J_{H-H}$ = 6.7 Hz, 3H, **CH$_3$**-Dioxane, H$_{22}$)

$^{19}$F NMR (proton decoupled) (376 MHz, CDCl$_3$) δ ppm: -61.98 (t, $^4J_{F-F}$ = 26.8 Hz, 2F, **F$_5$**), -108.14 (s, 2F, **F$_{16}$**), -110.52 (t, $^4J_{F-F}$ = 27.8 Hz, 2F, **F$_7$**), -111.70 (s, 2F, **F$_{11}$**), -132.42 (d, $^3J_{F-F}$ = 20.8 Hz, 2F, **F$_2$**), -163.06 (t, $^3J_{F-F}$ = 20.3 Hz, 1F, **F$_1$**)

$^{13}$C NMR (101 MHz, CDCl$_3$) δ ppm: 161.14 (dd, $^1J_{CF}$ = 259.0, $^3J_{CF}$ = 5.5 Hz, 2C), 161.32 – 158.47 (m$_{overlapping\,signals}$, 4C), 158.93 (s, apparent t, 1C), 151.44 (t, $^3J_{CF}$ = 14.2 Hz, 1C), 151.17 (ddd, $^1J_{CF}$ = 251.2, $^2J_{CF}$ = 10.7 Hz, $^3J_{CF}$ = 5.3 Hz, 2C), 146.13 (t, $^3J_{CF}$ = 10.0 Hz, 1C), 144.84 – 144.50 (m, 1C), 139.27 (dt, $^1J_{CF}$ = 250.4, $^2J_{CF}$ = 14.9 Hz, 1C), 134.42 (t, $^3J_{CF}$ = 11.4 Hz, 1C), 120.16 (t, $^1J_{CF}$ = 266.9 Hz, 1C), 114.90 (dd, $^2J_{CF}$ = 24.4, $^4J_{CF}$ = 2.3 Hz, 2C), 113.49 (t, $^2J_{CF}$ = 17.8 Hz, 1C, app dt), 110.55 (dd, $^2J_{CF}$ = 23.6 Hz, $^4J_{CF}$ = 3.4 Hz, 2C), 109.87 (t, $^2J_{CF}$ = 13.8 Hz, 1C), 109.04 (t, $^2J_{CF}$ = 16.7 Hz, 1C), 107.66 (dd, $^2J_{CF}$ = 23.8, $^4J_{CF}$ = 6.5 Hz, 2C), 106.86 (ddd, $^2J_{CF}$ = 21.7 Hz, $^3J_{CF}$ 8.4 Hz, $^4J_{CF}$ = 2.6 Hz, 2C), 98.64 (t, $^4J_{CF}$ = 2.3 Hz, 1C), 73.75 (2C), 29.40 (1C), 12.37 (1C)

IR $v_{max}$ (cm$^{-1}$): 3107 (C-H stretch, sp$^2$ hybridised), 2959 (C-H stretch, sp$^3$ hybridised), 2111 (C-H bend, overtones), 1744 (C=O stretch, ester)

**4'-(difluoro(3,4,5-trifluorophenoxy)methyl)-2,3',5',6-tetrafluoro-[1,1'-biphenyl]-4-yl 4-(5-ethyl-1,3-dioxan-2-yl)-2,6-difluorobenzoate (SR-2-Re)**

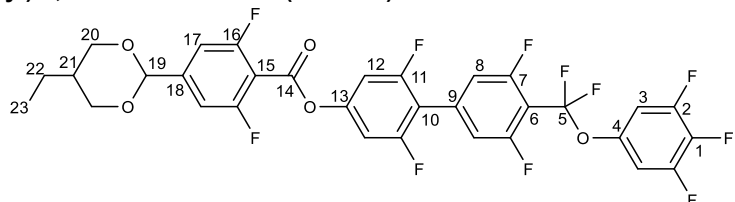

Yield: 0.033g (21%), R$_F$ 0.40 (1:9 ethyl acetate:petroleum ether (40-60))

m.p. 119 °C, (T$_{SmCF-NF}$ 89 °C), T$_{NF-N}$ 192 °C, T$_{N-I}$ 215 °C

$^1$H NMR (400 MHz, CDCl$_3$) δ ppm: 7.20 (d, $^3J_{HF}$ = 9.3 Hz, 2H, Ar-**H**, H$_{17}$), 7.16 (d, $^3J_{HF}$ = 10.5 Hz, 2H, Ar-**H**, H$_{12}$), 7.08 – 7.03 (m$_{overlapping\,d}$, 2H, Ar-**H**, H$_8$), 7.02 – 6.96 (m$_{overlapping\,d}$, 2H, Ar-**H**, H$_3$), 5.40 (s, 1H, R-CH-(CH$_2$)$_2$-(O)$_2$-**CH**-

Ar, H$_{19}$), 4.31 – 4.24 (m, 2H, R-CH-(**CH$_2$**)$_2$-(O)$_2$-CH-Ar, H$_{20}$), 3.58 – 3.49 (m, 2H, R-CH-(**CH$_2$**)$_2$-(O)$_2$-CH-Ar, H$_{20}$), 2.12 – 1.98 (m, 1H, R-**CH**-(CH$_2$)$_2$-(O)$_2$-CH-Ar, H$_{21}$), 1.18 (p, $^3J_{H-H}$ = 7.5 Hz, 2H, CH$_3$-**CH$_2$**-Dioxane, H$_{22}$), 0.95 (t, $J_{H-H}$ = 7.5 Hz, 3H, **CH$_3$**-CH$_2$-Dioxane, H$_{23}$)

$^{19}$F NMR (proton decoupled) (376 MHz, CDCl$_3$) δ ppm: -61.93 (t, $^4J_{F-F}$ = 27.5 Hz, 2F, **F$_5$**), -108.15 (s, 2F, **F$_{16}$**), -110.50 (t, $^4J_{F-F}$ = 27.8 Hz, 2F, **F$_7$**), -111.70 (s, 2F, **F$_{11}$**), -132.42 (d, $^3J_{F-F}$ = 20.7 Hz, 2F, **F$_2$**), -163.06 (t, $^3J_{F-F}$ = 20.4 Hz, 1F, **F$_1$**)

$^{13}$C NMR (101 MHz, CDCl$_3$) δ ppm: 161.14 (dd, $^1J_{CF}$ = 259.0 Hz, $^1J_{CF}$ = 5.6 Hz, 2C), 161.29 – 158.46 (m$_{overlapping\ signals}$, 4C), 158.94 (s, apparent t, 1C), 151.44 (t, $^3J_{CF}$ = 14.3 Hz, 1C), 151.17 (ddd, $^1J_{CF}$ = 250.9 Hz, $^2J_{CF}$ =10.6 Hz, $^3J_{CF}$ = 5.2 Hz, 2C), 146.18 (t, $^3J_{CF}$ = 9.9 Hz, 1C), 144.88 – 144.47 (m, 1C), 138.65 (dt, $^1J_{CF}$ = 250.4, $^2J_{CF}$ = 15.1 Hz, 1C), 134.41 (t, $^3J_{CF}$ = 11.5 Hz, 1C), 120.17 (t, $^1J_{CF}$ = 267.1 Hz, 1C), 114.86 (dd, $^2J_{CF}$ = 24.1, $^4J_{CF}$ = 2.2 Hz, 2C), 113.59 (t, $^2J_{CF}$ = 17.7 Hz, 1C), 110.55 (dd, $^2J_{CF}$ = 23.7, $^4J_{CF}$ = 3.4 Hz, 2C), 109.90 (t, $^2J_{CF}$ = 14.2 Hz, 1C), 109.05 (t, $^2J_{CF}$ = 16.8 Hz, 1C), 107.67 (dd, $^2J_{CF}$ = 23.9, $^4J_{CF}$ = 6.1 Hz, 2C), 106.87 (ddd, $^2J_{CF}$ = 22.0, $^3J_{CF}$ = 8.3, $^4J_{CF}$ = 2.6 Hz, 2C), 98.88 (t, $^4J_{CF}$ = 2.2 Hz, 1C), 72.55 (2C), 35.86 (1C), 21.25 (1C), 11.03 (1C)

IR $v_{max}$ (cm$^{-1}$): 3072 (C-H stretch, sp$^2$ hybridised), 2917 (C-H stretch, sp$^3$ hybridised), 2106 (C-H bend, overtones), 1762 (C=O stretch, ester)

**4'-(difluoro(3,4,5-trifluorophenoxy)methyl)-2,3',5',6-tetrafluoro-[1,1'-biphenyl]-4-yl 4-(5-propyl-1,3-dioxan-2-yl)-2,6-difluorobenzoate (SR-3-Re)**

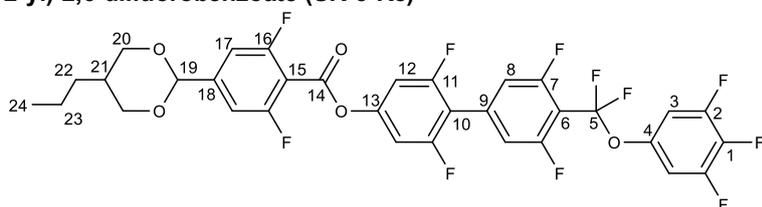

Yield: 0.040g (27%), R$_F$: 0.47 (1:9 ethyl acetate:petroleum ether (40-60))

m.p. 96 °C, (T$_{SmCF-NF}$ 91 °C), T$_{NF-Nx}$ 180 °C, T$_{Nx-N}$ 181 °C, T$_{N-I}$ 215 °C

$^1$H NMR (400 MHz, CDCl$_3$) δ ppm: 7.20 (d, $^3J_{HF}$ = 9.2 Hz, 2H, Ar-**H**, H$_{17}$), 7.16 (d, $^3J_{HF}$ = 10.4 Hz, 2H, Ar-**H**, H$_{12}$), 7.08 – 7.03 (m$_{overlapping\ d}$, 2H, Ar-**H**, H$_8$), 7.02 – 6.97 (m$_{overlapping\ d}$, 2H, Ar-**H**, H$_3$), 5.40 (s, 1H, R-CH-(CH$_2$)$_2$-(O)$_2$-**CH**-Ar, H$_{19}$), 4.29 – 4.23 (m, 2H, R-CH-(**CH$_2$**)$_2$-(O)$_2$-CH-Ar, H$_{20}$), 3.58 – 3.50 (m, 2H, R-CH-(**CH$_2$**)$_2$-(O)$_2$-CH-Ar, H$_{20}$), 2.19 – 2.09 (m, 1H, R-**CH**-(CH$_2$)$_2$-(O)$_2$-CH-Ar, H$_{21}$), 1.35 (h, $^3J_{H-H}$ = 7.3 Hz, 2H, CH$_3$-**CH$_2$**-CH$_2$-Dioxane, H$_{23}$), 1.11 (q, $^3J_{H-H}$ = 7.3 Hz, 2H, CH$_3$-CH$_2$-**CH$_2$**-Dioxane, H$_{22}$), 0.94 (t, $^3J_{H-H}$ = 7.3 Hz, 3H, **CH$_3$**-CH$_2$-CH$_2$-Dioxane, H$_{24}$)

$^{19}$F NMR (proton decoupled) (376 MHz, CDCl$_3$) δ ppm: -61.95 (t, $^4J_{F-F}$ = 26.5 Hz, 2F, **F$_5$**), -108.14 (s, 2F, **F$_{16}$**), -110.52 (t, $^4J_{F-F}$ = 27.0 Hz, 2F, **F$_7$**), -111.70 (s, 2F, **F$_{11}$**), -132.42 (d, $^3J_{F-F}$ = 20.7 Hz, 2F, **F$_2$**), -163.05 (t, $^3J_{F-F}$ = 20.5 Hz, 1F, **F$_1$**)

$^{13}$C NMR (101 MHz, CDCl$_3$) δ ppm: 161.01 (dd, $^1J_{CF}$ = 259.0 Hz, $^1J_{CF}$ = 5.5 Hz, 2C), 161.17 – 158.32 (m$_{overlapping\ signals}$, 4C), 158.80 (s, apparent t, 1C), 151.47 (ddd, $^1J_{CF}$ = 250.9 Hz, $^2J_{CF}$ = 10.6 Hz, $^3J_{CF}$ 5.3 Hz, 2C), 151.30 (t, $^3J_{CF}$ = 14.3 Hz, 1C), 146.05 (t, $^3J_{CF}$ = 10.0 Hz, 1C), 144.73 – 144.34 (m, 1C), 138.52 (dt, $^1J_{CF}$ = 250.4 Hz, $^2J_{CF}$ = 15.8 Hz, 1C), 134.28 (t, $^3J_{CF}$ = 11.4 Hz, 1C), 120.03 (t, $^1J_{CF}$ = 266.2 Hz, 1C), 114.77 (dd, $^2J_{CF}$ = 24.4, $^4J_{CF}$ = 2.0 Hz, 2C), 113.62 – 113.11 (m, 1C), 110.40 (dd, $^2J_{CF}$ = 23.5 Hz, $^4J_{CF}$ = 3.4 Hz, 2C), 109.74 (t, $^2J_{CF}$ = 13.5 Hz, 1C), 108.91 (t, $^2J_{CF}$ = 16.4 Hz, 1C), 107.53 (dd, $^2J_{CF}$ = 23.9, $^4J_{CF}$ = 6.4 Hz, 2C), 106.72 (ddd, $^2J_{CF}$ = 21.9 Hz, $^3J_{CF}$ = 8.5 Hz, $^4J_{CF}$ = 2.6 Hz, 2C), 98.73 (t, $^4J_{CF}$ = 2.3 Hz, 1C), 72.60 (2C), 33.90 (1C), 30.22 (1C), 19.53 (1C), 14.19 (1C)

IR $v_{max}$ (cm$^{-1}$): 3109 (C-H stretch, sp$^2$ hybridised), 2926 (C-H stretch, sp$^3$ hybridised), 2121 (C-H bend, overtones), 1750 (C=O stretch, ester)

**4'-(difluoro(3,4,5-trifluorophenoxy)methyl)-2,3',5',6-tetrafluoro-[1,1'-biphenyl]-4-yl 4-(5-butyl-1,3-dioxan-2-yl)-2,6-difluorobenzoate (SR-4-Re)**

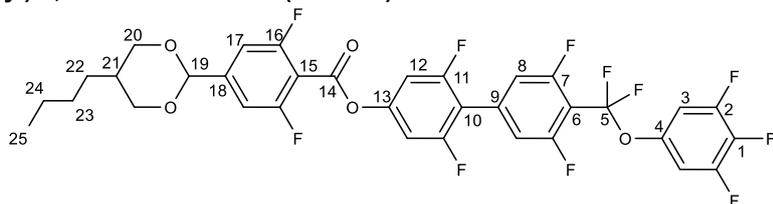

Yield: 0.055 g (35%), R$_F$: 0.50 (1:9 ethyl acetate:petroleum ether (40-60))

m.p. 89 °C, $T_{SmCF-NF}$ 102 °C, $T_{NF-N}$ 163 °C, $T_{N-I}$ 210 °C

$^1$H NMR (400 MHz, CDCl$_3$) δ ppm: 7.23 (d, $^3J_{HF}$ = 9.5 Hz, 2H, Ar-**H**, H$_{17}$), 7.19 (d, $^3J_{HF}$ = 10.5 Hz, 2H, Ar-**H**, H$_{12}$), 7.11 – 7.05 (m$_{overlapping\,d}$, 2H, Ar-**H**, H$_8$), 7.05 – 7.00 (m$_{overlapping\,d}$, 2H, Ar-**H**, H$_3$), 5.43 (s, 1H, R-CH-(CH$_2$)$_2$-(O)$_2$-C**H**-Ar, H$_{19}$), 4.33 – 4.23 (m, 2H, R-CH-(**CH$_2$**)$_2$-(O)$_2$-CH-Ar, H$_{20}$), 3.63 – 3.51 (m, 2H, R-CH-(**CH$_2$**)$_2$-(O)$_2$-CH-Ar, H$_{20}$), 2.23 – 2.08 (m, 1H, R-C**H**-(CH$_2$)$_2$-(O)$_2$-CH-Ar, H$_{21}$), 1.40 – 1.28 (m, 4H, CH$_3$-**CH$_2$**-**CH$_2$**-CH$_2$-Dioxane, H$_{23-24}$), 1.15 (q, $^3J_{H-H}$ = 7.0 Hz, 2H, CH$_3$-CH$_2$-CH$_2$-**CH$_2$**-Dioxane, H$_{22}$), 0.94 (t, $^3J_{H-H}$ = 6.9 Hz, 3H, **CH$_3$**-CH$_2$-CH$_2$-CH$_2$-Dioxane, H$_{25}$)

$^{19}$F NMR (proton decoupled) (376 MHz, CDCl$_3$) δ ppm: -61.95 (t, $^4J_{F-F}$ = 26.6 Hz, 2F, **F$_5$**), -108.14 (s, 2F, **F$_{16}$**), -110.52 (t, $^4J_{F-F}$ = 26.7 Hz, 2F, **F$_7$**), -111.70 (s, 2F, **F$_{11}$**), -132.42 (d, $^3J_{F-F}$ = 20.9 Hz, 2F, **F$_2$**), -163.03 (t, $^3J_{F-F}$ = 20.6 Hz, 1F, **F$_1$**)

$^{13}$C NMR (101 MHz, CDCl$_3$) δ ppm: 161.14 (dd, $^1J_{CF}$ = 259.0, $^3J_{CF}$ = 5.5 Hz, 2C), 161.28 – 158.46 (m$_{overlapping}$, 4C), 158.94 (s, apparent t, 1C), 151.44 (t, $^3J_{CF}$ = 14.4 Hz, 1C), 151.17 (ddd, $^1J_{CF}$ = 250.7 Hz, $^2J_{CF}$ = 10.7 Hz, $^3J_{CF}$ = 5.1 Hz, 2C), 146.18 (t, $^3J_{CF}$ = 10.0 Hz, 1C), 144.87 – 144.46 (m, 1C), 139.90 (dt, $^1J_{CF}$ = 250.5, $^2J_{CF}$ = 15.2 Hz, 1C), 134.41 (t, $^3J_{CF}$ = 11.4 Hz, 1C), 120.17 (t, $^1J_{CF}$ = 266.7 Hz, 1C), 114.90 (dd, $^2J_{CF}$ = 24.2, $^4J_{CF}$ = 2.3 Hz, 2C), 113.50 (t, $^2J_{CF}$ = 17.4 Hz, 1C), 110.54 (dd, $^2J_{CF}$ = 23.6, $^4J_{CF}$ = 3.4 Hz, 2C), 109.88 (t, $^2J_{CF}$ = 14.0 Hz, 1C), 109.05 (t, $^2J_{CF}$ = 16.6 Hz, 1C), 107.66 (dd, $^2J_{CF}$ = 23.7, $^4J_{CF}$ = 6.5 Hz, 2C), 106.84 (ddd, $^2J_{CF}$ = 21.8, $^3J_{CF}$ = 8.2, $^4J_{CF}$ = 2.5 Hz, 2C), 98.87 ($^4J_{CF}$, J = 2.2 Hz, 1C), 72.77 (2C), 34.27 (1C), 28.60 (1C), 27.92 (1C), 22.95 (1C), 14.02 (1C)

IR $v_{max}$ (cm$^{-1}$): 3107 (C-H stretch, sp$^2$ hybridised), 2925 (C-H stretch, sp$^3$ hybridised), 2109 (C-H bend, overtones), 1747 (C=O stretch, ester)

**4'-(difluoro(3,4,5-trifluorophenoxy)methyl)-2,3',5',6-tetrafluoro-[1,1'-biphenyl]-4-yl 4-(1,3-dioxan-2-pentyl)-2,6-difluorobenzoate (SR-5-Re)**

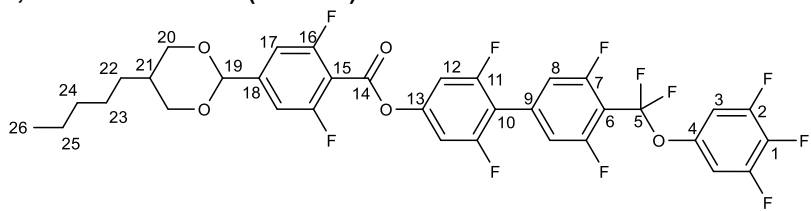

Yield: 0.038 g (24%), R$_F$: 0.52 (1:9 ethyl acetate:petroleum ether (40-60))

m.p. 86 °C, $T_{SmCF-NF}$ 97 °C, $T_{NF-Nx}$ 151 °C, $T_{Nx-N}$ 165 °C, $T_{N-I}$ 210 °C

$^1$H NMR (400 MHz, CDCl$_3$) δ ppm: 7.20 (d, $^3J_{HF}$ = 9.3 Hz, 2H, Ar-**H**, H$_{17}$), 7.16 (d, $^3J_{HF}$ = 10.5 Hz, 2H, Ar-**H**, H$_{12}$), 7.08 – 7.03 (m$_{overlapping\,d}$, 2H, Ar-**H**, H$_8$), 7.03 – 6.97 (m$_{overlapping\,d}$, 2H, Ar-**H**, H$_3$), 5.40 (s, 1H, R-CH-(CH$_2$)$_2$-(O)$_2$-C**H**-Ar, H$_{19}$), 4.30 – 4.22 (m, 2H, R-CH-(**CH$_2$**)$_2$-(O)$_2$-CH-Ar, H$_{20}$), 3.59 – 3.48 (m, 2H, R-CH-(**CH$_2$**)$_2$-(O)$_2$-CH-Ar, H$_{20}$), 2.19 – 2.06 (m, 1H, R-C**H**-(CH$_2$)$_2$-(O)$_2$-CH-Ar, H$_{21}$), 1.40 – 1.23 (m, 6H, CH$_3$-CH$_2$-**CH$_2$**-**CH$_2$**-**CH$_2$**-Dioxane, H$_{22-24}$), 1.17 – 1.07 (m, 2H, CH$_3$-**CH$_2$**-CH$_2$-CH$_2$-CH$_2$-Dioxane, H$_{25}$), 0.94 – 0.86 (m, 3H, **CH$_3$**-CH$_2$-CH$_2$-CH$_2$-CH$_2$-Dioxane, H$_{26}$)

$^{19}$F NMR (proton decoupled) (376 MHz, CDCl$_3$) δ ppm: -61.95 (t, $^4J_{F-F}$ = 26.6 Hz, 2F, **F$_5$**), -108.14 (s, 2F, **F$_{16}$**), -110.52 (t, $^4J_{F-F}$ = 26.7 Hz, 2F, **F$_7$**), -111.70 (s, 2F, **F$_{11}$**), -132.42 (d, $^3J_{F-F}$ = 20.9 Hz, 2F, **F$_2$**), -163.06 (t, $^3J_{F-F}$ = 20.7 Hz, 1F, **F$_1$**)

$^{13}$C NMR (101 MHz, CDCl$_3$) δ ppm: 161.14 (dd, $^1J_{CF}$ = 258.9 Hz, $^3J_{CF}$ = 5.5 Hz, 2C), 161.27 – 158.46 (m$_{overlapping\,signals}$, 4C), 158.93 (s, apparent t, 1C), 151.44 (t, $^3J_{CF}$ = 14.3 Hz, 1C), 151.17 (ddd, $^1J_{CF}$ = 251.0 Hz, $^2J_{CF}$ = 10.7 Hz, $^3J_{CF}$ = 5.2 Hz, 2C), 146.19 (t, $^3J_{CF}$ = 10.1 Hz, 1C), 144.87 – 144.48 (m, 1C), 138.65 (dt, $^1J_{CF}$ = 250.5 Hz, $^2J_{CF}$ = 15.0 Hz, 1C), 134.41 (t, $^3J_{CF}$ = 11.4 Hz, 1C), 120.16 (t, $^1J_{CF}$ = 267.0 Hz, 1C), 114.90 (dd, $^2J_{CF}$ = 24.4, $^4J_{CF}$ = 2.4 Hz, 2C), 113.76 – 113.25 (m, 1C), 110.54 (dd, $^2J_{CF}$ = 23.7 Hz, $^4J_{CF}$ = 3.4 Hz, 2C), 109.81 (t, $^2J_{CF}$ = 14.1 Hz, 1C), 109.04 ($^2J_{CF}$, J = 16.6 Hz, 1C), 107.66 (dd, $^2J_{CF}$ = 24.0 Hz, $^4J_{CF}$ = 6.5 Hz, 2C), 106.86 (ddd, $^2J_{CF}$ = 21.9 Hz, $^3J_{CF}$ = 8.4 Hz, $^4J_{CF}$ = 2.7 Hz, 2C), 98.87 (t, $^4J_{CF}$ = 2.2 Hz, 4C), 72.77 (2C), 34.27 (1C), 32.05 (1C), 28.17 (1C), 26.10 (1C), 22.60 (1C), 14.15 (1C)

IR $v_{max}$ (cm$^{-1}$): 3077 (C-H stretch, sp$^2$ hybridised), 2925 (C-H stretch, sp$^3$ hybridised), 2033 (C-H bend, overtones), 1750 (C=O stretch, ester)

**4'-(difluoro(3,4,5-trifluorophenoxy)methyl)-2,3',5',6-tetrafluoro-[1,1'-biphenyl]-4-yl 4-(1,3-dioxan-2-hexyl)-2,6-difluorobenzoate (SR-6-Re)**

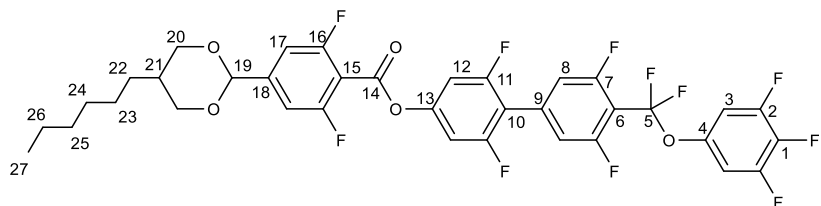

Yield: 0.036 g (23%), $R_F$: 0.53 (1:9 ethyl acetate:petroleum ether (40-60))

m.p. 75 °C, $T_{SmCF-NF}$ 95 °C, $T_{NF-SmA}$ 123 °C, $T_{SmA-N}$ 151 °C, $T_{N-I}$ 204 °C

$^1$H NMR (400 MHz, CDCl$_3$) δ ppm: 7.20 (d, $^3J_{HF}$ = 9.3 Hz, 2H, Ar-**H**, H$_{17}$), 7.16 (d, $^3J_{HF}$ = 10.5 Hz, 2H, Ar-**H**, H$_{12}$), 7.08 – 7.03 (m$_{overlapping\ d}$, 2H, Ar-**H**, H$_8$), 7.03 – 6.97 (m$_{overlapping\ d}$, 2H, Ar-**H**, H$_3$), 5.40 (s, 1H, R-CH-(CH$_2$)$_2$-(O)$_2$-C**H**-Ar, H$_{19}$), 4.29 – 4.21 (m, 2H, R-CH-(**CH$_2$**)$_2$-(O)$_2$-CH-Ar, H$_{20}$), 3.59 – 3.50 (m, 2H, R-CH-(**CH$_2$**)$_2$-(O)$_2$-CH-Ar, H$_{20}$), 2.19 – 2.05 (m, 1H, R-**CH**-(CH$_2$)$_2$-(O)$_2$-CH-Ar, H$_{21}$), 1.37 – 1.23 (m, 8H, CH$_3$-CH$_2$-**CH$_2$**-**CH$_2$**-**CH$_2$**-**CH$_2$**-Dioxane, H$_{22-25}$), 1.16 – 1.07 (m, 2H, CH$_3$-**CH$_2$**-CH$_2$-CH$_2$-CH$_2$-CH$_2$-Dioxane, H$_{26}$), 0.95 – 0.85 (m, 3H, **CH$_3$**-CH$_2$-CH$_2$-CH$_2$-CH$_2$-CH$_2$-Dioxane, H$_{27}$)

$^{19}$F NMR (proton decoupled) (376 MHz, CDCl$_3$) δ ppm: -61.95 (t, $^4J_{F-F}$ = 27.0 Hz, 2F, **F$_5$**), -108.14 (s, 2F, **F$_{16}$**), -110.52 (t, $^4J_{F-F}$ = 27.0 Hz, 2F, **F$_7$**), -111.70 (s, 2F, **F$_{11}$**), -132.42 (d, $^3J_{F-F}$ = 21.2 Hz, 2F, **F$_2$**), -163.05 (t, $^3J_{F-F}$ = 20.9 Hz, 1F, **F$_1$**)

$^{13}$C NMR (101 MHz, CDCl$_3$) δ ppm: 161.14 (dd, $^1J_{CF}$ = 258.9 Hz, $^3J_{CF}$ = 5.6 Hz, 2C), 161.27 – 158.46 (m$_{overlapping\ signals}$, 4C), 158.94 (s, apparent t, 1C), 151.58 (ddd, $^1J_{CF}$ = 250.9 Hz, $^2J_{CF}$ = 10.4 Hz, $^3J_{CF}$ = 4.9 Hz, 2C), 151.44 (t, $^3J_{CF}$ = 14.3 Hz, 1C), 146.19 (t, $^3J_{CF}$ = 9.9 Hz, 1C), 144.91 – 144.42 (m, 1C), 138.65 (dt, $^1J_{CF}$ = 250.6 Hz, $^2J_{CF}$ = 15.2 Hz, 1C), 134.41 (t, $^3J_{CF}$ = 11.5 Hz, 1C), 118.84 (t, $^1J_{CF}$ = 266.0 Hz, 1C), 114.91 (dd, $^2J_{CF}$ = 24.3 Hz, $^4J_{CF}$ = 2.4 Hz, 2C), 113.50 (t, $^2J_{CF}$ = 17.1 Hz, 1C), 110.54 (dd, $^2J_{CF}$ = 23.6 Hz, $^4J_{CF}$ = 3.4 Hz, 2C), 109.88 (t, $^2J_{CF}$ = 14.1 Hz, 1C), 108.96 (t, $^2J_{CF}$ = 16.8 Hz, 1C), 107.67 (dd, $^2J_{CF}$ = 23.8 Hz, $^4J_{CF}$ = 6.1 Hz), 106.87 (ddd, $^2J_{CF}$ = 21.7 Hz, $^3J_{CF}$ = 8.7 Hz, $^4J_{CF}$ = 2.6 Hz, 2C), 98.87 (t, $^4J_{CF}$ = 2.1 Hz, 1C), 72.77 (2C), 34.27 (1C), 31.78 (1C), 29.54 (1C), 28.21 (1C), 26.40 (1C), 22.74 (1C), 14.20 (1C)

IR $\nu_{max}$ (cm$^{-1}$): 3110 (C-H stretch, sp$^2$ hybridised), 2924 (C-H stretch, sp$^3$ hybridised), 2110 (C-H bend, overtones), 1755 (C=O stretch, ester)

**4'-(difluoro(3,4,5-trifluorophenoxy)methyl)-2,3',5',6-tetrafluoro-[1,1'-biphenyl]-4-yl 4-(1,3-dioxan-2-heptyl)-2,6-difluorobenzoate (SR-7-Re)**

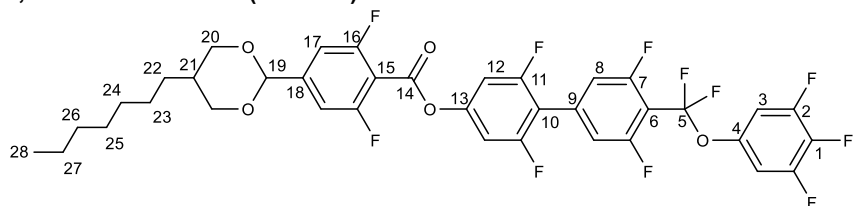

Yield: 0.059 g (34%), $R_F$: 0.60 (1:9 ethyl acetate:petroleum ether (40-60))

m.p. 95 °C, ($T_{SmCF-NF}$ 85 °C), $T_{NF-SmA}$ 88 °C, $T_{SmA-N}$ 161 °C, $T_{N-I}$ 206 °C

$^1$H NMR (400 MHz, CDCl$_3$) δ ppm: 7.20 (d, $^3J_{HF}$ = 9.4 Hz, 2H, Ar-**H**, H$_{17}$), 7.16 (d, $^3J_{HF}$ = 10.5 Hz, 2H, Ar-**H**, H$_{12}$), 7.08 – 7.03 (m$_{overlapping\ d}$, 2H, Ar-**H**, H$_8$), 7.03 – 6.97 (m$_{overlapping\ d}$, 2H, Ar-**H**, H$_3$), 5.40 (s, 1H, R-CH-(CH$_2$)$_2$-(O)$_2$-C**H**-Ar, H$_{19}$), 4.29 – 4.22 (m, 2H, R-CH-(**CH$_2$**)$_2$-(O)$_2$-CH-Ar, H$_{20}$), 3.58 – 3.49 (m, 2H, R-CH-(**CH$_2$**)$_2$-(O)$_2$-CH-Ar, H$_{20}$), 2.19 – 2.05 (m, 1H, R-**CH**-(CH$_2$)$_2$-(O)$_2$-CH-Ar, H$_{21}$), 1.34 – 1.23 (m, 10H, CH$_3$-CH$_2$-**CH$_2$**-**CH$_2$**-**CH$_2$**-**CH$_2$**-**CH$_2$**-Dioxane, H$_{22-26}$), 1.15 – 1.06 (m, 2H, CH$_3$-**CH$_2$**-CH$_2$-CH$_2$-CH$_2$-CH$_2$-CH$_2$-Dioxane, H$_{27}$), 0.94 – 0.85 (m, 3H, **CH$_3$**-CH$_2$-CH$_2$-CH$_2$-CH$_2$-CH$_2$-CH$_2$-Dioxane, H$_{28}$)

$^{19}$F NMR (proton decoupled) (376 MHz, CDCl$_3$) δ ppm: -61.97 (t, $^4J_{F-F}$ = 26.7 Hz, 2F, **F$_5$**), -108.14 (s, 2F, **F$_{16}$**), -110.48 (t, $^4J_{F-F}$ = 26.6 Hz, 2F, **F$_7$**), -111.70 (s, 2F, **F$_{11}$**), -132.42 (d, $^3J_{F-F}$ = 20.8 Hz, 2F, **F$_2$**), -163.04 (t, $^3J_{F-F}$ = 20.7 Hz, 1F, **F$_1$**)

$^{13}$C NMR (101 MHz, CDCl$_3$) δ ppm: 161.14 (dd, $^1J_{CF}$ = 258.9 Hz, $^3J_{CF}$ = 5.6 Hz, 2C), 161.30 – 158.46 (m$_{overlapping\ signals}$, 4C), 158.93 (s, apparent t, 1C), 151.44 (t, $^3J_{CF}$ = 14.3 Hz, 1C), 151.17 (ddd, $^1J_{CF}$ = 251.1 Hz, $^2J_{CF}$ = 10.8 Hz, $^3J_{CF}$ = 5.2 Hz, 2C), 146.19 (t, $^3J_{CF}$ = 9.9 Hz, 1C), 144.97 – 144.26 (m, 1C), 138.43 (dt, $^1J_{CF}$ = 250.3 Hz, $^2J_{CF}$ = 15.4 Hz, 1C), 134.41 (t, $^3J_{CF}$ = 11.5 Hz, 1C), 120.17 ($^1J_{CF}$ = 266.7 Hz, 1C), 114.90 (dd, $^2J_{CF}$ = 24.4 Hz, $^4J_{CF}$ = 2.4 Hz, 2C), 113.49 (t, $^2J_{CF}$ = 18.0 Hz, 1C), 110.54 (dd, $^2J_{CF}$ = 23.6 Hz, $^4J_{CF}$ = 3.4 Hz, 2C), 109.87 (t, $^2J_{CF}$ = 14.4 Hz, 1C), 109.04 (t, $^2J_{CF}$ = 16.7 Hz, 1C), 107.66 (dd, $^2J_{CF}$ = 23.9 Hz, $^4J_{CF}$ = 6.4 Hz, 2C), 106.85 (ddd, $^2J_{CF}$ = 21.7 Hz, $^3J_{CF}$ =

8.5 Hz, $^4J_{CF}$ = 2.6 Hz, 2C), 98.87 (t, $^4J_{CF}$ = 2.2 Hz, 1C), 72.77 (2C), 34.27 (1C), 31.93 (1C), 29.84 (1C), 29.24 (1C), 28.21 (1C), 26.44 (1C), 22.78 (1C), 14.22 (1C)

IR $v_{max}$ (cm$^{-1}$): 3109 (C-H stretch, sp$^2$ hybridised), 2922 (C-H stretch, sp$^3$ hybridised), 1981 (C-H bend, overtones), 1757 (C=O stretch, ester)

# NMR Spectra SR-*n*-Re
## SR-1-Re

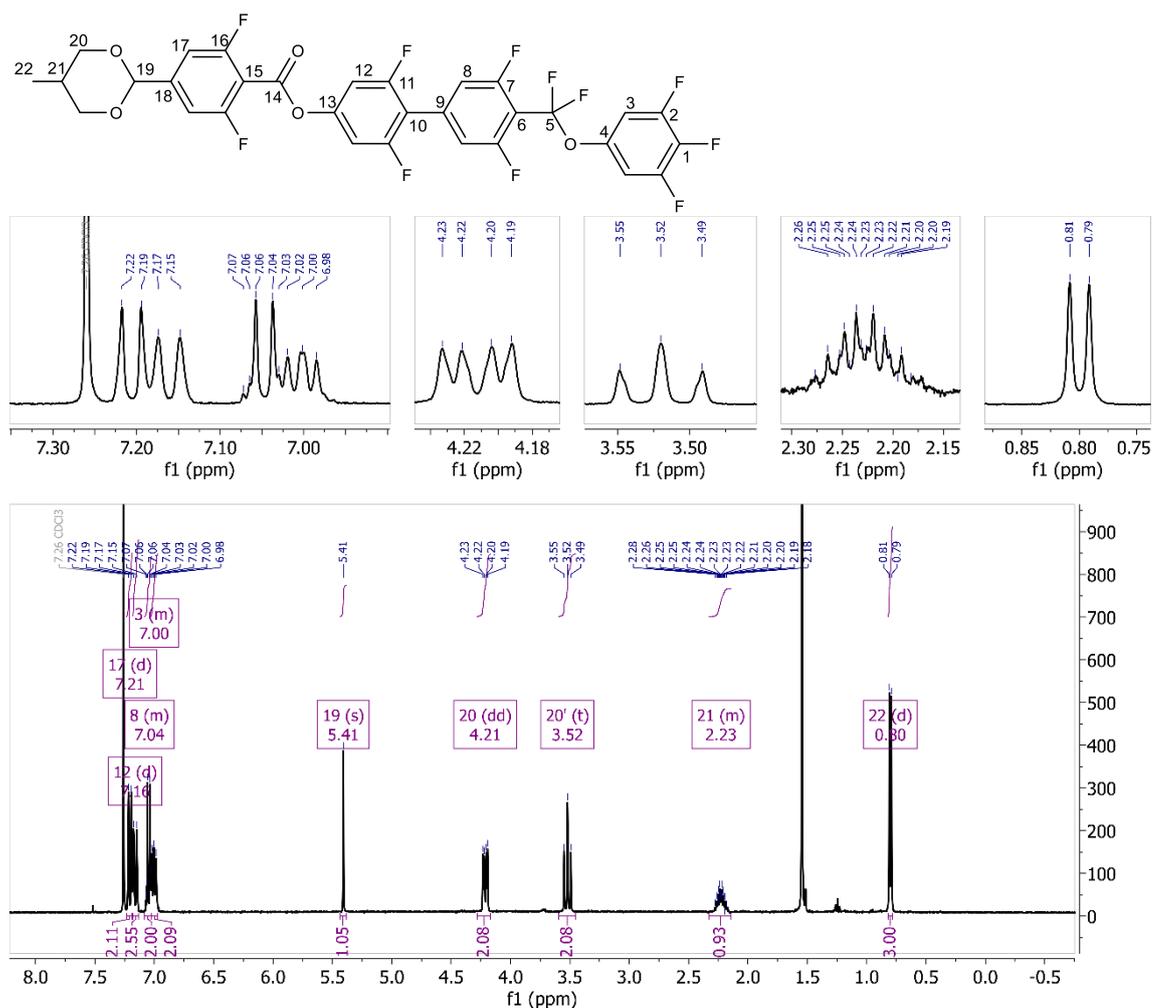

Figure S6: ¹H NMR spectrum of SR-1-Re in CDCl₃

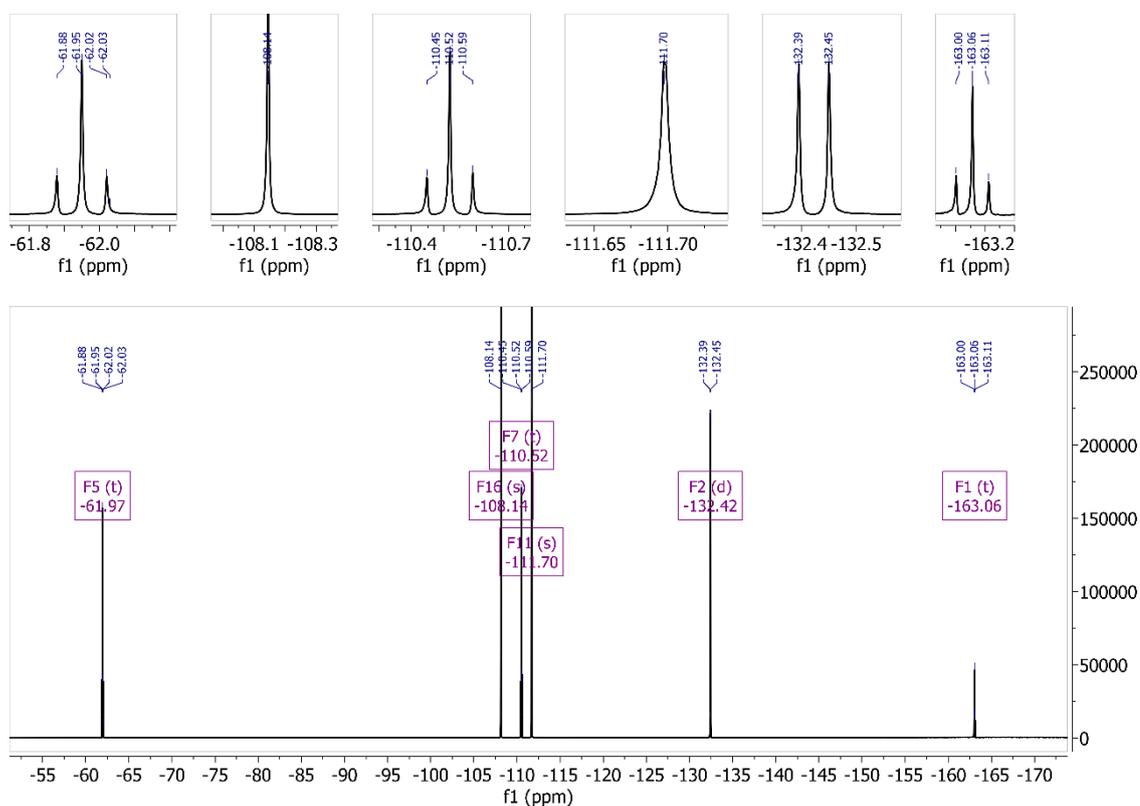

Figure S7: ¹H-decoupled ¹⁹F NMR spectrum of SR-1-Re in CDCl₃

**SR-2-Re**

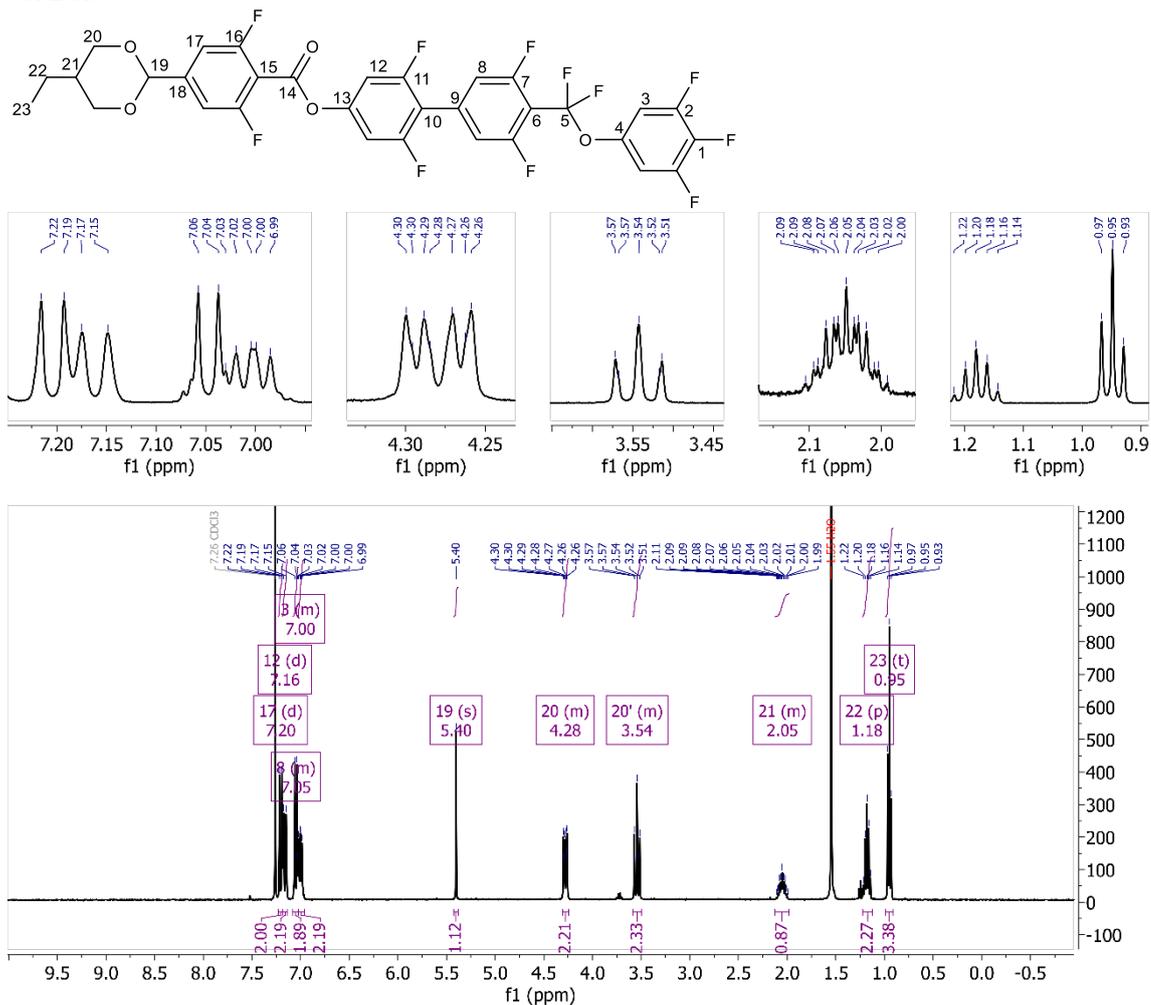

Figure S8: $^1$H NMR spectrum of SR-2-Re in CDCl$_3$

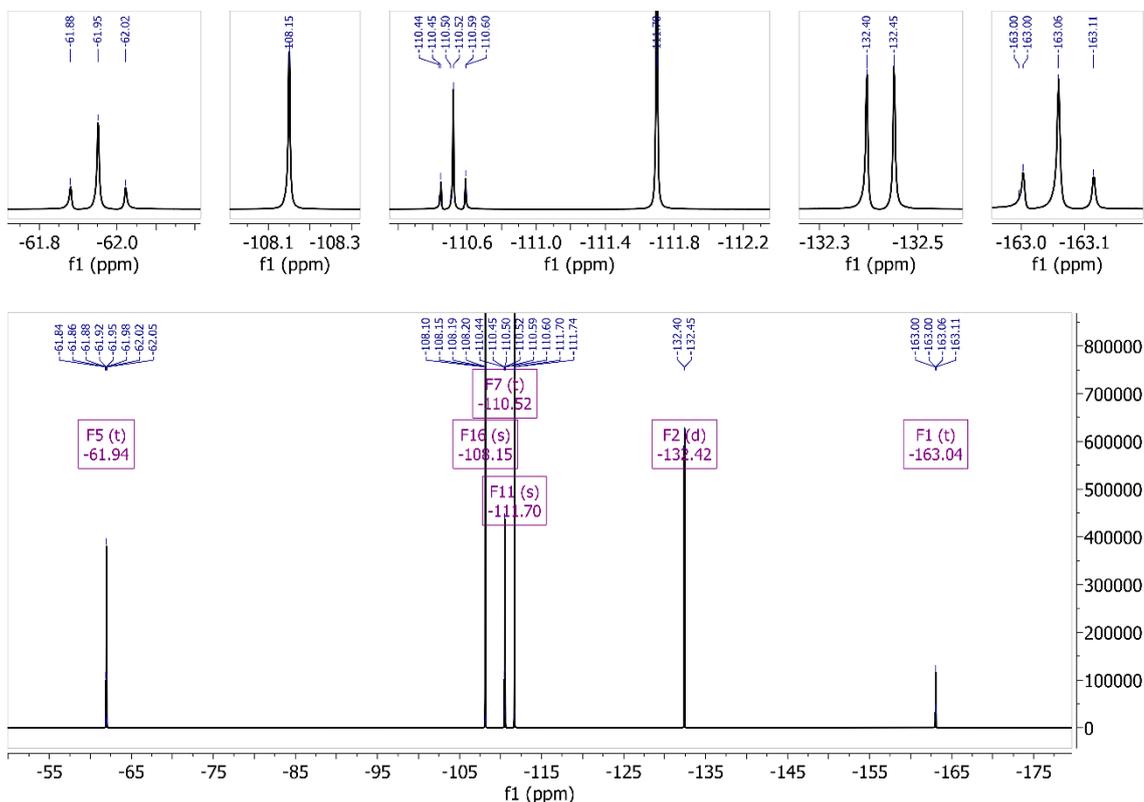

Figure S9: $^1$H-decoupled $^{19}$F NMR spectrum of SR-2-Re in CDCl$_3$

**SR-3-Re**

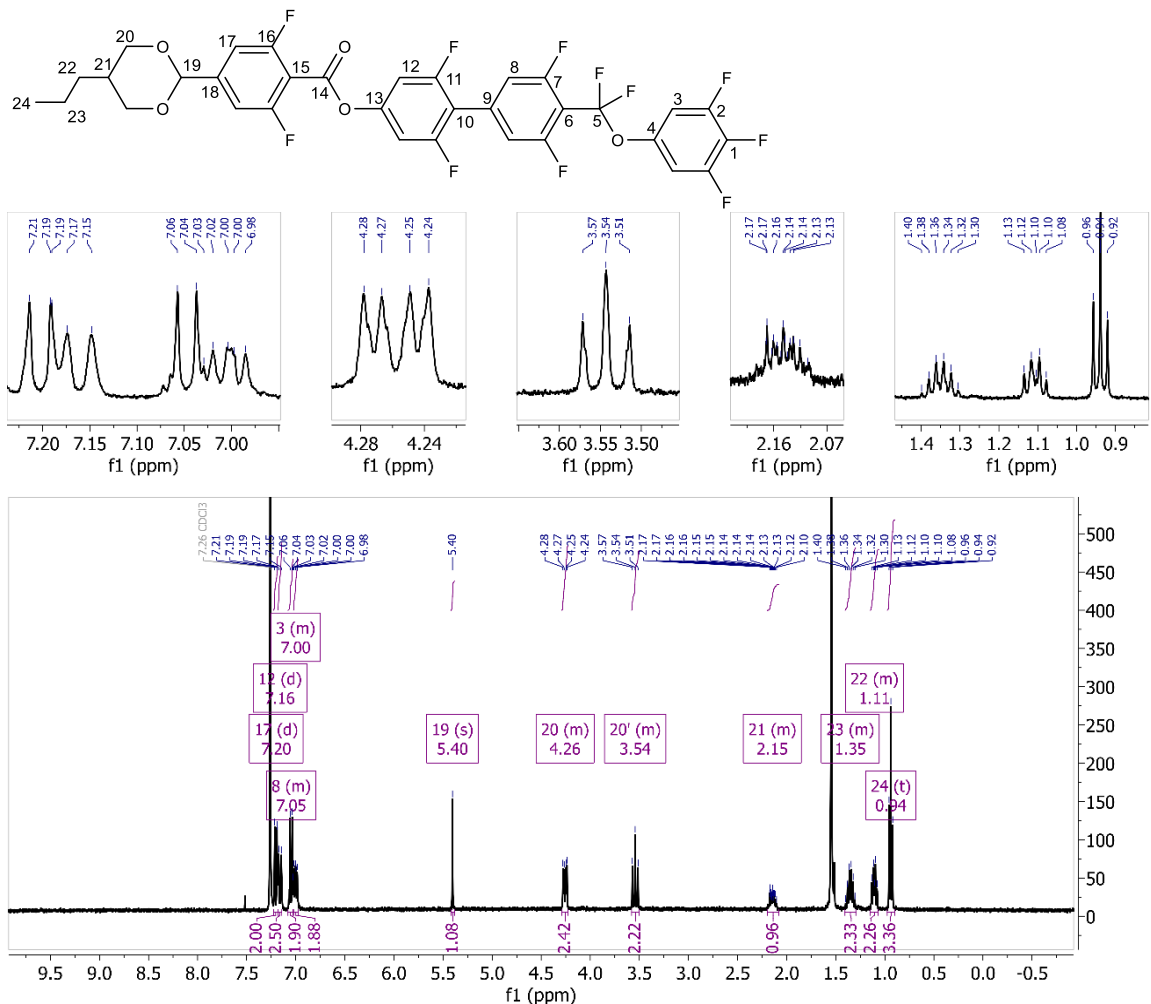

Figure S10: ¹H NMR spectrum of SR-3-Re in CDCl₃

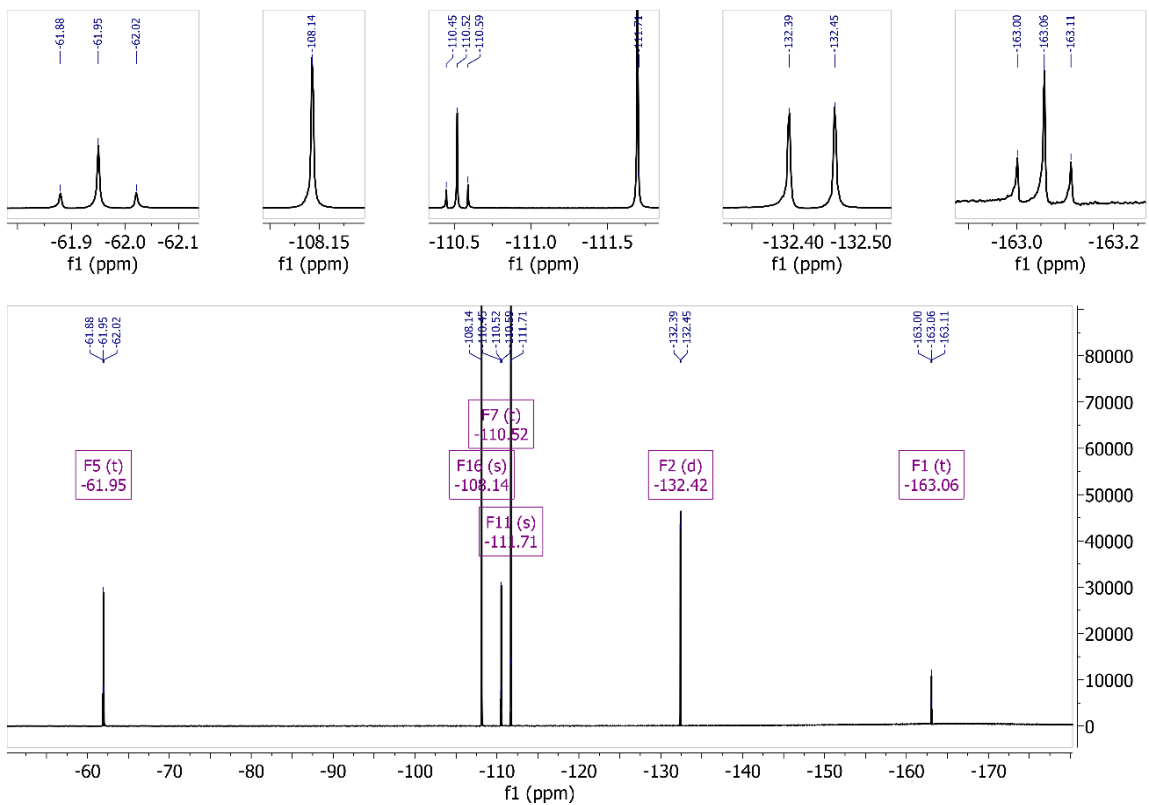

Figure S11: ¹H-decoupled ¹⁹F NMR spectrum of SR-3-Re in CDCl₃

**SR-4-Re**

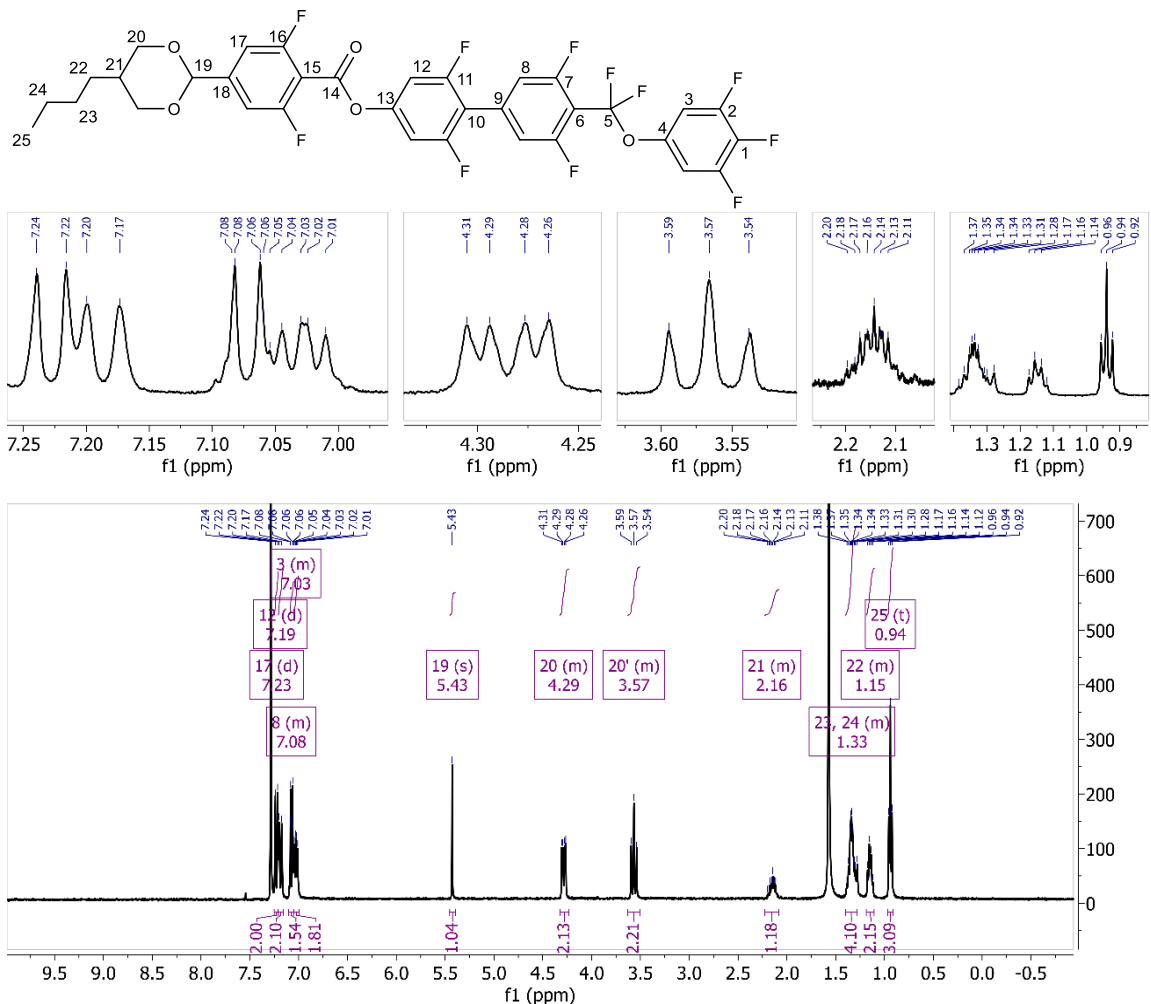

Figure S12: $^1$H NMR spectrum of SR-4-Re in CDCl$_3$

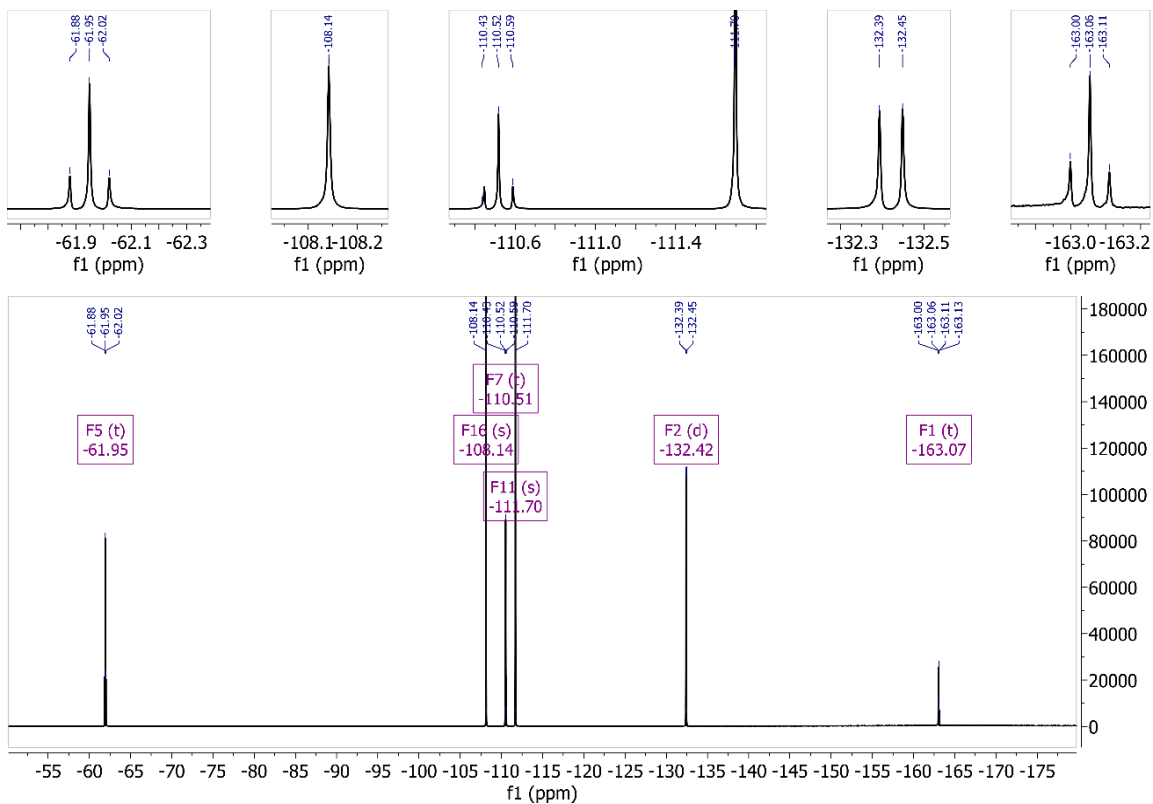

Figure S13: $^1$H-decoupled $^{19}$F NMR spectrum of SR-4-Re in CDCl$_3$

**SR-5-Re**

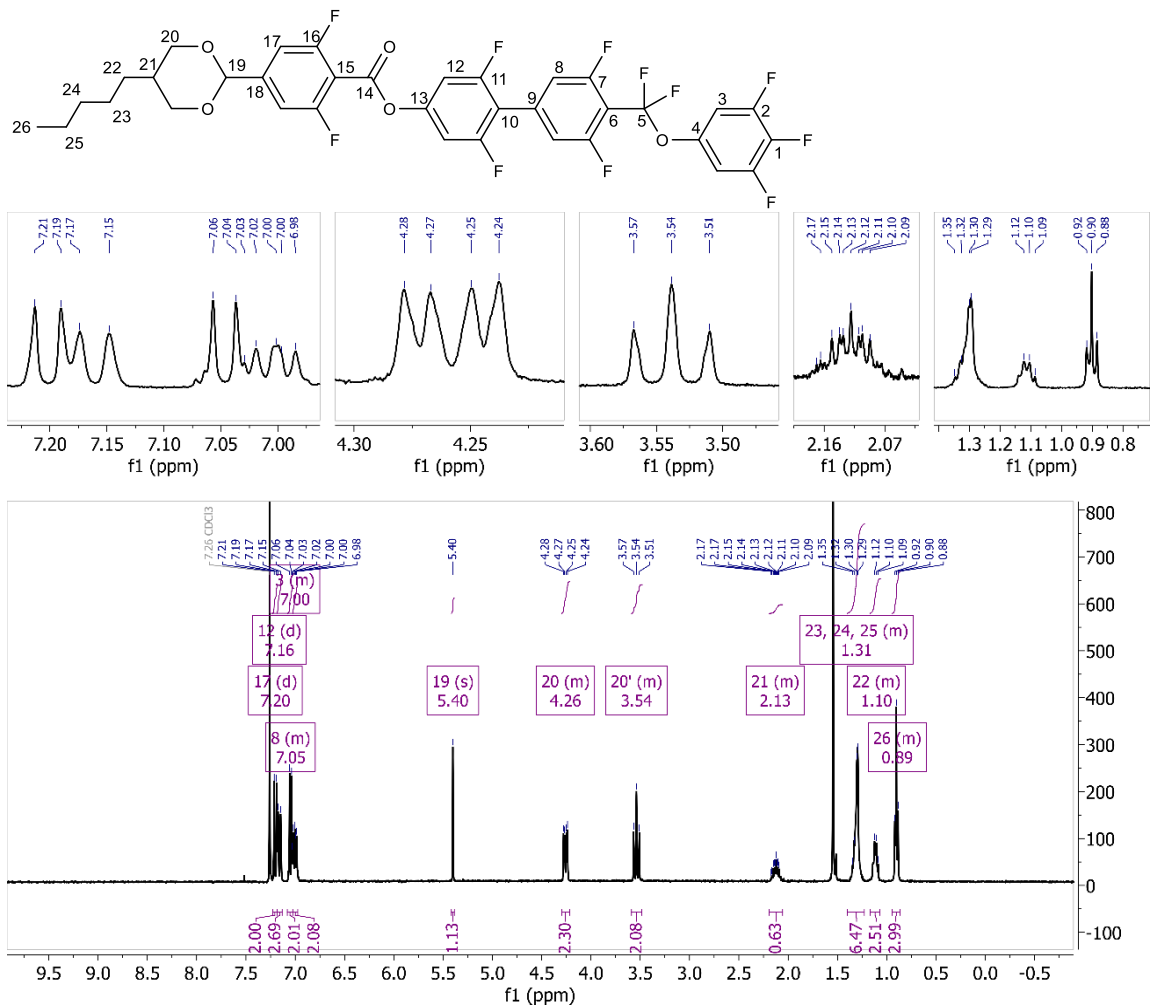

Figure S14: ¹H NMR spectrum of SR-5-Re in CDCl₃

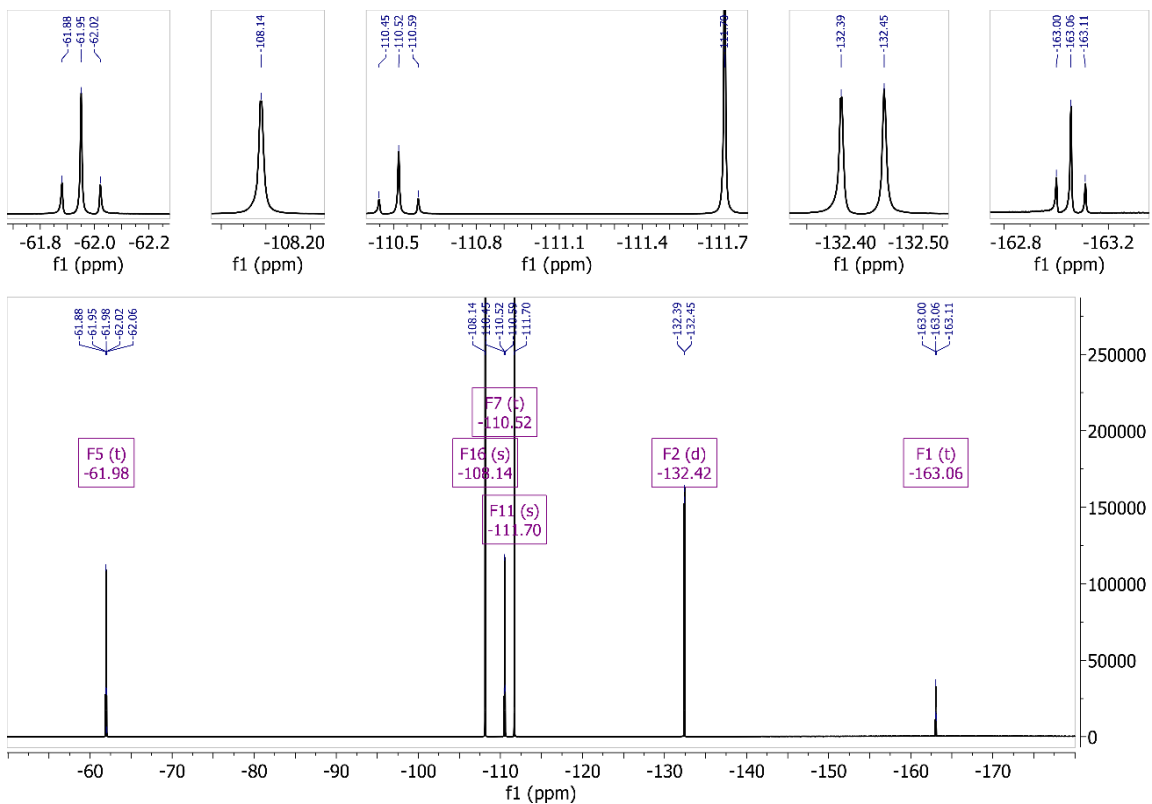

Figure S15: ¹H-decoupled ¹⁹F NMR spectrum of SR-5-Re in CDCl₃

**SR-6-Re**

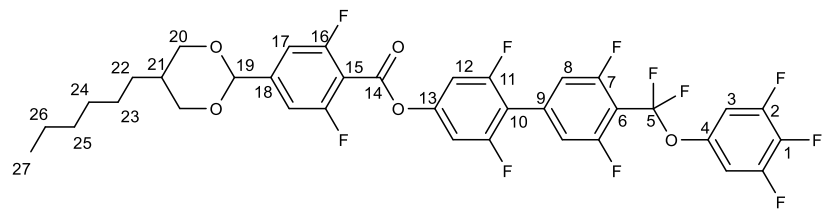

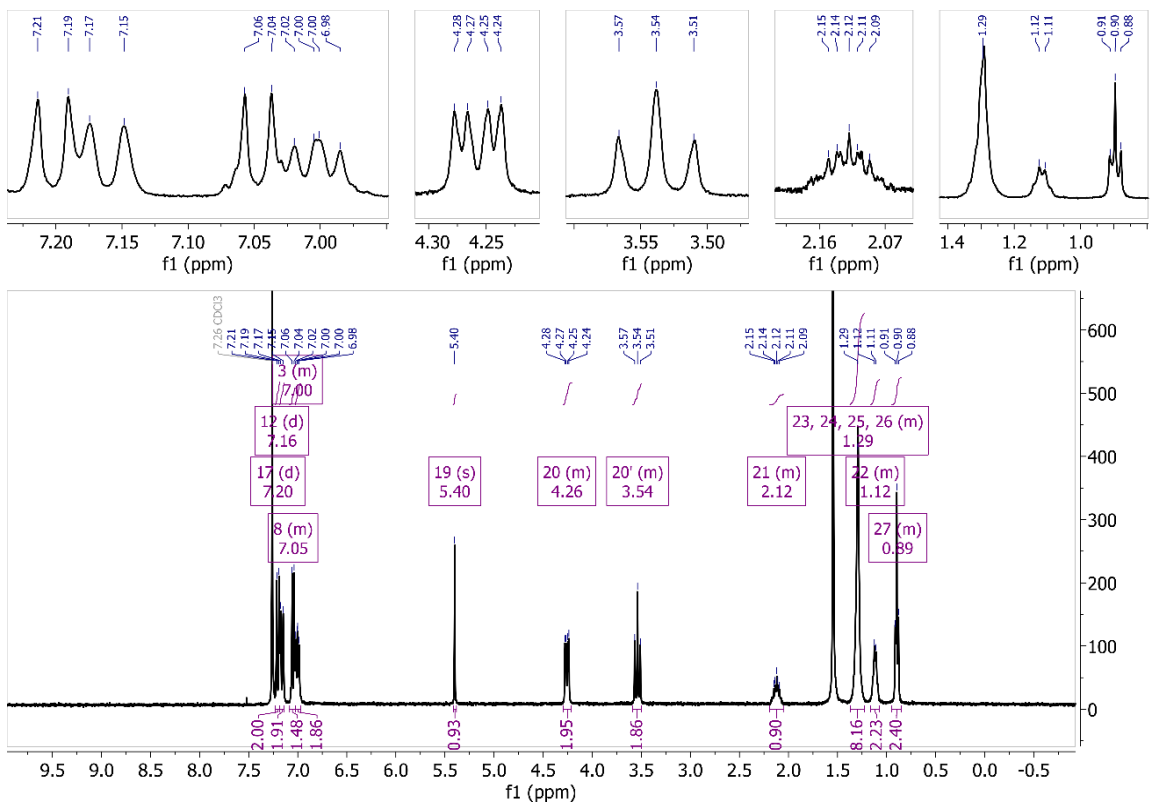

Figure S16: $^1$H NMR spectrum of SR-6-Re in CDCl$_3$

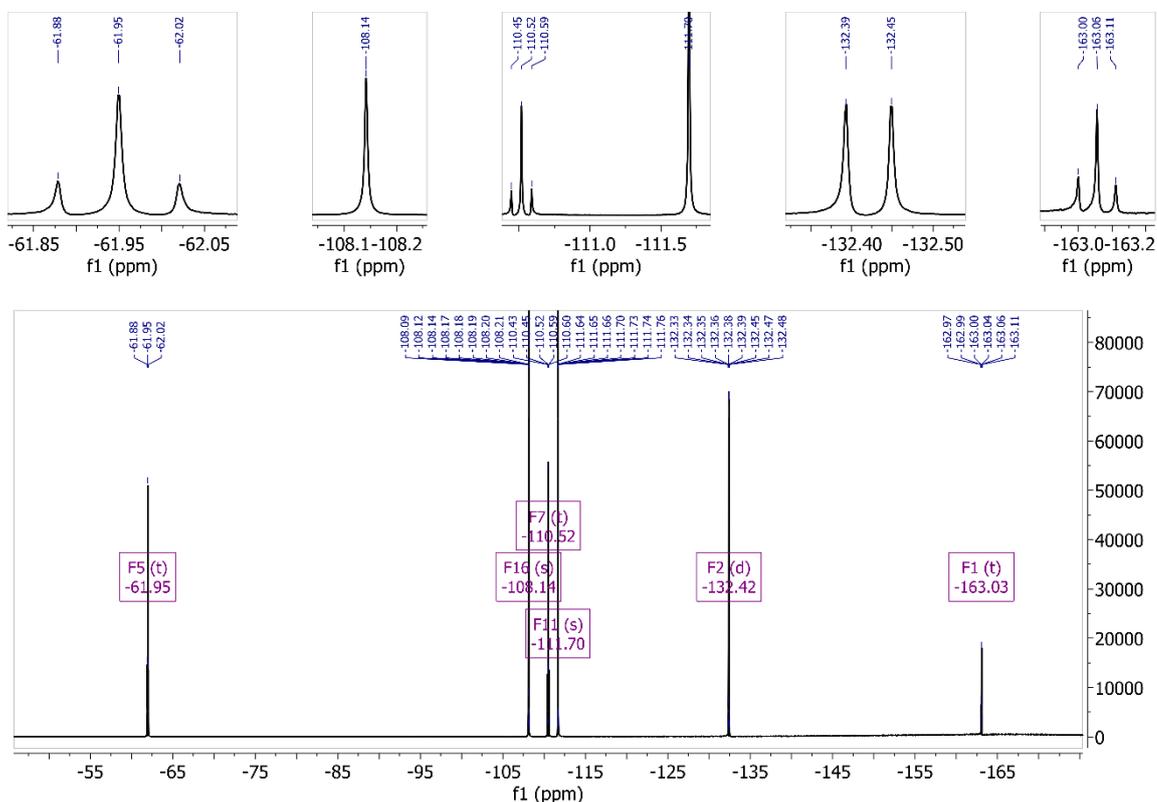

Figure S17: $^1$H-decoupled $^{19}$F NMR spectrum of SR-6-Re in CDCl$_3$

**SR-7-Re**

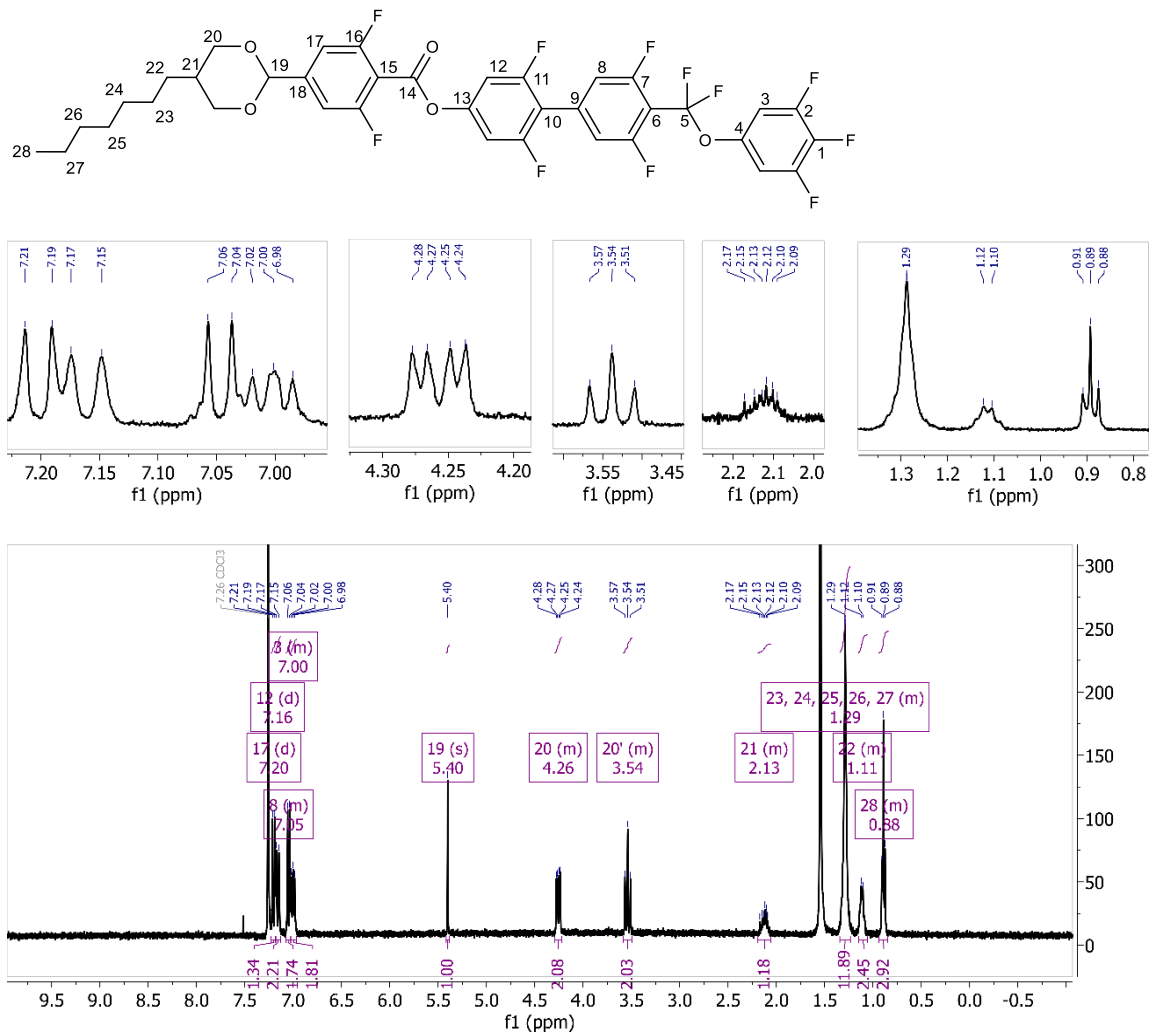

Figure S18: ¹H NMR spectrum of SR-7-Re in CDCl₃

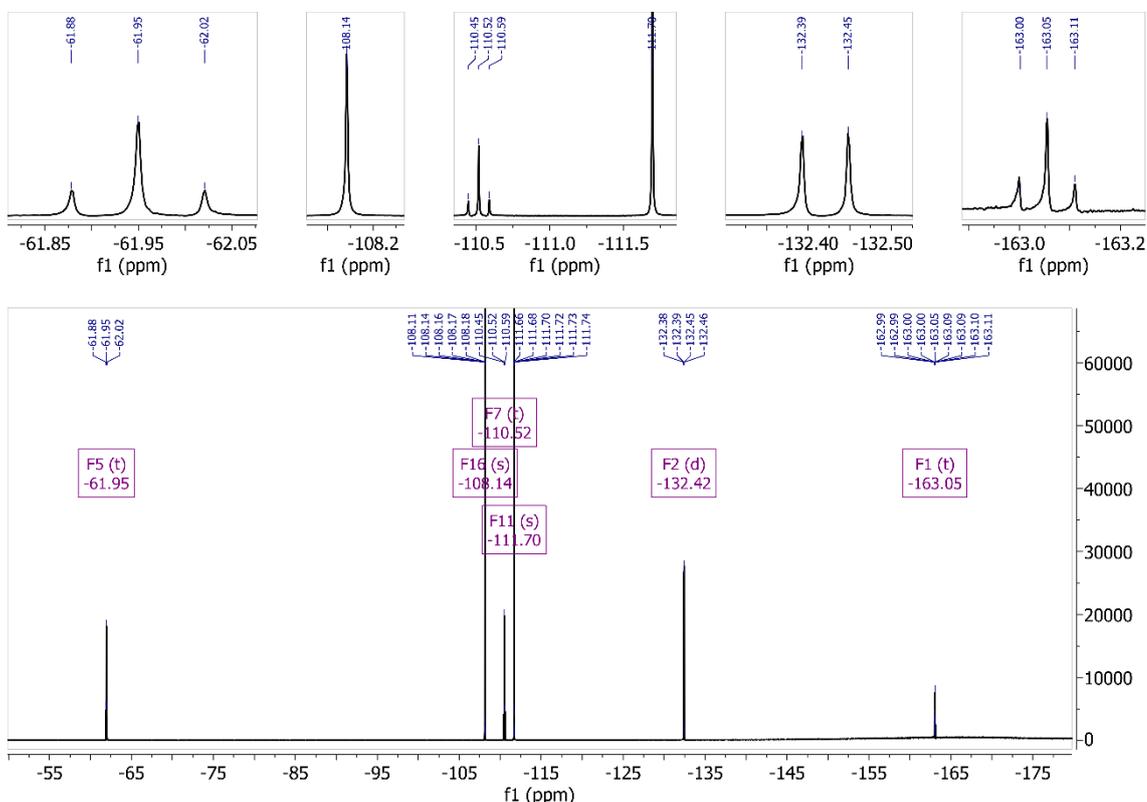

Figure S19: ¹H-decoupled ¹⁹F NMR spectrum of SR-7-Re in CDCl₃

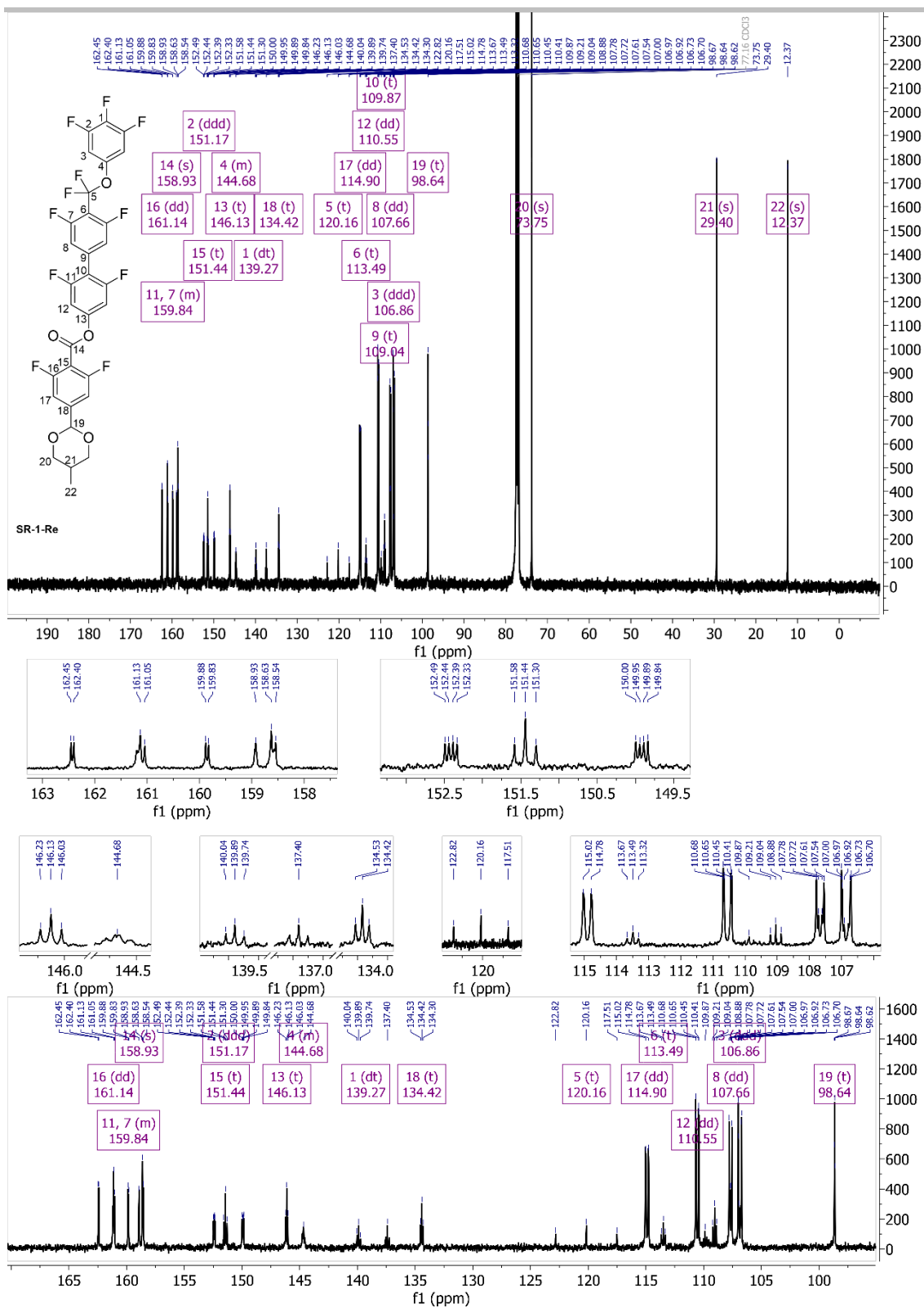

*Figure S20: Representative assigned $^{13}$C NMR spectrum of SR-1-Re in CDCl$_3$. Top: Full spectrum; bottom: enlarged aromatic region*

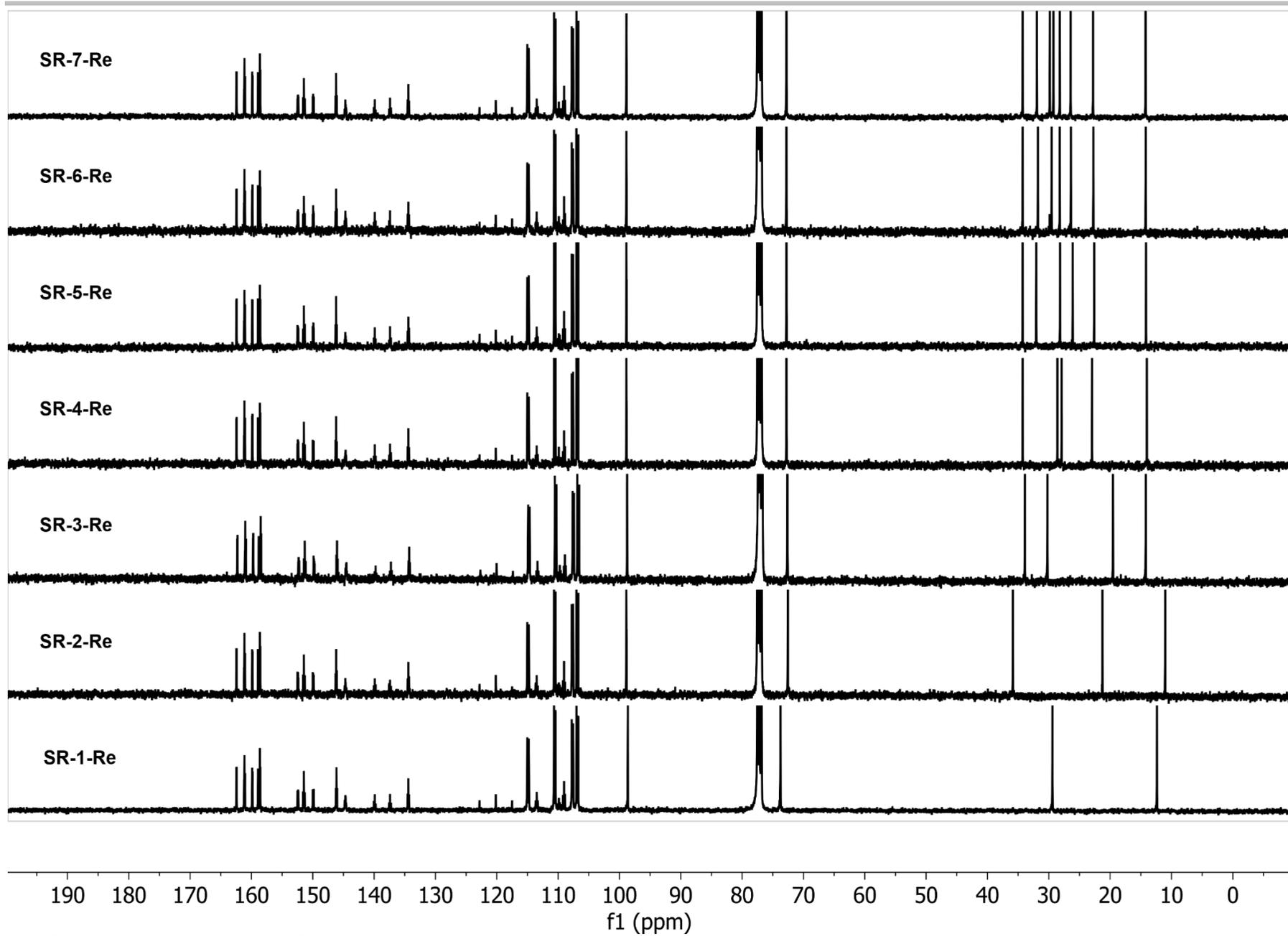

Figure S21: Stacked $^{13}$C NMR spectra for series SR-n-Re.

### 3.2 Synthesis of GS-*n*-Re series

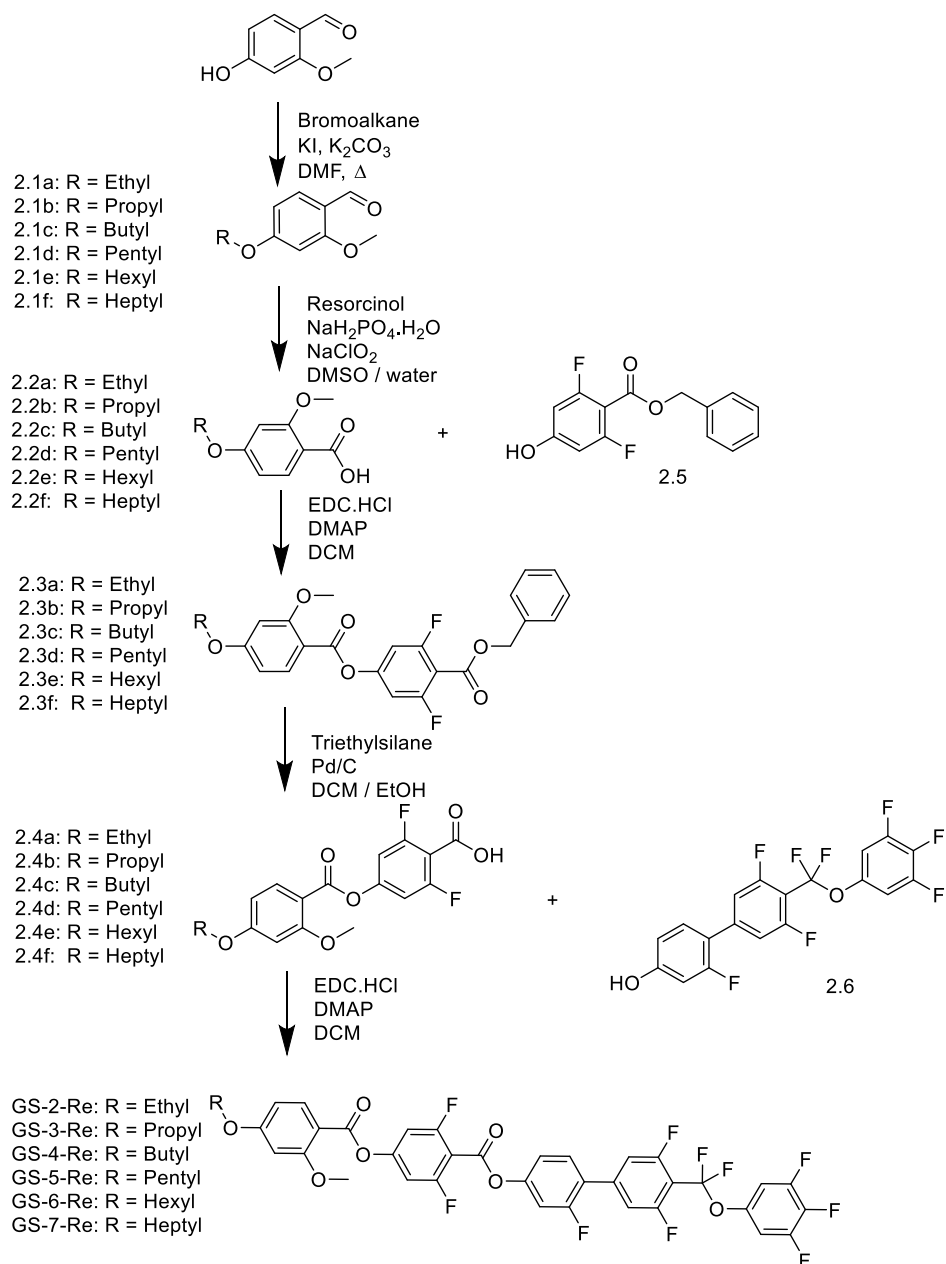

2.1a: R = Ethyl
2.1b: R = Propyl
2.1c: R = Butyl
2.1d: R = Pentyl
2.1e: R = Hexyl
2.1f: R = Heptyl

2.2a: R = Ethyl
2.2b: R = Propyl
2.2c: R = Butyl
2.2d: R = Pentyl
2.2e: R = Hexyl
2.2f: R = Heptyl

2.3a: R = Ethyl
2.3b: R = Propyl
2.3c: R = Butyl
2.3d: R = Pentyl
2.3e: R = Hexyl
2.3f: R = Heptyl

2.4a: R = Ethyl
2.4b: R = Propyl
2.4c: R = Butyl
2.4d: R = Pentyl
2.4e: R = Hexyl
2.4f: R = Heptyl

GS-2-Re: R = Ethyl
GS-3-Re: R = Propyl
GS-4-Re: R = Butyl
GS-5-Re: R = Pentyl
GS-6-Re: R = Hexyl
GS-7-Re: R = Heptyl

*Scheme 2. The synthetic route to materials GS-n-Re. Intermediates 1a[3], 2a[3], 5[4], and 6[5], and mesogens GS-1-Re[4] have been reported previously.*

**General Methods:**

**2A. Williamson ether**

Potassium carbonate, potassium iodide, and 4-hydroxy-2-methoxybenzaldehyde were added to dry DMF under an argon atmosphere and stirred for 10 minutes. The appropriate alkyl bromide was added and the reaction mixture heated to 70 °C overnight. The mixture was cooled, added to 500 ml of water, and stirred until an orange precipitate formed. This was collected by vacuum filtration and the crude product was carried forward without further purification.

**2B. Oxidation**

The aldehyde and resorcinol were dissolved in DMSO. Sodium chlorite and sodium dihydrogen phosphate were dissolved in water, and this was added carefully to the organics and the reaction left stirring overnight. The reaction mixture was diluted with 100 ml water, acidified, and the resulting precipitate collected by vacuum filtration.

### 2C. Esterification

The appropriate benzoic acid and EDC.HCl were dissolved in DCM and stirred for 10 minutes. The corresponding phenol and 4-dimethylaminopyridine (DMAP) were added and the reaction was left stirring at room temperature overnight. The reaction was washed 3x with water and the solvent removed *in vacuo*. The crude product was recrystallised from ethanol to yield the product as a white solid.

### 2D. Deprotection

Under an argon atmosphere triethylsilane was added dropwise to a stirred solution of 2.3(a-f) and 5 % Pd/C in a 1:1 mix of ethanol and DCM. The reaction was stirred for 15 minutes after addition was complete, then filtered through celite and the solvent removed *in vacuo*. The crude product was washed with hexane to yield the product as a white powder.

**4-propoxy-2-methoxybenzaldehyde 2.1b.**

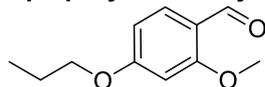

Method 2A

| | | | |
|---|---|---|---|
| 4-hydroxy-2-methoxybenzaldehyde | 2.067 g, | 13.6 mmol, | 1 eq. |
| 1-bromopropane | 1.4 ml, | 15 mmol, | 1.1 eq. |
| $K_2CO_3$ | 3.776 g, | 27 mmol, | 2 eq. |
| KI | 600 mg, | 3.6 mmol, | 0.3 eq. |
| DMF | 30 ml | | |
| 20 hrs | | | |
| Yield 2.89 g. | | | |

$^1$H NMR (400 MHz, CDCl$_3$) δ = 10.27 (s, 1H), 7.78 (d, *J*=8.6, 1H), 6.52 (dd, *J*=8.6, $^4J$=2.2, 1H), 6.43 (d, $^4J$=2.2, 1H), 3.98 (t, *J*=6.6, 2H), 3.88 (s, 3H), 1.89 – 1.75 (m, 2H), 1.04 (t, *J*=7.4, 3H).

**4-butyloxy-2-methoxybenzaldehyde 2.1c.**

Method 2A

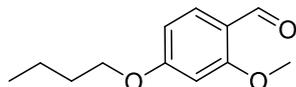

| | | | |
|---|---|---|---|
| 4-hydroxy-2-methoxybenzaldehyde | 2.063 g, | 13.6 mmol, | 1 eq. |
| 1-bromobutane | 1.7 ml, | 15 mmol, | 1.1 eq. |
| $K_2CO_3$ | 3.998 g, | 29 mmol, | 2.1 eq. |
| KI | 360 mg, | 2.1 mmol, | 0.2 eq. |
| DMF | 30 ml | | |
| 15 hrs | | | |
| Yield 2.82 g. | | | |

$^1$H NMR (400 MHz, CDCl$_3$) δ = 10.28 (s, 1H), 7.79 (d, *J*=8.6, 1H), 6.53 (dd, *J*=8.6, $^4J$=2.2, 1H), 6.44 (d, $^4J$=2.2, 1H), 4.03 (t, *J*=6.5, 2H), 3.90 (s, 2H), 1.85 – 1.73 (m, 2H), 1.57 – 1.43 (m, 2H), 0.99 (t, *J*=7.4, 3H).

**4-pentyloxy-2-methoxybenzaldehyde 2.1d.**

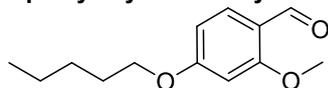

Method 2A

| | | | |
|---|---|---|---|
| 4-hydroxy-2-methoxybenzaldehyde | 2.085 g, | 13.7 mmol, | 1 eq. |
| 1-bromopentane | 1.9 ml, | 15 mmol, | 1.1 eq. |
| $K_2CO_3$ | 3.785 g, | 27 mmol, | 2 eq. |
| KI | 200 mg, | 1.2 mmol, | 0.1 eq. |
| DMF | 30 ml | | |
| 16 hours | | | |
| Yield 2.92 g. | | | |

$^1$H NMR (400 MHz, CDCl$_3$) δ = 10.28 (s, 1H), 7.79 (d, *J*=8.7, 1H), 7.26 (s, 1H), 6.53 (dd, *J*=8.7, $^4J$=2.1, 1H), 6.44 (d, $^4J$=2.1, 1H), 4.02 (t, *J*=6.5, 2H), 3.90 (s, 2H), 1.81 (p, *J*=6.5, 2H), 1.51 – 1.32 (m, 4H), 0.94 (t, *J*=7.0, 3H).

**4-hexyloxy-2-methoxybenzaldehyde 2.1e.**

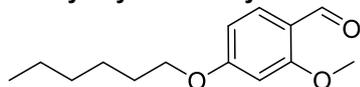

Method 2A

| | | | |
|---|---|---|---|
| 4-hydroxy-2-methoxybenzaldehyde | 2.072 g, | 13.6 mmol, | 1 eq. |

1-bromohexane      2.1 ml,      15 mmol,      1.1 eq.
$K_2CO_3$      3.777 g,      27 mmol,      2 eq.
KI      150 mg,      1.2 mmol,      0.06 eq.
DMF      30 ml
15 hours
Yield 3.11 g.
$^1$H NMR (400 MHz, CDCl$_3$) δ = 10.31 (s, 1H), 7.82 (d, $J$=8.7, 1H), 6.56 (dd, $J$=8.7, $^4J$=2.2, 1H), 6.46 (d, $^4J$=2.2, 1H), 4.05 (t, $J$=6.5, 2H), 3.92 (s, 3H), 1.83 (p, $J$=6.7, 2H), 1.55 – 1.44 (m, 2H), 1.44 – 1.33 (m, 4H), 0.98 – 0.90 (m, 3H).

**4-heptyloxy-2-methoxybenzaldehyde 2.1f.**

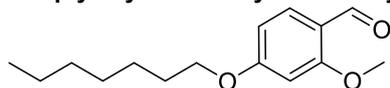

Method 2A
4-hydroxy-2-methoxybenzaldehyde      2.072 g,      13.6 mmol,      1 eq.
1-bromoheptane      2.4 ml,      15 mmol,      1.1 eq.
$K_2CO_3$      3.774 g,      27 mmol,      2 eq.
KI      100 mg,      1.2 mmol,      0.04 eq.
DMF      30 ml
15 hours
Yield 3.16 g.
$^1$H NMR (400 MHz, CDCl$_3$) δ = 10.31 (s, 1H), 7.82 (d, $J$=8.7, 1H), 6.56 (dd, $J$=8.7, $^4J$=2.2, 1H), 6.47 (d, $^4J$=2.2, 1H), 4.05 (t, $J$=6.5, 2H), 3.92 (s, 3H), 1.89 – 1.75 (m, 2H), 1.55 – 1.39 (m, 2H), 1.42 – 1.29 (m, $J$=3.5, 6H), 0.96 – 0.88 (m, 3H).

**4-propoxy-2-methoxybenzoic acid 2.2b.**

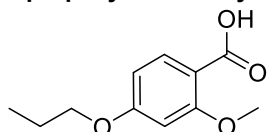

Method 2B
Aldehyde 2.1c      2.546 g,      11.4 mmol,      1 eq.
Resorcinol      4.751 g,      43 mmol,      4 eq.
$NaClO_2$      5.140 g,      57 mmol,      5 eq.
$NaH_2PO_4 \cdot H_2O$      6.860 g,      50 mmol,      4.4 eq.
DMSO      60 ml
Water      45 ml
23 hrs
Yield 1.68 g.
$^1$H NMR (400 MHz, CDCl$_3$) δ = 8.11 (d, $J$=8.8, 1H), 6.63 (dd, $J$=8.8, $^4J$=2.2, 1H), 6.53 (d, $^4J$=2.2, 1H), 4.04 (s, 3H), 3.99 (t, $J$=6.5, 2H), 1.90 – 1.77 (m, 2H), 1.05 (t, $J$=7.4, 3H).

**4-Butyloxy-2-methoxybenzoic acid 2.2c.**

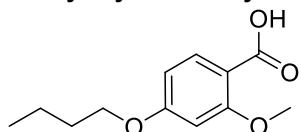

Method 2B
Aldehyde 2.1b      2.600 g,      12.5 mmol,      1 eq.
Resorcinol      3.975 g,      36 mmol,      3 eq.
$NaClO_2$      4.369 g,      48 mmol,      4 eq.
$NaH_2PO_4 \cdot H_2O$      5.810 g,      42 mmol,      3.3 eq.
DMSO      50 ml
Water      40 ml
23 hrs
Yield 2.84 g.
$^1$H NMR (400 MHz, CDCl$_3$) δ = 8.12 (d, $J$=8.8, 1H), 6.63 (dd, $J$=8.8, $^4J$=2.2, 1H), 6.53 (d, $^4J$=2.2, 1H), 4.04 (s, 3H), 4.03 (t, $J$=6.5, 2H), 1.85 – 1.73 (m, 2H), 1.57 – 1.43 (m, 2H), 0.99 (t, $J$=7.4, 3H).

**4-Pentyloxy-2-methoxybenzoic acid 2.2d.**

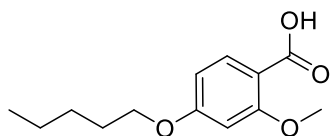

Method 2B
Aldehyde 2.1d        2.90 g,      13 mmol,    1 eq.
Resorcinol           5.10 g,      46 mmol,    3.5 eq.
NaClO$_2$            5.62 g,      62 mmol,    4.7 eq.
NaH$_2$PO$_4$.H$_2$O 7.49 g,      55 mmol,    4.2 eq.
DMSO                 60 ml
Water                45 ml
21 hrs
Yield 3.24 g.
$^1$H NMR (400 MHz, CDCl$_3$) δ = 8.14 (d, $J$=8.8, 1H), 6.66 (dd, $J$=8.8, $^4J$=2.2, 1H), 6.55 (d, $^4J$=2.2, 1H), 4.06 (overlapping singlet and triplet, 5H), 1.89 – 1.78 (m, 2H), 1.54 – 1.35 (m, 4H), 0.97 (t, $J$=7.0, 3H).

### 4-Hexyloxy-2-methoxybenzoic acid 2.2e.

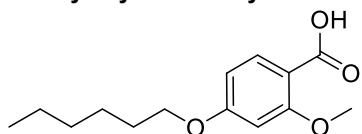

Method 2B
Aldehyde 2.1e        3.109 g,     13 mmol,    1 eq.
Resorcinol           2.17 g,      20 mmol,    1.5 eq.
NaClO$_2$            4.79 g,      53 mmol,    4 eq.
NaH$_2$PO$_4$.H$_2$O 6.50 g,      47 mmol,    3.6 eq.
DMSO                 60 ml
Water                40 ml
50 hrs
Yield 3.98 g.
$^1$H NMR (400 MHz, CDCl$_3$) δ = 8.14 (d, $J$=8.8, 1H), 6.66 (dd, $J$=8.8, $^4J$=2.2, 1H), 6.55 (d, $^4J$=2.2, 1H), 4.06 (overlapping singlet and triplet, 5H), 1.88 – 1.77 (m, 2H), 1.55 – 1.43 (m, 2H), 1.42 – 1.33 (m, 4H), 0.98 – 0.90 (m, 3H).

### 4-Heptyloxy-2-methoxybenzoic acid 2.2f.

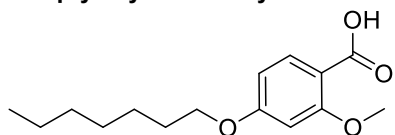

Method 2B
Aldehyde 2.1f        3.168 g,     13 mmol,    1 eq.
Resorcinol           2.609 g,     24 mmol,    1.8 eq.
NaClO$_2$            4.963 g,     55 mmol,    4.2 eq.
NaH$_2$PO$_4$.H$_2$O 6.370 g,     46 mmol,    3.5 eq.
DMSO                 60 ml
Water                40 ml
50 hrs
Yield 2.80 g.
$^1$H NMR (400 MHz, CDCl$_3$) δ = 8.15 (d, $J$=8.8, 1H), 6.66 (dd, $J$=8.8, $^4J$=2.3, 1H), 6.55 (d, $^4J$=2.3, 1H), 4.06 (overlapping singlet and triplet, 5H), 1.95 – 1.77 (m, 2H), 1.54 – 1.39 (m, 2H), 1.36 – 1.29 (m, 6H), 0.96 – 0.88 (m, 3H).

### 4-((benzyloxy)carbonyl)-3,5-difluorophenyl-4'-ethoxy-2'-methoxybenzoate 2.3a

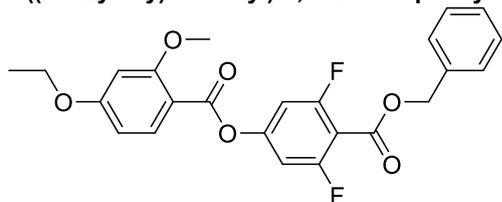

Method 2C
Benzoic acid 2.2a                             350 mg,      1.8 mmol,     1.1 eq.
Benzyl-4-hydroxy-2,6-difluorobenzoate         427 mg,      1.6 mmol,     1 eq.

| | | | |
|---|---|---|---|
| EDC.HCl | 636 mg, | 3.2 mmol, | 2 eq. |
| DMAP | 23 mg, | 0.18 mmol, | 0.1 eq. |
| DCM | 10 ml | | |
| 18 hrs | | | |
| Yield 170 mg. | | | |

$^1$H NMR (400 MHz, CDCl$_3$) δ = 8.05 – 7.97 (m, 1H), 7.47 – 7.32 (m, 5H), 6.91 (d, $J_{HF}$=9.6, 2H), 6.54 (overlapping signals, 2H), 5.40 (s, 2H), 4.17 – 4.07 (m, 2H), 3.91 (s, 3H), 1.51 – 1.42 (m, 3H).

### 4-((benzyloxy)carbonyl)-3,5-difluorophenyl-4'-propoxy-2'-methoxybenzoate 2.3b

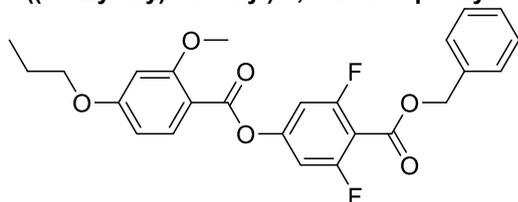

Method 2C

| | | | |
|---|---|---|---|
| Benzoic acid 2.2b | 962 mg, | 2.3 mmol, | 2.3 eq. |
| Benzyl-4-hydroxy-2,6-difluorobenzoate | 264 mg, | 1 mmol, | 1 eq. |
| EDC.HCl | 346 mg, | 1.8 mmol, | 1.8 eq. |
| DMAP | 10 mg, | 0.1 mmol, | 0.1 eq. |
| DCM | 10 ml | | |
| 18 hrs | | | |
| Yield 160 mg. | | | |

$^1$H NMR (400 MHz, CDCl$_3$) δ = 8.01 (d, $J$=8.6, 1H), 7.48 – 7.32 (m, 5H), 6.91 (d, $J_{HF}$=8.7, 1H), 6.58 – 6.46 (overlapping, 2H), 5.40 (s, 2H), 4.01 (t, $J$=6.7, 2H), 3.92 (s, 3H), 1.92 – 1.80 (m, 2H), 1.07 (t, $J$=7.4, 3H).

### 4-((benzyloxy)carbonyl)-3,5-difluorophenyl-4'-butoxy-2'-methoxybenzoate 2.3c

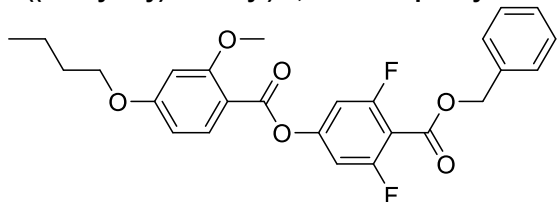

Method 2C

| | | | |
|---|---|---|---|
| Benzoic acid 2.2c | 436 mg, | 2 mmol, | 1 eq. |
| Benzyl-4-hydroxy-2,6-difluorobenzoate | 264 mg, | 2 mmol, | 1 eq. |
| EDC.HCl | 846 mg, | 4.3 mmol, | 2.1 eq. |
| DMAP | 22 mg, | 0.2 mmol, | 0.1 eq. |
| DCM | 10 ml | | |
| 18 hrs | | | |
| Yield 160 mg. | | | |

$^1$H NMR (400 MHz, CDCl$_3$) δ = 8.00 (d, $J$=8.8, 1H), 7.48 – 7.31 (m, 5H), 6.90 (d, $J_{HF}$=8.8, 2H), 6.57 – 6.49 (m, 2H), 5.39 (s, 2H), 4.04 (t, $J$=6.5, 2H), 3.92 (s, 3H), 1.86 – 1.74 (m, 2H), 1.51 (h, $J$=7.4, 2H), 0.99 (t, $J$=7.4, 3H).

### 4-((benzyloxy)carbonyl)-3,5-difluorophenyl-4'-pentyloxy-2'-methoxybenzoate 2.3d

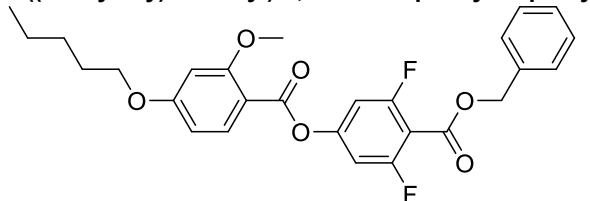

Method 2C

| | | | |
|---|---|---|---|
| Benzoic acid 2.2d | 514 mg, | 2.1 mmol, | 1 eq. |
| Benzyl-4-hydroxy-2,6-difluorobenzoate | 870 mg, | 2.1 mmol, | 1 eq. |
| EDC.HCl | 1.495 g, | 3.2 mmol, | 1.5 eq. |
| DMAP | 15 mg, | 0.1 mmol, | 0.05 eq. |
| DCM | 10 ml | | |
| 26 hrs | | | |
| Yield 508 mg. | | | |

$^1$H NMR (400 MHz, CDCl$_3$) δ = 8.00 (d, $J$=8.7, 1H), 7.48 – 7.30 (m, 5H), 6.94 – 6.87 (m, 2H), 6.58 – 6.45 (overlapping signals, 2H), 5.40 (s, 2H), 4.04 (t, $J$=6.5, 2H), 3.92 (s, 2H), 1.82 (p, $J$=6.8, 2H), 1.52 – 1.35 (m, 4H), 0.99 – 0.91 (m, 3H).

**4-((benzyloxy)carbonyl)-3,5-difluorophenyl-4'-hexyloxy-2'-methoxybenzoate 2.3e**

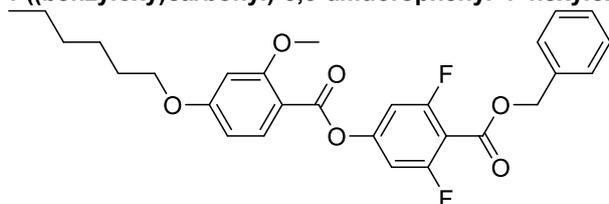

Method 2C
Benzoic acid 2.2e                        627 mg,       2.5 mmol,     1.1 eq.
Benzyl-4-hydroxy-2,6-difluorobenzoate    608mg,        2.3 mmol,     1 eq.
EDC.HCl                                  887 mg,       4.5 mmol,     2 eq.
DMAP                                     20 mg,        0.2 mmol,     0.1 eq.
DCM                                      20 ml
15 hrs
Colum chromatography: $R_F$ (DCM) 0.44.
Yield 250 mg.
$^1$H NMR (400 MHz, CDCl$_3$) δ = 8.00 (d, *J*=8.7, 1H), 7.49 – 7.30 (m, 5H), 6.95 – 6.86 (m, 2H), 6.58 – 6.49 (overlapping signals, 2H), 5.40 (s, 2H), 4.04 (t, *J*=6.6, 2H), 3.92 (s, 3H), 1.81 (p, *J*=6.7, 2H), 1.50 – 1.45 (m, 2H), 1.40 – 1.31 (m, 4H), 1.01 – 0.86 (m, 3H).

**4-((benzyloxy)carbonyl)-3,5-difluorophenyl-4'-heptyloxy-2'-methoxybenzoate 2.3f**

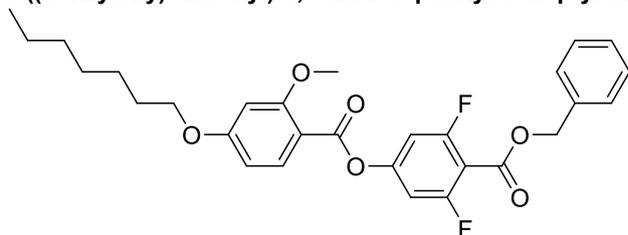

Method 2C
Benzoic acid 2.2f                        664 mg,       2.5 mmol,     1.1 eq.
Benzyl-4-hydroxy-2,6-difluorobenzoate    610 mg,       2.3 mmol,     1 eq.
EDC.HCl                                  903 mg,       4.6 mmol,     2 eq.
DMAP                                     10 mg,        0.2 mmol,     0.04 eq.
DCM                                      15 ml
15 hrs
Colum chromatography: $R_F$ (DCM) 0.5.
Yield 206 mg.
$^1$H NMR (400 MHz, CDCl$_3$) δ = 8.00 (d, *J*=9.0, 1H), 7.48 – 7.30 (m, 5H), 6.91 (d, $J_{HF}$=9.7, 2H), 6.52 (overlapping signals, 2H), 5.40 (s, 2H), 4.03 (t, *J*=6.1, 2H), 3.92 (s, 3H), 1.85 – 1.77 (m, 2H), 1.50 – 1.44 (m, 2H), 1.40 – 1.30 (m, 6H), 0.92 – 0.88 (m, 3H).

**4-carboxyl-3,5-difluorophenyl-4'-ethoxy-2'-methoxybenzoate 2.4a**

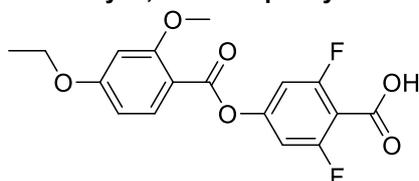

Method 2D
Benzyl ester 2.3a       165 mg,       0.4 mmol, 1 eq.
Triethylsilane          0.2 ml,       1.3 mmol, 3 eq.
5 % Pd/C                42mg,
Ethanol                 3 ml
DCM                     3 ml
Yield 114 mg.
$^1$H NMR (400 MHz, CDCl$_3$) δ = 8.02 (d, *J*=8.5, 1H), 6.96 (d, $J_{HF}$=9.9, 2H), 6.58 – 6.51 (overlapping signals, 2H), 4.13 (q, *J*=6.9, 2H), 3.93 (s, 3H), 1.46 (t, *J*=6.9, 3H).

**4-carboxyl-3,5-difluorophenyl-4'-propyloxy-2'-methoxybenzoate 2.4b**

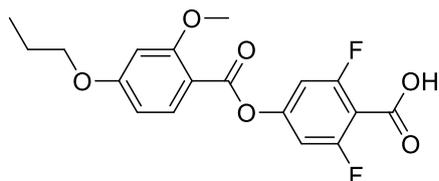

Method 2D
| | | | |
|---|---|---|---|
| Benzyl ester 2.3b | 158 mg, | 0.4 mmol, | 1 eq. |
| Triethylsilane | 0.2 ml, | 1.3 mmol, | 3 eq. |
| 5 % Pd/C | 47 mg, | | |
| Ethanol | 3 ml | | |
| DCM | 3 ml | | |

Yield 101 mg.
[1]H NMR (400 MHz, CDCl$_3$) δ = 8.01 (d, $J$=8.7, 1H), 6.96 (d, $J_{HF}$=9.9, 2H), 6.59 – 6.51 (overlapping signals, 2H), 4.01 (t, $J$=6.5, 2H), 3.93 (s, 3H), 1.86 (t, $J$=7.1, 2H), 1.07 (t, $J$=7.1, 3H).

**4-carboxyl-3,5-difluorophenyl-4'-butyloxy-2'-methoxybenzoate 2.4c**

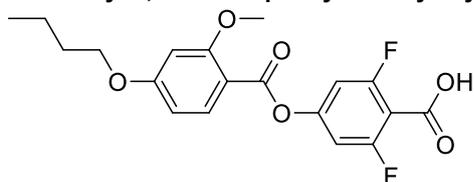

Method 2D
| | | | |
|---|---|---|---|
| Benzyl ester 2.3c | 154 mg, | 0.3 mmol, | 1 eq. |
| Triethylsilane | 0.2 ml, | 1.3 mmol, | 3 eq. |
| 5 % Pd/C | 42 mg, | | |
| Ethanol | 3 mL | | |
| DCM | 3 ml | | |

Yield 97 mg.
[1]H NMR (400 MHz, CDCl$_3$) δ = 8.01 (d, $J$=8.6, 1H), 6.95 (d, $J_{HF}$=9.7, 2H), 6.55 – 6.50 (overlapping signals, 1H), 4.05 (t, $J$=6.5, 2H), 3.93 (s, 3H), 1.83 – 1.77 (m, 2H), 1.55 – 1.49 (m, 2H), 1.00 (t, $J$=7.4, 3H).

**4-carboxyl-3,5-difluorophenyl-4'-pentyloxy-2'-methoxybenzoate 2.4d**

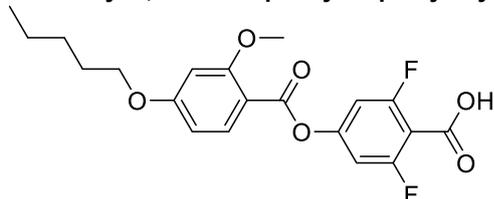

Method 2D
| | | | |
|---|---|---|---|
| Benzyl ester 2.3d | 500 mg, | 1 mmol, | 1 eq. |
| Triethylsilane | 0.55 ml, | 3.6 mmol, | 3 eq. |
| 5 % Pd/C | 116 mg, | | |
| Ethanol | 6 ml | | |
| DCM | 6 ml | | |

Yield 377 mg.
[1]H NMR (400 MHz, CDCl$_3$) δ = 8.01 (d, $J$=9.1, 1H), 6.99 – 6.92 (m, 2H), 6.58 – 6.50 (overlapping signals, 2H), 4.05 (t, $J$=6.5, 2H), 3.93 (s, 3H), 1.83 (p, $J$=6.7, 2H), 1.47 – 1.36 (m, 4H), 0.99 – 0.91 (m, 3H).

**4-carboxyl-3,5-difluorophenyl-4'-hexyloxy-2'-methoxybenzoate 2.4e**

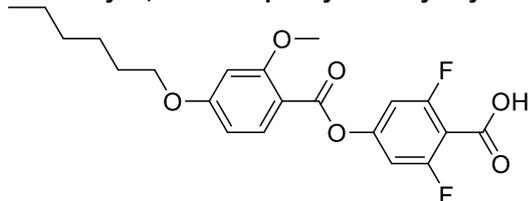

Method 2D
| | | | |
|---|---|---|---|
| Benzyl ester 2.3e | 230 mg, | 0.5 mmol, | 1 eq. |
| Triethylsilane | 0.22 ml, | 1.4 mmol, | 3 eq. |
| 5 % Pd/C | 45 mg, | | |
| Ethanol | 5 ml | | |

DCM                              5 ml
Yield 170 mg.

$^1$H NMR (400 MHz, CDCl$_3$) δ = 8.03 (d, *J*=8.8, 1H), 7.28 (s, 2H), 6.97 (d, *J*$_{HF}$=9.0, 1H), 6.60 – 6.52 (overlapping signals, 2H), 4.06 (t, *J*=6.6, 2H), 3.95 (s, 3H), 1.84 (p, *J*=6.7, 2H), 1.57 – 1.44 (m, 2H), 1.41 – 1.34 (m, 4H), 0.98 – 0.90 (m, 3H).

**4-carboxyl-3,5-difluorophenyl-4'-hexyloxy-2'-methoxybenzoate 2.4f**

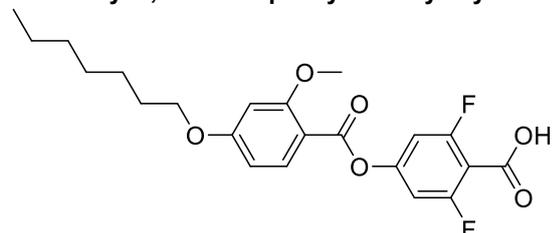

Method 2D
Benzyl ester 2.3f           230 mg,       0.5 mmol,     1 eq.
Triethylsilane              0.22 ml,      1.4 mmol,     3 eq.
5 % Pd/C                    45 mg,
Ethanol                     5 ml
DCM                         5 ml
Yield 170 mg.

$^1$H NMR (400 MHz, CDCl$_3$) δ = 8.04 (d, *J*=8.7, 1H), 6.98 (d, *J*$_{HF}$=9.0, 1H), 6.61 – 6.52 (overlapping signals, 2H), 4.07 (t, *J*=6.6, 2H), 3.95 (s, 3H), 1.90 – 1.79 (m, 2H), 1.55 – 1.41 (m, 2H), 1.44 – 1.30 (m, 6H), 0.96 – 0.89 (m, 3H).

**Ester GS-2-Re**

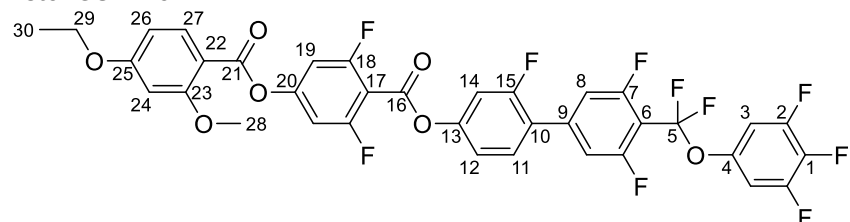

Method 2C
Benzoic acid 2.4a           57 mg,        0.16 mmol,    1.1 eq.
Phenol 2.6                  61 mg,        0.14 mmol,    1 eq.
EDC.HCl                     44 mg,        0.22 mmol,    1.5 eq.
DMAP                        2 mg,         0.016 mmol,   0.1 eq.
DCM                         10 ml
13 hrs
Yield 27 mg.

m.p.105 °C, T$_{SmCF-NF}$ 66 °C, T$_{NF-N}$ 183 °C, T$_{N-I}$ 237 °C

$^1$H NMR (400 MHz, CDCl$_3$) δ = 8.04 (d, *J*=8.8, 1H, Ar-**H**, H27), 7.50 (t, $^3$*J*$_{HF}$=8.5, 1H, Ar-**H,** H14), 7.29 – 7.18 (m$_{overlapping}$, 4H, Ar-**H**, H18, H11, H13), 7.06 – 6.96 (m$_{overlapping}$, 4H, Ar-**H**, H3, H19), 6.61 – 6.52 (m$_{overlapping}$, 2H, Ar-**H**, H24, H26), 4.14 (q, *J*=7.0, 2H, O-**CH$_2$**-CH$_3,$ H29), 3.94 (s, 3H, O-**CH$_3$**, H28), 1.47 (t, *J*=7.0, 3H, O-CH$_2$-**CH$_3$**, H30).

$^{19}$F NMR (376 MHz, CDCl$_3$) δ = -61.82 (t, $^4$*J*$_{FF}$=26.3, 2F, F5), -106.88 (d, $^3$*J*$_{HF}$=9.6, 2F, F18), -110.36 (td, $^4$*J*$_{FF}$=26.3, $^3$*J*$_{HF}$=10.6, 2F, F7), -113.63 (t, $^3$*J*$_{HF}$=9.8, 1F, F15), -132.45 (dd, $^3$*J*$_{FF}$=20.8, $^3$*J*$_{HF}$=8.0, 2F, F2), -163.11 (tt, $^3$*J*$_{FF}$=20.8, 5.8, 1F, F1).

$^{13}$C NMR (101 MHz, CDCl$_3$) δ = 165.19 (1C), 162.78 (1C), 161.84 (1C), 161.67 (dd, $^1$*J*$_{CF}$=259.2, $^3$*J*$_{CF}$=7.3, 1C), 161.34 – 158.46 (m, 2C), 159.50 (d, $^1$*J*$_{CF}$=252.5, 2C), 158.93 (s, apparent t, 1C), 155.43 (t, $^3$*J*$_{CF}$=14.2, 1C), 151.62 (d, $^3$*J*$_{CF}$=11.0, 1C, 152.53 – 149.50 (m, 2C), 144.85 – 144.36 (m, 1C), 140.87 – 140.35 (m, 1C), 134.72 (1C), 130.61 (d, $^3$*J*$_{CF}$=3.9, 1C), 123.59 (d, $^3$*J*$_{CF}$=13.6, 1C), 120.22 – 119.97 (m, 2C), 118.35 (d, $^4$*J*$_{CF}$=3.7, 1C), 113.12 (dt, $^2$*J*$_{CF}$=25.0, $^4$*J*$_{CF}$=3.4, 2C), 110.84 (d, $^2$*J*$_{CF}$=25.8, 1C), 109.23 (1C), 107.47 (dd, $^2$*J*$_{CF}$=17.8, $^3$*J*$_{CF}$=7.3, 2C), 107.14 (dd, $^2$*J*$_{CF}$=25.4, $^4$*J*$_{CF}$=3.8, 2C), 106.62 – 106.32 (m, 1C), 105.51 (1C), 99.41 (1C), 64.05 (1C), 56.03 (1C), 14.63(1C).

HRMS (ESI): *m/z* calcd for C$_{36}$H$_{20}$O$_7$F$_{10}$ [M+H]$^+$: 755.11221. Found: 755.11225. Difference 0.052 ppm

IR *v*max (cm$^{-1}$): 3100 (C-H stretch), 2983 (C-H stretch, 1747 (C=O stretch, ester), 1709.

**Ester GS-3-Re**

Method 2C
| | | | |
|---|---|---|---|
| Benzoic acid 2.4b | 50 mg, | 0.15 mmol, | 1.3 eq. |
| Phenol 2.6 | 51 mg, | 0.12 mmol, | 1 eq. |
| EDC.HCl | 34 mg, | 0.17 mmol, | 1.4 eq. |
| DMAP | 2 mg, | 0.016 mmol, | 0.1 eq. |
| DCM | 10 ml | | |

13 hrs
Yield 14 mg.

m.p.82 °C, $T_{SmCF-NF}$ 92 °C, $T_{NF-N}$ 167 °C, $T_{N-I}$ 225 °C

$^1$H NMR (400 MHz, CDCl$_3$) δ = 8.03 (d, *J*=8.6, 1H Ar-**H,** H27), 7.50 (t, $^3J_{HF}$=8.6, 1H, Ar-**H**, H14), 7.26 – 7.17 (m$_{overlapping}$, 4H, Ar-**H**, H18, H11, H13), 7.06 – 6.96 (m$_{overlapping}$, 4H, Ar-**H**, H3, H19), 6.60 – 6.52 (m$_{overlapping}$, 2H, Ar-**H**, H24, H26), 4.02 (t, *J*=6.5, 2H, O-**CH$_2$**-CH$_2$, H29), 3.94 (s, 2H, O-**CH$_3$**, H28), 1.93 – 1.80 (m, 2H, **CH$_2$**-CH$_3$, H30), 1.08 (t, *J*=7.4, 3H, CH$_2$-**CH$_3$**, H31).

$^{19}$F NMR (376 MHz, CDCl$_3$) δ = -61.82 (t, $^4J_{FF}$=26.3, 2F, F5), -106.90 (d, $^3J_{HF}$=9.8, 2F, F18), -110.37 (td, $^4J_{FF}$=26.3, $^3J_{HF}$=6.3, 2F, F7), -113.63 (t, $^3J_{HF}$ =9.8, 1F, F15), -132.45 (dd, $^3J_{FF}$ =21.0, $^3J_{HF}$=8.4, 2F, F2), -163.12 (tt, $^3J_{FF}$ =21.0, 6.0, 1F, F1).

$^{13}$C NMR (101 MHz, CDCl$_3$) δ = 165.51 (1C), 162.89 (1C), , 163.56 – 159.99 (m, 1C), 161.94, (1C), 161.58 – 158.84 (m, 2C), 159.58 (d, $^1J_{CF}$=251.1, 2C), 158.75 (1C), 156.05 – 155.27 (m, 1C), 151.71 (d, $^3J_{CF}$=9.9, 1C), 152.76 – 149.43 (m, 2C), 145.03 – 144.30 (m, 1C), 140.76 (1C), 134.78 (1C), 130.69 (1C), 123.91 – 123.46 (m, 1C123.07 – 116.98 (m, 1C), 118.42 (1C), 113.92 – 112.71 (m, 2C), 110.93 (d, $^2J_{CF}$=26.1, 1C), 109.24 (1C), 107.76 – 107.38 (m, 2C), 107.41 – 106.66 (m, 2C106.65 – 106.48 (m, 1C), 105.69 (1C), 99.48 (1C), 70.07 (1C), 56.11 (1C), 22.53 (1C), 10.55 (1C).

HRMS (ESI): *m/z* calcd for C$_{37}$H$_{22}$O$_7$F$_{10}$ [M+H]$^+$: 769.12786. Found: 769.12802. Difference 0.207 ppm

IR *v*$_{max}$ (cm$^{-1}$): 3109 (C-H stretch), 2945 (C-H stretch), 2880 (C-H stretch), 1746 (C=O stretch, ester), 1709.

**Ester GS-4-Re**

Method 2C
| | | | |
|---|---|---|---|
| Benzoic acid 2.4c | 168 mg, | 0.49 mmol, | 1.3 eq. |
| Phenol 2.6 | 152 mg, | 0.36 mmol, | 1 eq. |
| EDC.HCl | 158 mg, | 0.81 mmol, | 2.3 eq. |
| DMAP | 5 mg, | 0.05 mmol, | 0.1 eq. |
| DCM | 10 ml | | |

7 hrs, purified by column chromatography (DCM) R$_f$ 0.82, then recrystallized from EtOH
Yield 11 mg.

m.p.61 °C, $T_{SmCF-NF}$ 90 °C, $T_{NF-N}$ 147 °C, $T_{N-I}$ 219 °C

$^1$H NMR (400 MHz, CDCl$_3$) δ = 8.03 (d, *J*=8.7, 1H Ar-**H,** H27), 7.50 (t, $^3J_{HF}$=8.6, 1H, Ar-**H**, H14), 7.29 – 7.19 (m$_{overlapping}$, 4H, Ar-**H**, H18, H11, H13), 7.05 – 6.96 (m$_{overlapping}$, 4H, Ar-**H**, H3, H19), 6.60 – 6.52 (m$_{overlapping}$, 2H,

Ar-**H**, H24, H26), 4.06 (t, *J*=6.5, 2H, O-**CH$_2$**-CH$_2$, H29), 3.95 (s, 2H, O-**CH$_3$**, H28), 1.82 (p$_{apparent}$, *J*=7.0, 2H, O-CH$_2$-**CH$_2$**, H30), 1.59 – 1.46 (m, 2H, **CH$_2$**-CH$_3$, H31), 1.01 (t, *J*=7.4, 3H CH$_2$-**CH$_3$**, H32).

$^{19}$F NMR (376 MHz, CDCl$_3$) δ = -61.81 (t, $^4J_{FF}$=26.6, 2F, F5), -106.91 (d, $^3J_{HF}$=9.9, 2F, F18), -110.36 (td, $^4J_{FF}$=26.5, $^3J_{HF}$=10.8, 2F, F7), -113.63 (t, $^3J_{HF}$=9.6, 1F, F15), -132.45 (dd, $^3J_{FF}$=20.9, $^3J_{HF}$=8.0, 2F, F2), -163.12 (tt, $^3J_{FF}$=20.9, 6.0, 1F, F1).

$^{13}$C NMR (101 MHz, CDCl$_3$) δ = 165.57 (1C), 162.94 (1C), 162.01 (1C), 161.82 (dd, $^1J_{CF}$=259.0, $^3J_{CF}$=7.7, 1C), 160.07 (dd, $^1J_{CF}$=256.1, $^3J_{CF}$=4.7, 2C), 159.65 (d, $^1J_{CF}$=252.5, 2C), 159.08 (s, apparent t, 1C), 155.61 (t, $^3J_{CF}$=14.5, 1C), 151.77 (d, $^3J_{CF}$=11.1, 1C), 151.15 (ddd, $^1J_{CF}$=251.4, $^2J_{CF}$=10.4, $^3J_{CF}$=5.3, 2C), 145.13 – 144.33 (m, 1C), 141.35 – 140.51 (m, 1C), 138.61 (dt, $^1J_{CF}$=251.4, $^2J_{CF}$=16.6, 1C), 134.84 (1C), 130.76 (d, $^3J_{CF}$=3.8, 1C), 123.74 (d, $^3J_{CF}$=12.7, 1C), 120.26 (t, $^1J_{CF}$=272.1, 1C), 118.49 (d, $^4J_{CF}$=3.8, 1C), 113.27 (dt, $^2J_{CF}$=24.7, $^4J_{CF}$=3.4, 2C), 112.70 – 112.09 (m, 1C), 110.98 (d, $^2J_{CF}$=25.9, 1C), 109.29 (1C), 107.61 (dd, $^2J_{CF}$=17.8, $^3J_{CF}$=6.5, 2C), 107.28 (dd, $^2J_{CF}$=25.4, $^4J_{CF}$=3.9, 2C), 106.61 (t, $^2J_{CF}$=16.6, 1C), 105.75 (1C), 99.53 (1C), 68.36 (1C), 56.17 (1C), 31.24 (1C), 19.32 (1C), 13.94 (1C).

HRMS (ESI): *m/z* calcd for C$_{38}$H$_{24}$O$_7$F$_{10}$ [M+H]$^+$: 783.14351. Found: 783.14383. Difference 0.407 ppm

IR *v*max (cm$^{-1}$): 3110 (C-H stretch), 2945 (C-H stretch), 2880 (C-H stretch), 1746 (C=O stretch, ester), 1709.

**Ester GS-5-Re**

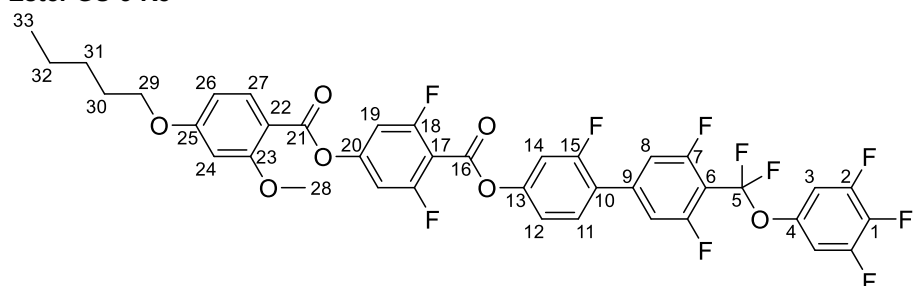

Method 2C
| | | | |
|---|---|---|---|
| Benzoic acid 2.4d | 88 mg, | 0.25 mmol, | 1.3 eq. |
| Phenol 2.6 | 85 mg, | 0.2 mmol, 1 eq. | |
| EDC.HCl | 62 mg, | 0.32 mmol, | 1.5 eq. |
| DMAP | 3 mg, | 0.02 mmol, | 0.2 eq. |
| DCM | 10 ml | | |
| 12 hrs | | | |
| Yield 27 mg. | | | |

m.p.73 °C, T$_{SmCF-NF}$ 95 °C, T$_{NF-SmAP}$ 126 °C T$_{SmAP-SmA}$ 129 °C T$_{SmA-N}$ 136 °C, T$_{N-I}$ 209 °C

$^1$H NMR (400 MHz, CDCl$_3$) δ = 8.03 (d, *J*=8.6, 1H Ar-**H**, H27), 7.50 (t, $^3J_{HF}$=8.5, 1H, Ar-**H**, H14), 7.25 – 7.18 (m$_{overlapping}$, 4H, Ar-**H**, H18, H11, H13), 7.06 – 6.96 (m$_{overlapping}$, 4H, Ar-**H**, H3, H19), 6.60 – 6.52 (m$_{overlapping}$, 2H, Ar-**H**, H24, H26), 4.06 (t, *J*=6.6, 2H, O-**CH$_2$**-CH$_2$, H29), 3.94 (s, 2H, O-**CH$_3$**, H28), 1.84 (p$_{apparent}$, *J*=6.6, 2H, O-CH$_2$-**CH$_2$**, H30), 1.53 – 1.36 (m, 4H, **CH$_2$**-**CH$_2$**-CH$_3$, H31, H32), 0.96 (t, *J*=7.0, 3H CH$_2$-**CH$_3$**, H33).

$^{19}$F NMR (376 MHz, CDCl$_3$) δ = -61.82 (t, $^4J_{FF}$=26.3, 2F, F5), -106.90 (d, $^3J_{HF}$=9.8, 2F, F18), -110.36 (td, $^4J_{FF}$=26.3, $^3J_{HF}$=10.6, 2F, F7), -113.63 (t, $^3J_{HF}$=9.8, 1F, F15), -132.45 (dd, $^3J_{FF}$=20.8, $^3J_{HF}$=8.1, 2F, F2), -163.12 (tt, $^3J_{FF}$=20.8, 5.7, 1F, F1).

$^{13}$C NMR (101 MHz, CDCl$_3$) δ = 165.41 (1C), 162.94 (1C), 161.85 (1C), 161.67 (dd, $^1J_{CF}$=258.9, $^3J_{CF}$=7.7, 1C), 160.85 – 157.89 (m, 2C), 161.31 – 158.52 (m, 2C), 158.94 (s, apparent t, 1C), 155.76 – 155.22 (m, 1C), 151.62 (d, $^3J_{CF}$=11.1, 1C), 151.00 (ddd, $^1J_{CF}$=251.4, $^2J_{CF}$=10.8, $^3J_{CF}$=5.6, 2C), 144.72 – 144.30 (m, 1C), 140.89 – 140.45 (m, 1C), 139.94 – 136.89 (m, 1C), 134.70 (1C), 130.61 (d, $^3J_{CF}$=3.8, 1C), 123.64 (d, $^3J_{CF}$=12.1, 1C), 120.11 (t, $^1J_{CF}$=266.3, 1C), 118.35 (d, $^4J_{CF}$=3.9, 1C), 113.13 (dt, $^2J_{CF}$=23.5, $^4J_{CF}$=3.1, 2C), 110.84 (d, $^2J_{CF}$=25.8, 1C), 109.14 (1C), 107.75 – 107.26 (m, 2C), 107.14 (dd, $^2J_{CF}$=25.3, $^4J_{CF}$=3.9, 2C), 106.47 (t, $^2J_{CF}$=16.3, 1C), 105.58(1C), 99.38(1C), 68.51 (1C), 56.03 (1C), 28.77 (1C), 28.11 (1C), 22.42 (1C), 14.00 (1C).

HRMS (ESI): *m/z* calcd for C$_{39}$H$_{26}$O$_7$F$_{10}$ [M+H]$^+$: 797.15916. Found: 797.15946. Difference 0.375 ppm

IR *v*max (cm$^{-1}$): 3099 (C-H stretch), 2939 (C-H stretch), 2875 (C-H stretch), 1734 (C=O stretch, ester).

**Ester GS-6-Re**

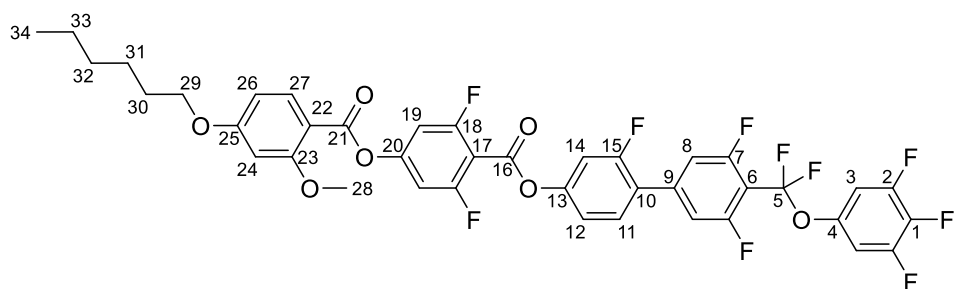

Method 2C
| | | | |
|---|---|---|---|
| Benzoic acid 2.4e | 122 mg, | 0.3 mmol, | 1.1 eq. |
| Phenol 2.6 | 112 mg, | 0.27 mmol, | 1 eq. |
| EDC.HCl | 113 mg, | 0.58 mmol, | 2 eq. |
| DMAP | 3 mg, | 0.02 mmol, | 0.1 eq. |
| DCM | 10 ml | | |

14 hrs, purified by column chromatography (2:1 DCM:Hexane) then recrystallized from EtOH. $R_f$ (DCM) 0.5. Yield 37 mg.

m.p.79 °C, $T_{SmCF-SmA}$ 78 °C, $T_{SmA-N}$ 152 °C, $T_{N-I}$ 204 °C

$^1$H NMR (400 MHz, CDCl$_3$) δ = 8.02 (d, $J$=8.7, 1H Ar-**H,** H27), 7.50 (t, $^3J_{HF}$=8.9, 1H, Ar-**H,** H14), 7.26 – 7.17 (m$_{overlapping}$, 4H, Ar-**H**, H18, H11, H13), 7.06 – 6.96 (m$_{overlapping}$, 4H, Ar-**H**, H3, H19), 6.60 – 6.51 (m$_{overlapping}$, 2H, Ar-**H**, H24, H26), 4.05 (t, $J$=6.5, 2H, O-**CH$_2$**-CH$_2$, H29), 3.94 (s, 2H, O-**CH$_3$**, H28), 1.88 – 1.77 (m, 2H, O-CH$_2$-**CH$_2$**, H30), 1.55 – 1.43 (m, 2H, O-CH$_2$-CH$_2$-**CH$_2$**, H31, H32), 1.41 – 1.32 (m, 4H, **CH$_2$-CH$_2$**-CH$_3$, H32, H33) 0.97 – 0.88 (m, 3H CH$_2$-

$^{19}$F NMR (282 MHz, CDCl$_3$) δ = -61.78 (t, $^4J_{FF}$=26.6, 2F, F5), -106.88 (d, $^3J_{HF}$=9.2, 2F, F18), -110.33 (td, $^4J_{FF}$=26.7, $^3J_{HF}$=10.4, 2F, F7), -113.60 (t, $^3J_{HF}$=9.6, 1F, F15), -132.43 (dd, $^3J_{FF}$=20.9, $^3J_{HF}$=7.5, 2F, F2), -163.12 (tt, $^3J_{FF}$=20.9, 5.8, 1F, F1).

$^{13}$C NMR (101 MHz, CDCl$_3$) δ = 165.46 (1C), 162.83 (1C), 161.90 (1C), 161.72 (dd, $^1J_{CF}$=259.2, $^3J_{CF}$=7.5, 1C), 161.53 – 158.43 (m, 2C), 159.55 (d, $^1J_{CF}$=252.6, 2C), 158.97 (s, apparent t, 1C), 155.73 – 155.19 (m, 2C), 151.67 (d, $^3J_{CF}$=11.1, 1C), 152.57 – 149.59 (m, 2C), 144.96 – 144.39 m, 1C), 140.96 – 140.51 (m, 1C), 134.73 (1C), 130.65 (d, $^3J_{CF}$=3.9, 1C), 123.64 (d, $^3J_{CF}$=11.6, 1C), 118.39 (d, $^4J_{CF}$=4.0, 1C), 113.17 (dt, $^2J_{CF}$=23.7, $^4J_{CF}$=2.2, 2C), 110.88 (d, $^2J_{CF}$=25.8, 1C), 109.21 (1C), 107.76 – 107.30 (m, 2C), 107.18 (dd, $^2J_{CF}$=25.3, $^4J_{CF}$=3.9, 2C), 106.89 – 106.42 (m, 1C), 105.65 (1C), 99.44 (1C), 68.58 (1C), 56.08 (1C), 29.08 (1C), 25.69 (1C), 22.62 (1C), 14.06 (1C).

HRMS (ESI): $m/z$ calcd for C$_{40}$H$_{28}$O$_7$F$_{10}$ [M+H]$^+$: 811.17481. Found: 811.17446. Difference -0.433 ppm

IR $v_{max}$ (cm$^{-1}$): 3101 (C-H stretch), 2940 (C-H stretch), 2873 (C-H stretch), 1738 (C=O stretch, ester).

**Ester GS-7-Re**

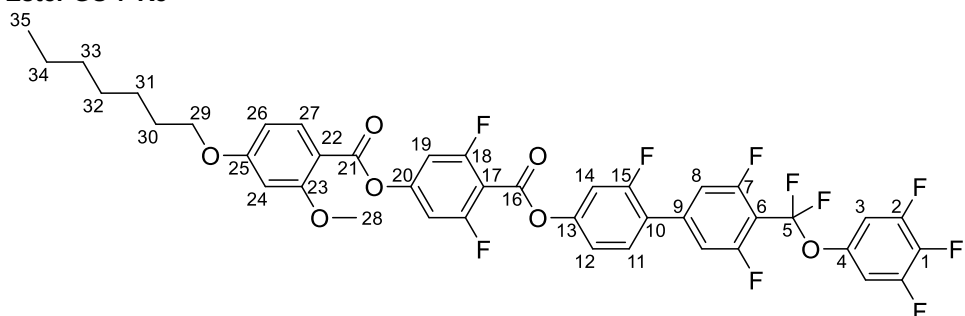

Method 2C
| | | | |
|---|---|---|---|
| Benzoic acid 2.4f | 104 mg, | 0.25 mmol, | 1.1 eq. |
| Phenol 2.6 | 95 mg, | 0.23 mmol, | 1 eq. |
| EDC.HCl | 105 mg, | 0.54 mmol, | 2.3 eq. |
| DMAP | 3 mg, | 0.02 mmol, | 0.1 eq. |
| DCM | 15 ml | | |

14 hrs, purified by column chromatography (2:1 DCM:Hexane) then recrystallized from EtOH. $R_F$ (DCM) 0.5. Yield 24 mg.

m.p.73 °C, $T_{SmCF-SmA}$ 46 °C, $T_{SmA-N}$ 160 °C, $T_{N-I}$ 196 °C

$^1$H NMR (400 MHz, CDCl$_3$) δ = 8.03 (d, *J*=8.7, 1H Ar-**H,** H27), 7.50 (t, $^3J_{HF}$=8.6, 1H, Ar-**H,** H14), 7.29 – 7.17 (m$_{overlapping}$, 4H, Ar-**H**, H18, H11, H13), 7.06 – 6.96 (m$_{overlapping}$, 4H, Ar-**H**, H3, H19), 6.60 – 6.51 (m$_{overlapping}$, 2H, Ar-**H**, H24, H26), 4.05 (t, *J*=6.6, 2H, O-**CH$_2$**-CH$_2$, H29), 3.94 (s, 2H, O-**CH$_3$**, H28), 1.83 (p$_{apparent}$, *J*=6.8, 2H, O-CH$_2$-**CH$_2$**, H30), 1.48 (p$_{apparent}$, *J*=7.0, 2H, O-CH$_2$-CH$_2$-**CH$_2$** H31, H32), 1.42 – 1.28 (m, 6H, **CH$_2$-CH$_2$-CH$_2$**-CH$_3$, H32, H33, H34) 0.95 – 0.87 (m, 3H CH$_2$-**CH$_3$**, H35).

$^{19}$F NMR (282 MHz, CDCl$_3$) δ = -61.78 (t, $^4J_{FF}$=26.5, 2F, F5), -106.89 (d, $^3J_{HF}$=9.2, 2F, F18), -110.34 (td, $^4J_{FF}$=26.5, $^3J_{HF}$=10.3, 2F, F7), -113.60 (t, $^3J_{HF}$=10.0, 1F, F15), -132.44 (dd, $^3J_{FF}$=20.9, $^3J_{HF}$=8.1, 2F, F2), -163.11 (tt, $^3J_{FF}$=20.9, 5.8, 1F, F1).

$^{13}$C NMR (101 MHz, CDCl$_3$) δ = 165.55 (1C), 162.92 (1C), 162.08 – 161.88 (m, 1C), 161.80 (dd, $^1J_{CF}$=258.2, $^3J_{CF}$=8.8, 1C), 161.46 – 158.63 (m, 2C), 161.07 – 158.04 (m, 2C), 159.04 (s, apparent t, 1C155.78 – 155.33 (m, 1C), 151.73 (d, $^3J_{CF}$=7.4, 1C), 152.60 – 149.55 (m, 2C), 144.92 – 144.40 (m, 1C), 134.80 (1C), 130.91 – 130.46 (m, 2C), 123.70 (d, $^3J_{CF}$=15.9, 1C), 120.29 – 120.06 (m, 1C), 118.63 – 118.36 (m, 1C), 113.23 (m, 2C), 110.96 (d, $^2J_{CF}$=26.4, 1C), 109.30 (1C), 107.90 – 107.39 (m, 2C), 107.41 – 106.93 (m, 2C), 105.73 (1C), 99.52 (1C), 68.66 (1C), 56.16 (1C), 31.88 (1C), 29.21 (1C), 29.1 (1C), 26.06 (1C), 22.73 (1C), 14.20 (1C).

HRMS (ESI): *m/z* calcd for C$_{41}$H$_{30}$O$_7$F$_{10}$ [M+H]$^+$: 825.19046. Found: 825.18986. Difference -0.729 ppm

IR *v*max (cm$^{-1}$): 3113 (C-H stretch), 2928 (C-H stretch), 2857 (C-H stretch), 1747 (C=O stretch, ester), 1736.

## NMR Spectra GS-*n*-Re
### GS-2-Re

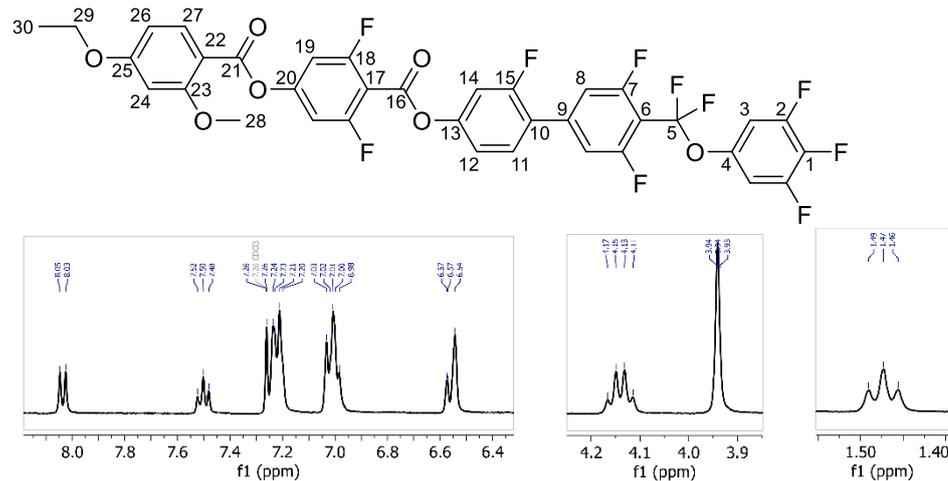

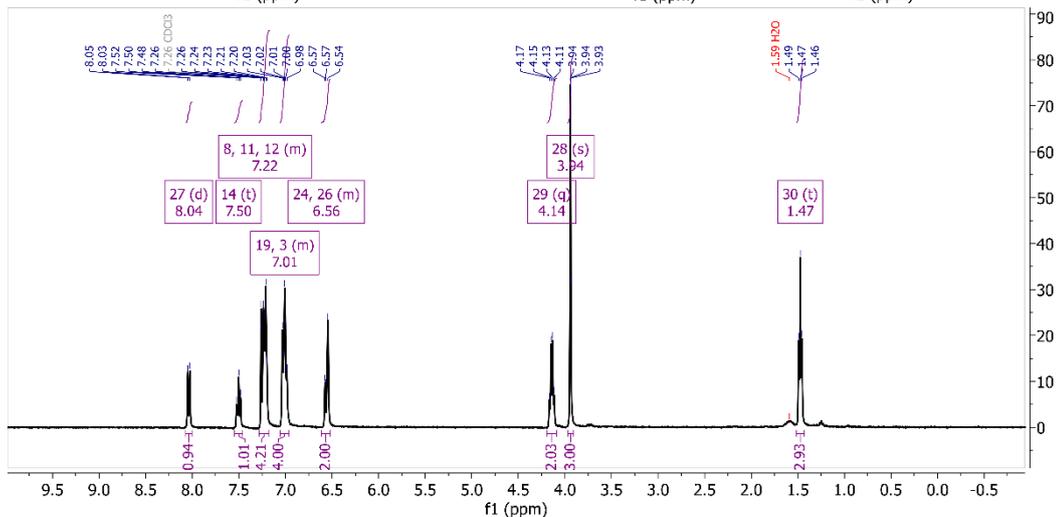

Figure S22: ¹H NMR spectrum of GS-2-Re in CDCl₃

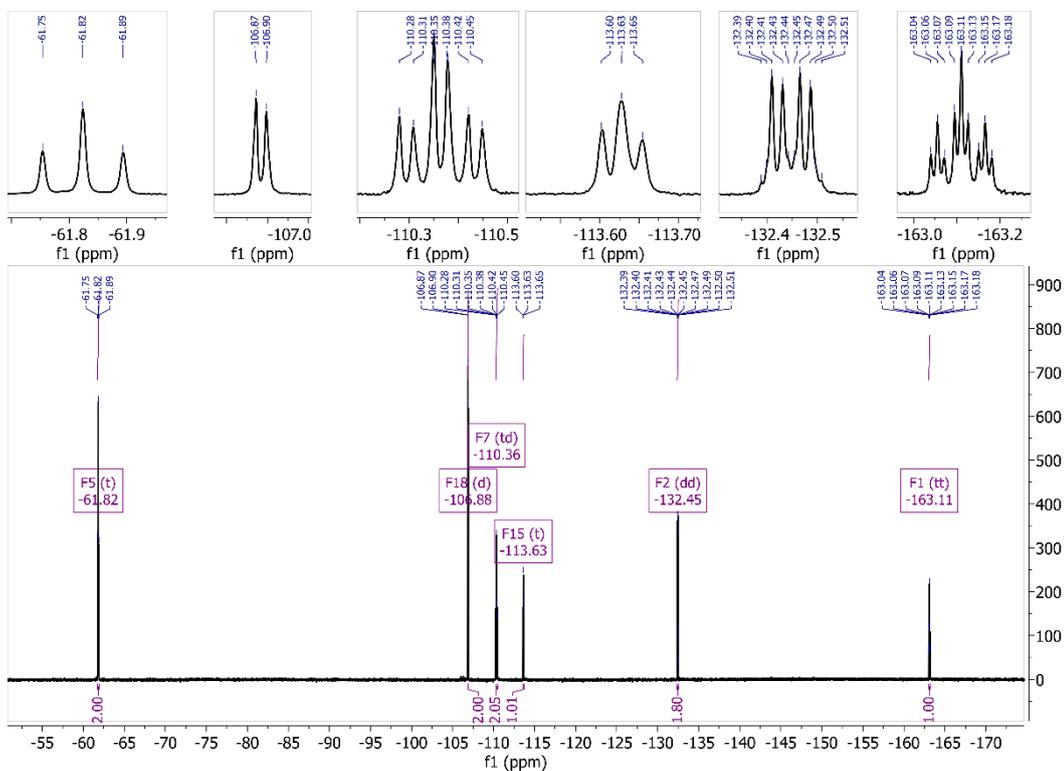

Figure S23: ¹⁹F NMR spectrum of GS-2-Re in CDCl₃

**Ester GS-3-Re**

Figure S24: $^1$H NMR spectrum of GS-3-Re in CDCl$_3$

Figure S25: $^{19}$F NMR spectrum of GS-3-Re in CDCl$_3$

**Ester GS-4-Re**

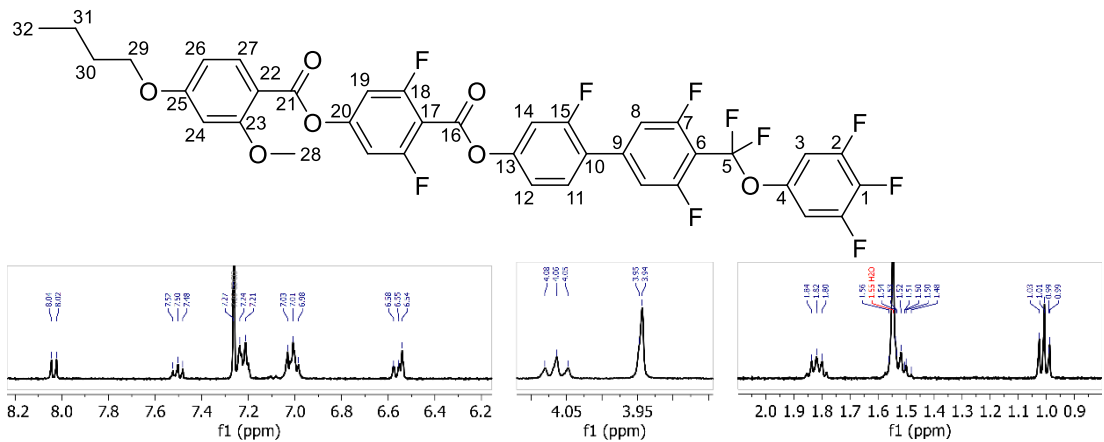

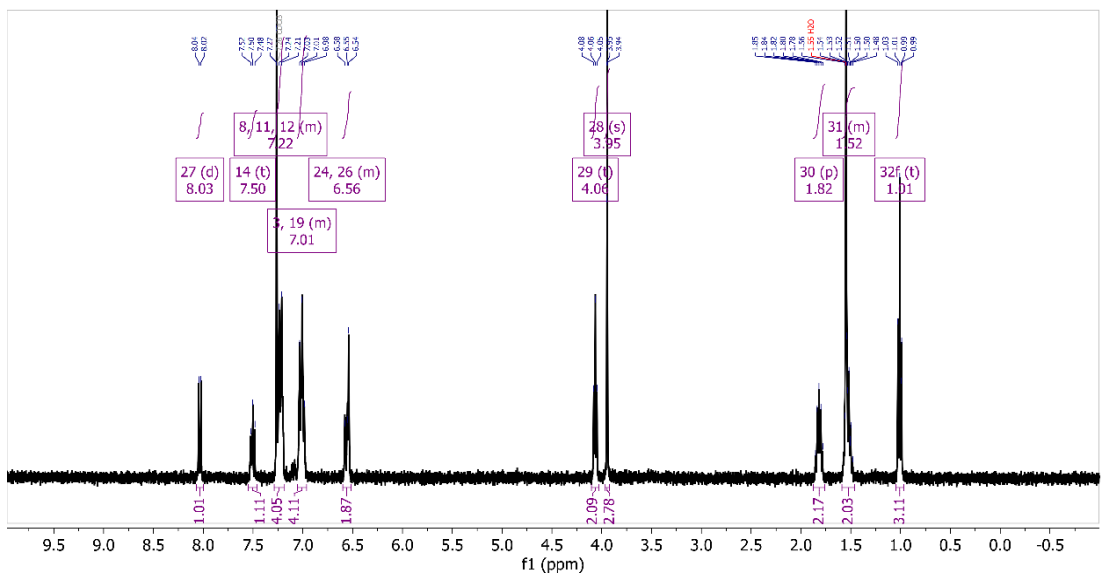

Figure S26: ¹H NMR spectrum of GS-4-Re in CDCl₃

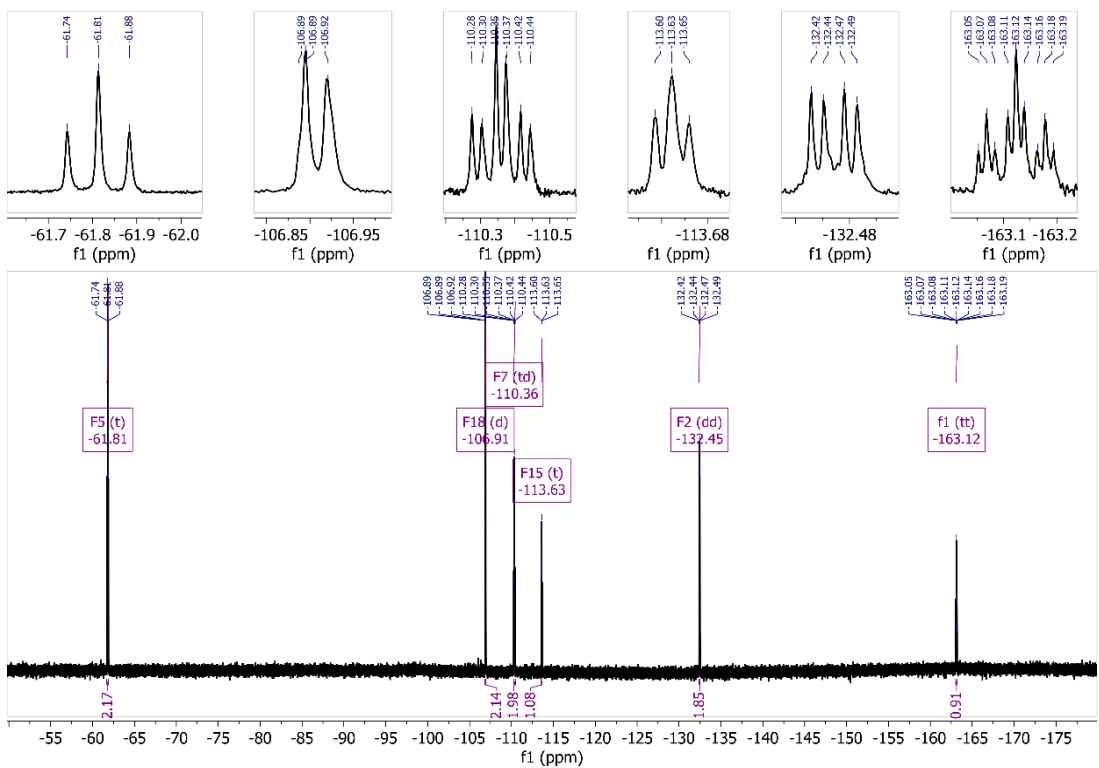

Figure S27: ¹⁹F NMR spectrum of GS-4-Re in CDCl₃

**Ester GS-5-Re**

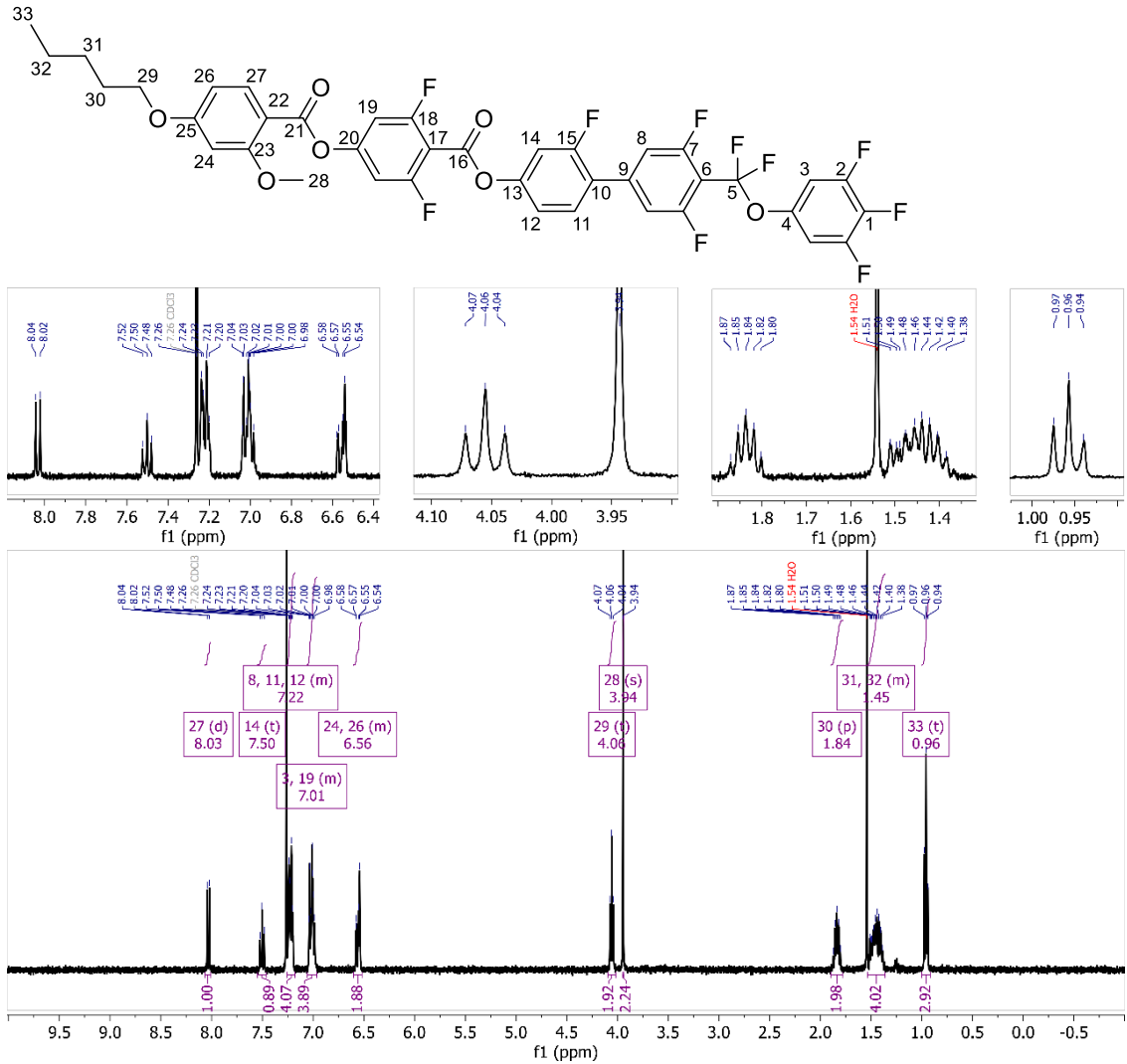

Figure S28: [1]H NMR spectrum of GS-5-Re in CDCl[3]

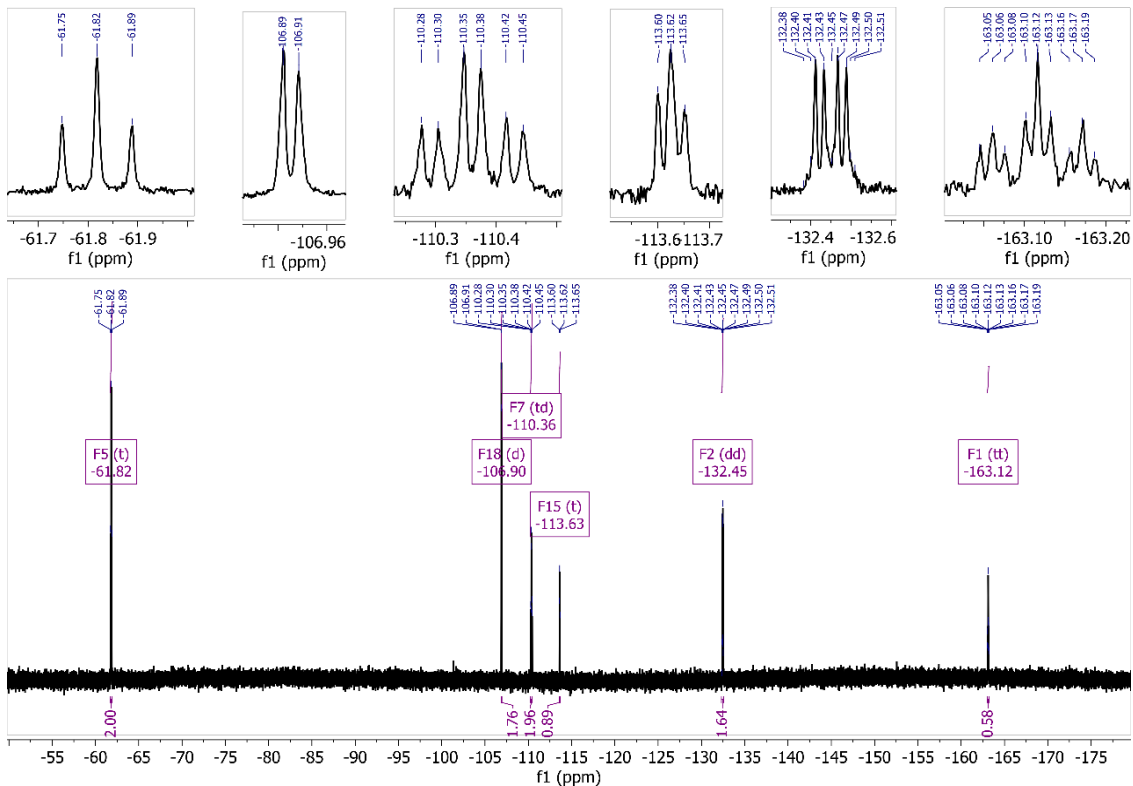

Figure S29: [19]F NMR spectrum of GS-5-Re in CDCl[3]

**Ester GS-6-Re**

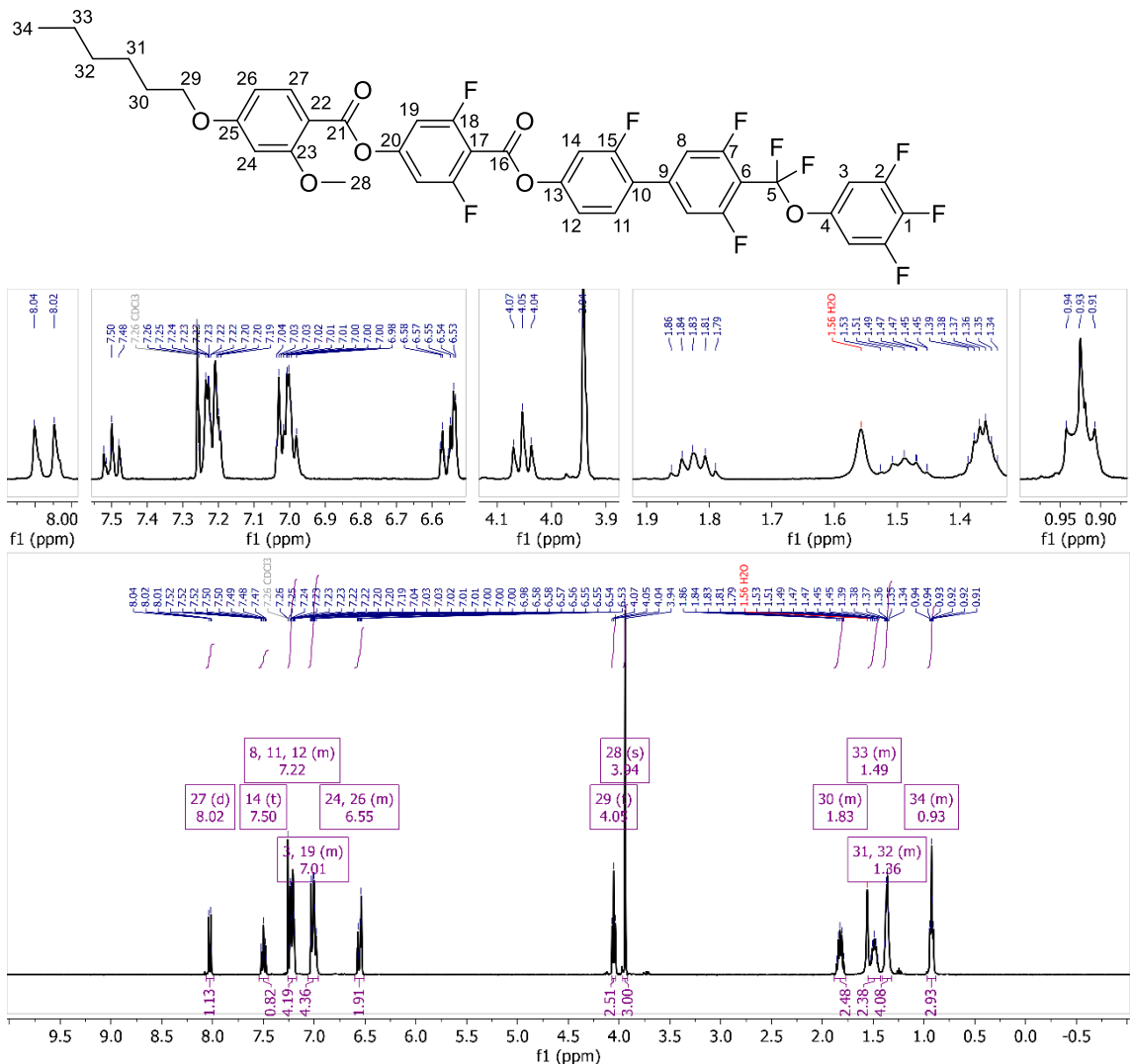

Figure S30: ¹H NMR spectrum of GS-6-Re in CDCl₃

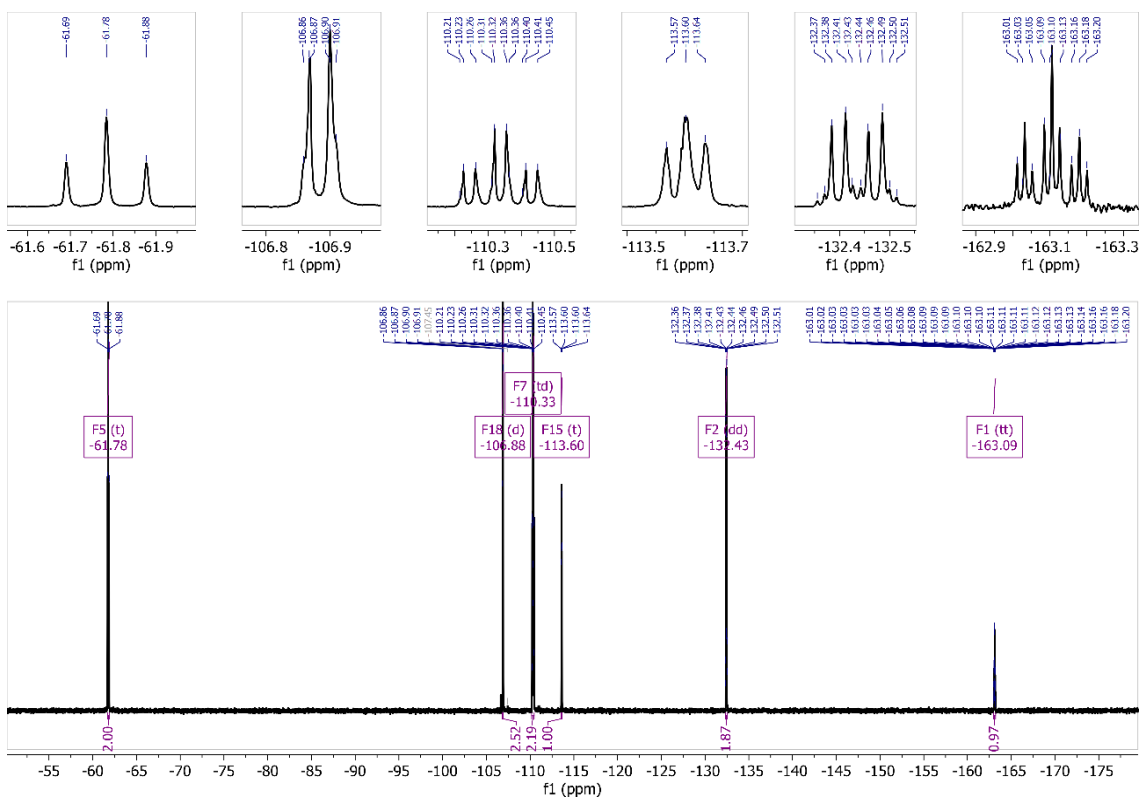

Figure S31: ¹⁹F NMR spectrum of GS-6-Re in CDCl₃

**Ester GS-7-Re**

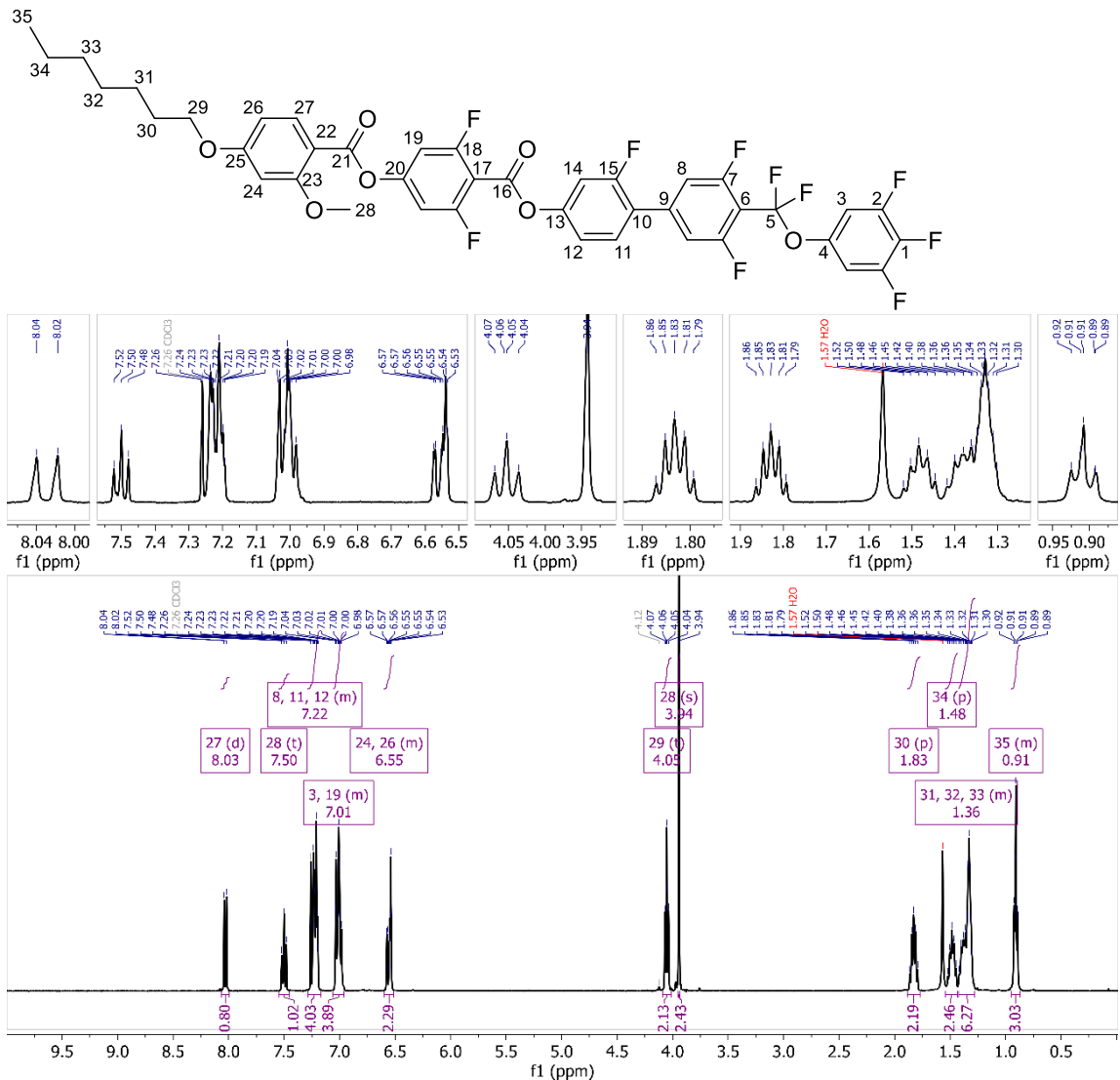

Figure S32: $^1$H NMR spectrum of GS-7-Re in CDCl$_3$

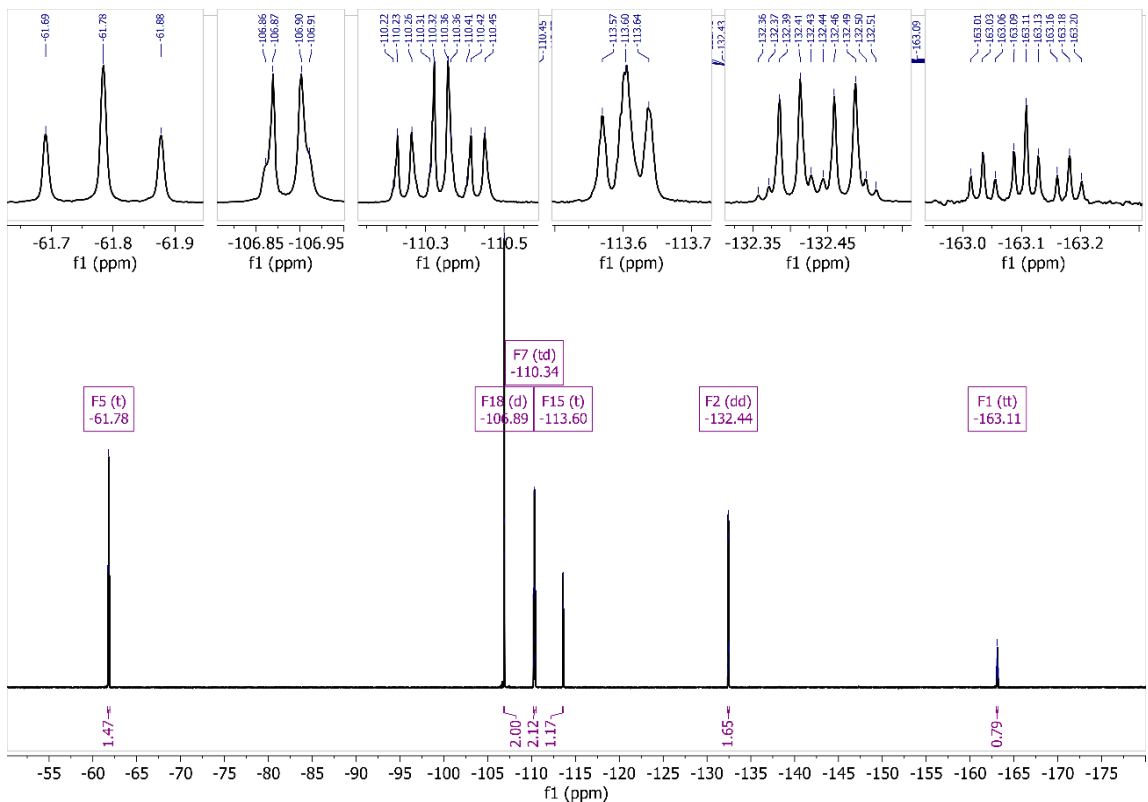

Figure S33: $^{19}$F NMR spectrum of GS-7-Re in CDCl$_3$

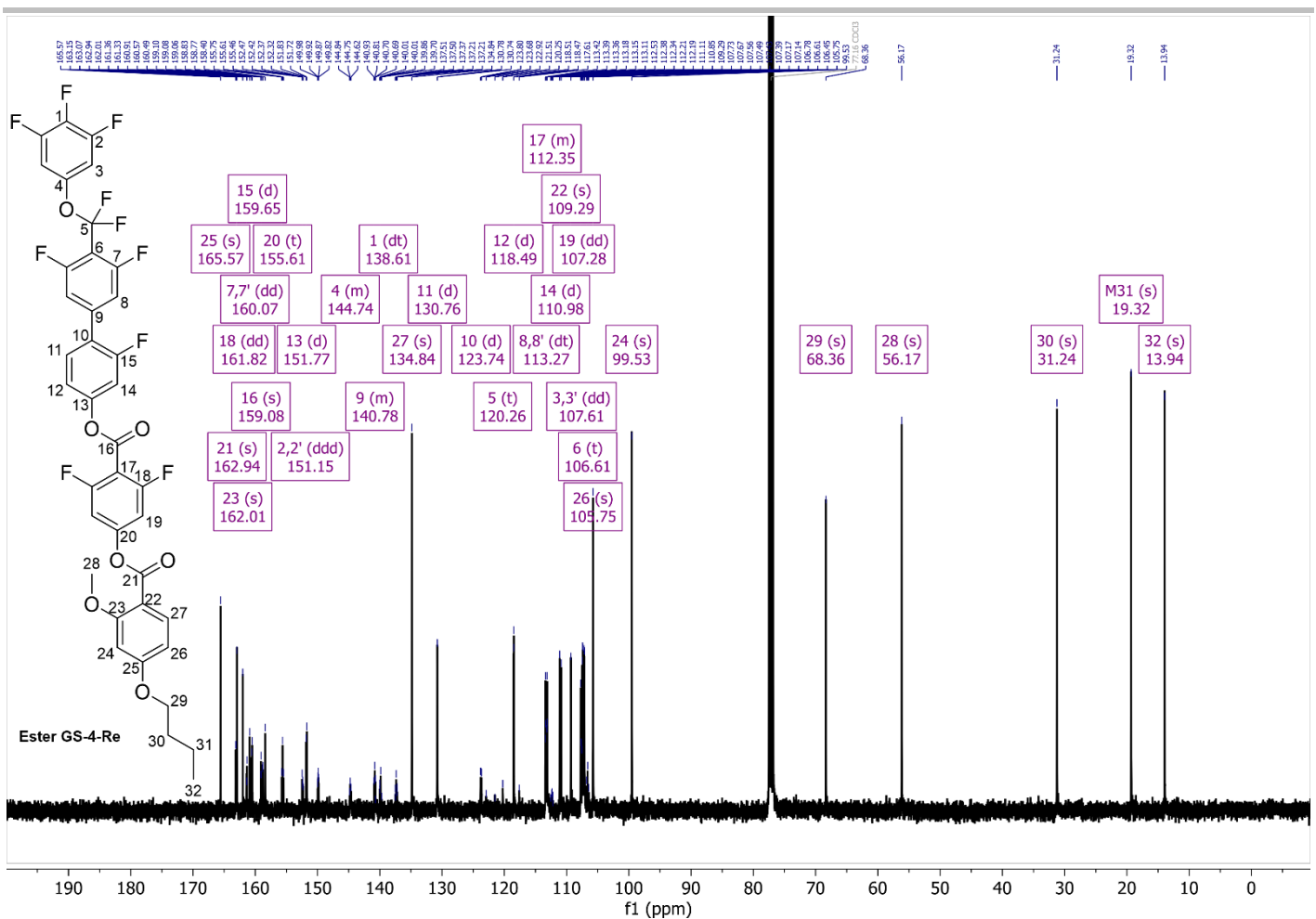
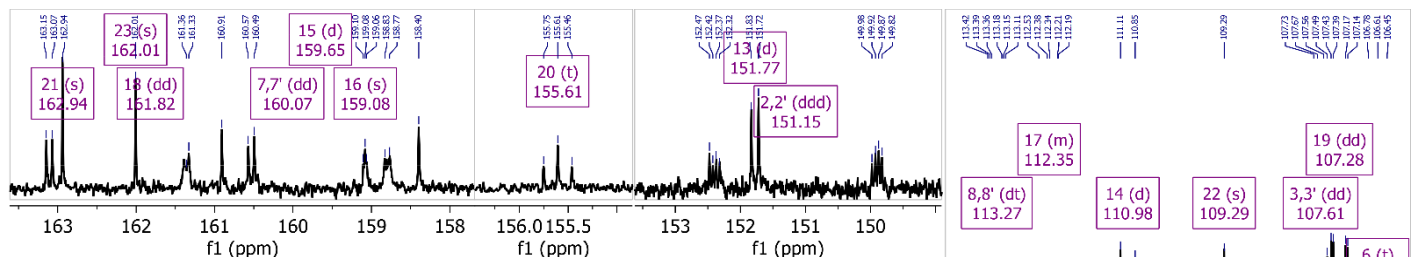
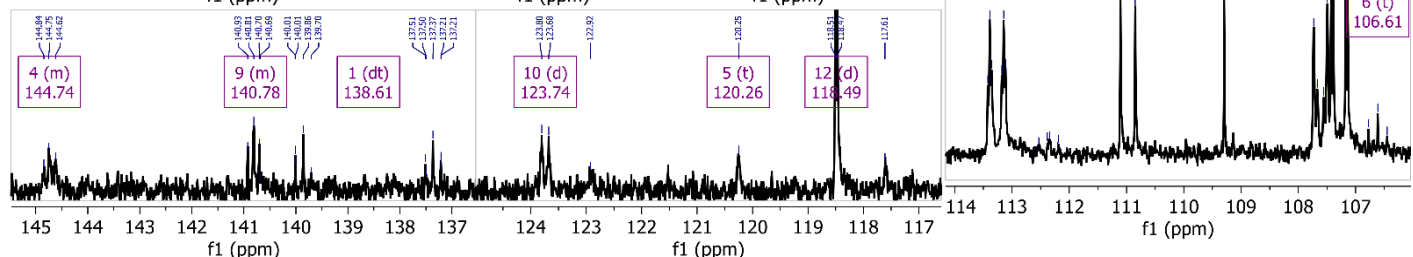
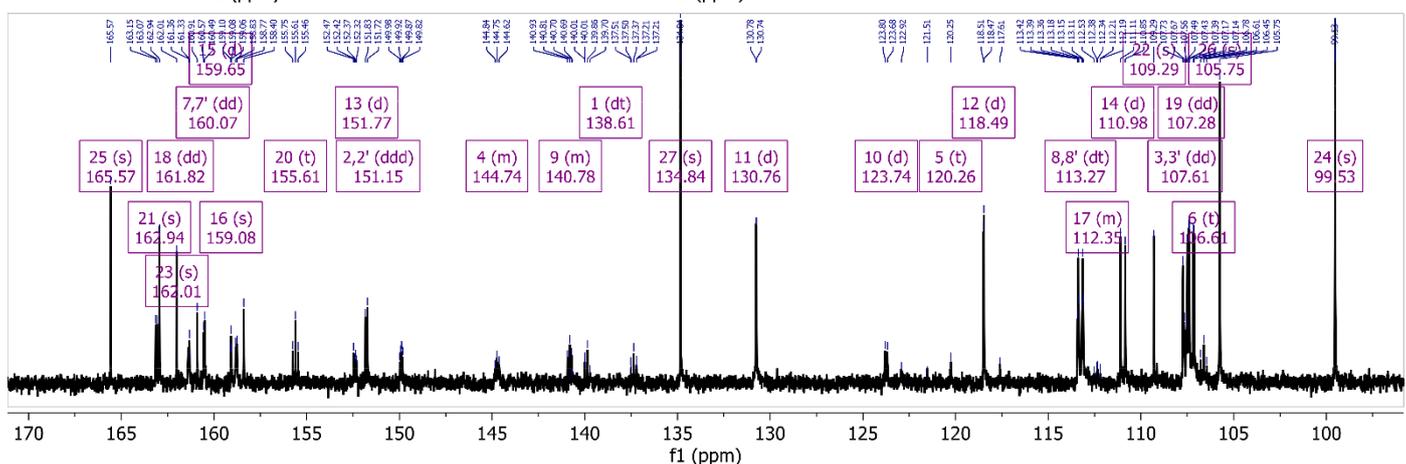

Figure S34: Representative assigned $^{13}$C NMR spectrum of GS-4-Re in CDCl$_3$. Top: Full spectrum; bottom: enlarged aromatic region.

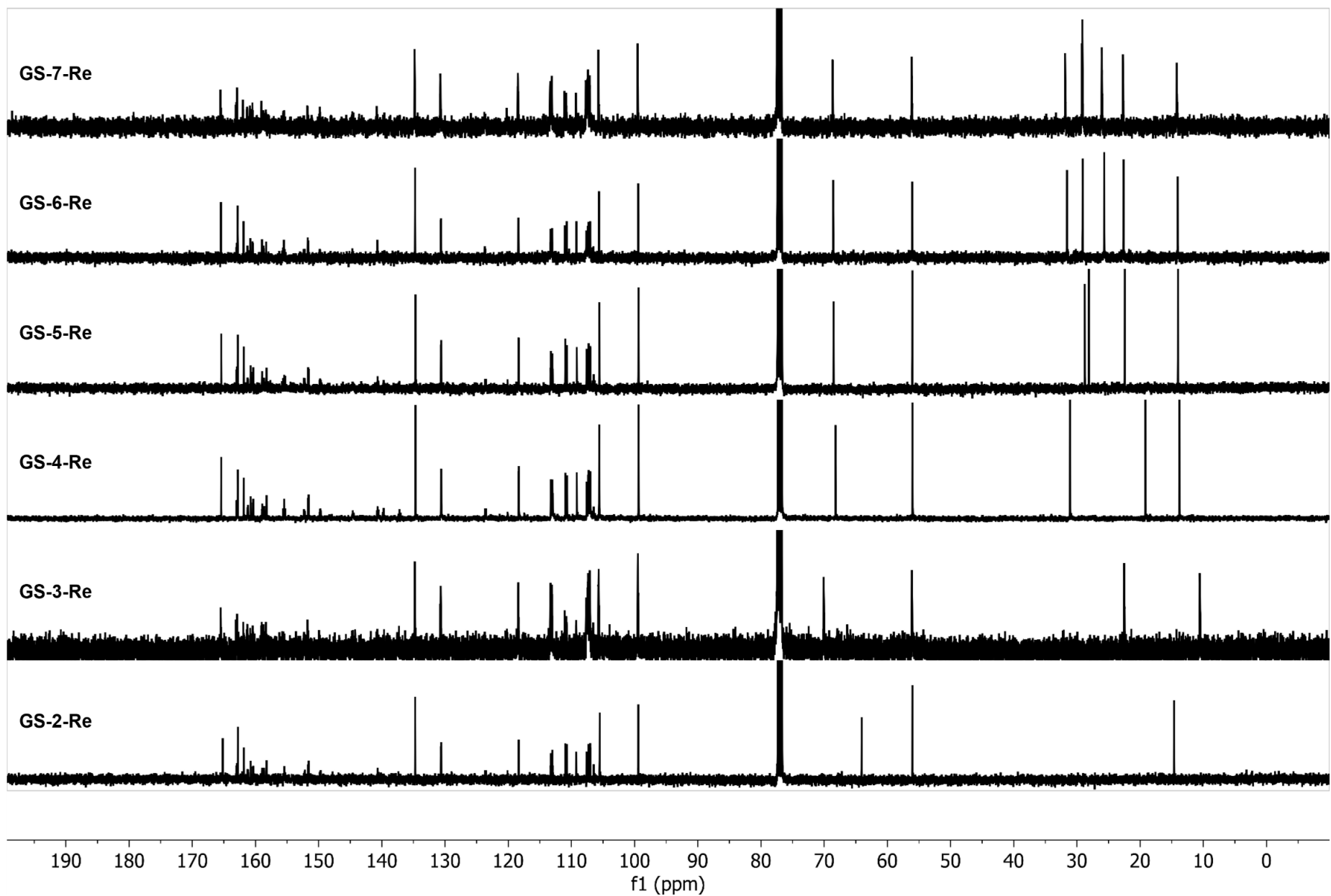

*Figure S35: Stacked $^{13}C$ NMR spectra for series GS-n-Re.*